%% file: PMSTM_arXiv.tex
\documentclass[12pt]{article}
\usepackage{adjustbox,lipsum}
\usepackage{float}
\usepackage{amssymb}
\usepackage{amsthm}
\usepackage{amsmath}
\usepackage{latexsym}
\usepackage{epsfig}
\usepackage{graphicx}
\usepackage{wasysym}
\usepackage{threeparttable}
\usepackage{natbib}
\usepackage{color}
\usepackage{epstopdf}
\usepackage{caption}
\usepackage{subcaption}

\newcommand{\ind}{\stackrel{\mathrm{ind}}{\sim}}

\usepackage[breaklinks]{hyperref}

\usepackage{enumitem}

\def\boxit#1{\vbox{\hrule\hbox{\vrule\kern6pt
          \vbox{\kern6pt#1\kern6pt}\kern6pt\vrule}\hrule}}

\def\bse{\begin{eqnarray*}}
\def\ese{\end{eqnarray*}}
\def\be{\begin{eqnarray}}
\def\ee{\end{eqnarray}}
\def\bq{\begin{equation}}
\def\eq{\end{equation}}
\def\bse{\begin{eqnarray*}}
\def\ese{\end{eqnarray*}}

\usepackage[ruled]{algorithm2e}
\usepackage{mathptmx}      
\usepackage{bm}

 {\begin{list}{}%
         {\setlength{\leftmargin}{#1}}%
         \item[]%
 }
 {\end{list}}
\usepackage{setspace}

\include{notations4}

\begin{document}
\part*{}
\thispagestyle{empty} \baselineskip=28pt

\begin{center}
{\LARGE{\bf Computationally Efficient Distribution Theory for Bayesian Inference of High-Dimensional Dependent Count-Valued Data}}
\end{center}

\baselineskip=12pt

\vskip 2mm
\begin{center}
Jonathan R. Bradley\footnote{(\baselineskip=10pt to whom correspondence should be addressed) Department of Statistics, University of Missouri, 146 Middlebush Hall, Columbia, MO 65211, bradleyjr@missouri.edu},
Scott H. Holan\footnote{\baselineskip=10pt  Department of Statistics, University of Missouri, 146 Middlebush Hall, Columbia, MO 65211-6100},
Christopher K. Wikle$^2$
\end{center}
%
%
%
%
\vskip 4mm

\begin{center}
\large{{\bf Abstract}}
\end{center}
We introduce a Bayesian approach for multivariate spatio-temporal prediction for high-dimensional count-valued data. Our primary interest is when there are possibly millions of data points referenced over different variables, geographic regions, and times. This problem requires extensive methodological advancements, as jointly modeling correlated data of this size leads to the so-called ``big $n$ problem.'' The computational complexity of prediction in this setting is further exacerbated by acknowledging that count-valued data are naturally non-Gaussian. Thus, we develop a new computationally efficient distribution theory for this setting. In particular, we introduce a multivariate log-gamma distribution and provide substantial theoretical development including: results regarding conditional distributions, marginal distributions, an asymptotic relationship with the multivariate normal distribution, and full-conditional distributions for a Gibbs sampler. To incorporate dependence between variables, regions, and time points, a multivariate spatio-temporal mixed effects model (MSTM) is used. The results in this manuscript are extremely general, and can be used for data that exhibit fewer sources of dependency than what we consider (e.g., multivariate, spatial-only, or spatio-temporal-only data). Hence, the implications of our modeling framework may have a large impact on the general problem of jointly modeling correlated count-valued data. We show the effectiveness of our approach through a simulation study. Additionally, we demonstrate our proposed methodology with an important application analyzing data obtained from the Longitudinal Employer-Household Dynamics (LEHD) program, which is administered by the U.S. Census Bureau.
\baselineskip=12pt

%
%
%

\baselineskip=12pt
\par\vfill\noindent
{\bf Keywords:} American Community Survey; Big Data; Aggregation; Quarterly Workforce Indicators; Bayesian hierarchical model; Longitudinal Employer-Household Dynamics (LEHD) program; Markov chain Monte Carlo; Non-Gaussian.
\par\medskip\noindent
\clearpage\pagebreak\newpage \pagenumbering{arabic}
\baselineskip=24pt
\onehalfspacing
\section{Introduction}The multivariate normal distribution has become a fundamental tool for statisticians, as it provides a way to incorporate dependence for Gaussian and non-Gaussian data alike. Notice that many statistical models are defined hierarchically, where the joint distribution of the data, latent processes, and unknown parameters are written as the product of a data model, a Gaussian process model, and a parameter model \citep[e.g., see][among others]{cressie-wikle-book,banerjee-etal-2004}. Thus, the switch from jointly modeling Gaussian data to, say, Poisson count data may be seen as straightforward to some - one can simply exchange a Gaussian data model with a Poisson data model using the hierarchical modeling framework. Models of this form are typically referred to as \textit{latent Gaussian process} (LGP) models; see \citet{diggle}, \citet{rue}, Sections 4.1.2 and 7.1.5 of \citet{cressie-wikle-book}, and \citet{holan_glm}, among others.

Unfortunately, LGPs are difficult to apply to high-dimensional correlated count-valued data. The difficulty arises because tuning and convergence of Markov chain Monte Carlo (MCMC) algorithms becomes non-trivial in the big correlated data setting since parameters are often highly correlated \citep[e.g., see][for a discussion on convergence issues of MCMC algorithms for LGPs]{rue}. Additionally, LGPs are not necessarily realistic for \textit{every} dataset. For example, \citet{Oliveira} shows that there are parametric limitations to the LGP paradigm for count-valued data (e.g., when spatial overdispersion is small). Thus, the primary goal of this article is to layout new computationally efficient distribution theory to jointly model correlated count-valued data that are possibly referenced over multiple variables, regions, and times.

To demonstrate the difficulties in fitting a LGP to high-dimensional count-valued data, consider the US Census Bureau's Longitudinal Employer Household Dynamics program (LEHD), which provides timely estimates of important US economic variables referred to as Quarterly Workforce Indicators (QWI). Recently, \citet{bradleyMSTM} efficiently modeled 7,530,037 jointly \textit{Gaussian} QWIs, over 40 variables (20 industries and two genders), 92 time-points, and 3,145 different counties. To do this, they develop a type of dynamic spatio-temporal model (DSTM) that is referred to as the multivariate spatio-temporal mixed effects model (MSTM). Now, suppose one is interested in predicting the mean number of people employed at the beginning of a quarter (i.e., a count-valued QWI), over all 3,145 US counties, 20 industries, and 96 quarters. This results in a dataset of size 4,089,755 (see Figure 1(a)). The LGP version of the MSTM to analyze this count-valued dataset seems trivial, as one can simply replace the Gaussian data model in \citet{bradleyMSTM} with a Poisson data model using the log-link. For illustration, say we have Poisson data with a log-link and use the same specification of the Gaussian process model and parameter models from Section 4 of \citet{bradleyMSTM}. Then, consider performing inference by implementing a Gibbs sampler with Metropolis-Hastings updates. There are many choices that one can use to tune a MCMC algorithm like this. For example, consider tuning using the Metropolis Adjusted Langevin (MALA) \citep{Tweedie}, adaptive proposals based on the Robbins-Monroe process \citep{RM}, and Log-Adaptive Proposals (LAP) \citep{Shaby}.

Figure 1(b,c,d) displays the trace plots associated with the intercept parameter using each of the aforementioned methods for tuning a Metropolis-Hastings algorithm. It is clear that convergence is not achieved; furthermore, convergence is also problematic for other parameters and random effects (not shown). In addition, the average wait-time (in seconds) for one MCMC replicate is 51 seconds when using Matlab (Version 8.0) on a dual 10 core 2.8 GHz Intel Xeon E5-2680 v2 processor, with 256 GB of RAM. This is an extremely long wait-time for one replicate and for 10,000 replicates it takes approximately 6 days (separately) to produce each of the three Markov chains in Figure 1. Of course, it is important to note that there are many confounding factors involved with citing computation times, including the software, computer, and code used for computation. It is not our goal to say that these tuning procedures are not useful, or even to say that it is not possible to calibrate them. The main point is that one willfully fails until these algorithms are calibrated appropriately.

Ideally, we would like to obtain convergence of the Markov chain Monte Carlo (MCMC) algorithm the first time it is executed, low prediction error, and a wait-time that rivals the Gaussian data setting (e.g., in \citet{bradleyMSTM} the MCMC computations took approximately 1.2 days). A multivariate spatio-temporal model for count data that achieves this wish-list is unprecedented, and requires significant development of the distribution theory for multivariate spatio-temporal count-valued data to avoid the computational issues that naturally arise when using various Metropolis-Hastings algorithms. Specifically, we introduce a class of Poisson log-gamma random variables to jointly model high-dimensional multivariate spatio-temporal count data. The main motivating feature of this modeling framework is that it incorporates dependency and results in full-conditional distributions (within a Gibbs sampler) that are easy to simulate from. To achieve this, we introduce a \textit{multivariate log-gamma distribution}, which provides these computational advantages.

This computationally efficient distribution theory could have an important impact on a number of different communities within and outside statistics. For example, count-data are ubiquitous datasets within the official statistics setting. Approximately 70$\%$ of the (aforementioned) available QWIs are count-valued. Additionally, a clear majority of the US Census Bureau's American Community Survey (ACS) period estimates are count-valued (e.g., see \url{http://factfinder.census.gov/}). High-dimensional count-valued data are also pervasive in the ecological setting \citep[e.g., see][among others]{wuJABES,Hooten} and climatology \citep[e.g., see][]{wikleanderson}. Despite the need for analyzing high-dimensional correlated (over possibly different variables, regions, and times) count data, a majority of the dependent data literature is focused on the Gaussian data setting \citep[e.g., see][among many others]{stcar,royle1999,wikle2001,banerjee,johan,finley,finley2,lindgren-2011,hughes,nychkaLK,steinr,bradleyMSTM,banerjeeNN}. Hence, the methodology presented here offers an exciting avenue that makes new research for modeling correlated count-valued data practical for modern big datasets.

%
%

There are other choices besides the LGP strategy available in the literature. For instance, an important alternative to the LGP paradigm was proposed by \citet{wolpert}, who introduced a (spatial) convolution of gamma random variables, and provide a data augmentation scheme for Gibbs sampling that yields predictions in the spatial-only setting. However, their framework can only be applied for smaller-dimensional settings (e.g., their examples include at most 19,600 observations). Additionally, in the non-spatially referenced settings some have used different types of multivariate log-gamma (and gamma) distributions as an alternative to the multivariate normal distribution \citep[e.g., see][]{Lee,multgamma,loggama}. However, there has been no distribution theory proposed to handle \textit{high-dimensional} dependent count-valued data that are possibly correlated across multiple variables, geographic regions, and times.

This mindset of developing distribution theory that is computationally efficient is clearly ideal from a practical perspective, and would promote much needed statistical research on dependent count-data. However, in some settings, deviating from LGP may result in a distributional form that does not reflect the true behavior of the underlying process. Thus, we provide a result that shows that one can specify the multivariate log-gamma distribution so that it is arbitrarily close to a multivariate normal distribution. Furthermore, we incorporate the Moran's I (MI) basis functions, MI propagator matrix, and MI prior distribution from \citet{bradleyMSTM} to better describe the dependency of latent processes. The resulting hierarchical statistical model is called the \textit{Poisson multivariate spatio-temoral mixed effects model} (P-MSTM).

The remainder of this article is organized as follows. In Section 2, we introduce a multivariate log-gamma distribution and provide the necessary  technical development of this distribution. Then, in Section 3, we use this new distribution theory to define the P-MSTM. In Section 4, we use a simulation study that is calibrated towards count-valued QWIs, so that simulated values are similar to QWIs. This empirical simulation study is used to evaluate the effectiveness of the P-MSTM at recovering unobserved ``true values.'' Additionally, we use the P-MSTM to jointly analyze 4,089,755 count-valued QWIs obtained from the US Census Bureau's LEHD program partially presented in Figure 1(a). Finally, Section 5 contains discussion. For convenience of exposition, proofs of the technical results are given in an appendix, and supporting materials, including a spatial-only example involving the American Community Survey (ACS), are provided in Supplemental Materials.

\section{Distribution Theory: The Multivariate Log-Gamma Distribution} The rudimentary quantity in our development of the multivariate log-gamma distribution is the (univariate) log-gamma random variable $q$ \citep{Prentice,multgamma,Crooks}, where
\begin{equation}\label{simp_lin}
q\equiv \mathrm{log}(\gamma),
\end{equation}
\noindent
and $\gamma$ is a gamma random variable with shape parameter $\alpha>0$ and scale parameter $\kappa>0$. There are many relationships between the log-gamma distribution and other distributions including the Gumbel distribution, the Amoroso distribution, and the normal distribution \citep[e.g., see][]{Crooks}. These relationships are derived by considering special cases of the probability density function (pdf) associated with $q$ in (\ref{simp_lin}). Straightforward change-of-variable techniques lead to the following expression for the pdf of $q$,
\begin{equation}
\label{univ_LG}
f(q\vert \alpha,\kappa) = \frac{1}{\Gamma(\alpha)\kappa^{\alpha}}\mathrm{exp}\left\lbrace\alpha q - \frac{1}{\kappa}\mathrm{exp}(q)\right\rbrace; \hspace{5pt} q \in \mathbb{R},
\end{equation}
\noindent
where $f$ will be used to denote a generic pdf and $ \mathrm{LG}(\alpha,\kappa)$ denotes a shorthand for the pdf in (\ref{univ_LG}).

The importance of the log-gamma random variable for our purpose of modeling count-valued data is transparent in the univariate setting. Let $Z\vert q\sim \mathrm{Pois}\{\mathrm{exp}(q)\}$, where ``Pois'' stands for ``Poisson,'' and notice that
\begin{equation}\label{propPois}
f(Z\vert q) \propto \mathrm{exp}\left\lbrace Zq - \mathrm{exp}(q)\right\rbrace.
\end{equation}
\noindent
It is immediate from (\ref{univ_LG}) and (\ref{propPois}) that 
\begin{equation}\label{univ_congj}
q\vert Z,\alpha,\kappa \sim \mathrm{LG}\left\lbrace Z+\alpha, \left(1+\frac{1}{\kappa}\right)^{-1}\right\rbrace.
\end{equation}
\noindent
This conjugacy between the Poisson distribution and the log-gamma distribution motivates us to develop a multivariate version of the log-gamma distribution to model multivariate spatio-temporal count data. Thus, in this section, we define a multivariate log-gamma distribution and develop a distribution theory that will prove to be extremely useful for fully Bayesian analysis in this more complex setting.

\subsection{The Multivariate Log-Gamma Distribution} \citet{wolpert} incorporate dependence in latent stages of a random spatial process by using a kernel convolution of independent gamma random variables. We take a similar approach by using a linear combination of independent log-gamma random variables. Specifically, let the $m$-dimensional random vector $\textbf{w} = (w_{1},\ldots.,w_{m})^{\prime}$ consist of $m$ mutually independent log-gamma random variables such that $w_{i}\sim\mathrm{LG}(\alpha_{i},\kappa_{i})$ for $i = 1,\ldots,m$. Then, define
\begin{equation}\label{linear_comb}
\textbf{q} = \textbf{c} + \textbf{V}\textbf{w},
\end{equation}
\noindent
where the matrix $\textbf{V} \in \mathbb{R}^{m}\times \mathbb{R}^{m}$ and $\textbf{c} \in\mathbb{R}^{m}$. Call $\textbf{q}$ in (\ref{linear_comb}) a multivariate log-gamma (MLG) random vector. The linear combination in (\ref{linear_comb}) is similar to the derivation of the multivariate normal distribution; that is, if one replaces $\textbf{w}$ with an $m$-dimensional random vector consisting of independent and identically standard normal random variables, one obtains the multivariate normal distribution \citep[e.g., see][among others]{anderson,Johnson} with mean $\bm{c}$ and covariance $\textbf{V}\textbf{V}^{\prime}$.

To use the MLG distribution in a Bayesian context, we require its pdf, which is formally stated below.\\

\noindent
\textit{Theorem 1: Let $\textbf{q}=\textbf{c} + \textbf{V}\textbf{w}$, where $\textbf{c}\in \mathbb{R}^{m}$, the $m\times m$ real valued matrix $\textbf{V}$ is invertible, and the $m$-dimensional random vector $\textbf{w} = (w_{1},\ldots,w_{m})^{\prime}$ consists of $m$ mutually independent log-gamma random variables such that $w_{i}\sim\mathrm{LG}(\alpha_{i},\kappa_{i})$ for $i = 1,\ldots,m$.
\begin{enumerate}[label=\roman*.]
 \item Then $\textbf{q}$ has the following pdf:
\begin{align}
\label{mlg_pdf}
\nonumber
& f(\textbf{q}\vert \textbf{c},\textbf{V},\bm{\alpha},\bm{\kappa}) =\\ &\frac{1}{\mathrm{det}(\textbf{V}\textbf{V}^{\prime})^{1/2}}\left(\prod_{i = 1}^{m}\frac{1}{\Gamma(\alpha_{i})\kappa_{i}^{\alpha_{i}}}\right)\mathrm{exp}\left[\bm{\alpha}^{\prime}\textbf{V}^{-1}(\textbf{q} - \textbf{c}) - \bm{\kappa}^{(-1)\prime}\mathrm{exp}\left\lbrace\textbf{V}^{-1}(\textbf{q} - \textbf{c})\right\rbrace\right];\hspace{5pt} \textbf{q} \in \mathbb{R}^{m},
\end{align}
\noindent
where ``det'' represents the determinant function, $\bm{\alpha}\equiv (\alpha_{1},\ldots,\alpha_{m})^{\prime}$, $\bm{\kappa} \equiv (\kappa_{1},\ldots,\kappa_{m})^{\prime}$, and $\bm{\kappa}^{(-1)}\equiv \left(\frac{1}{\kappa_{1}},\ldots,\frac{1}{\kappa_{m}}\right)^{\prime}$.
\item The mean and variance of $\textbf{q}$ is given by,
\begin{align}
\label{meancovar}
\nonumber
& E(\textbf{q}\vert \bm{\alpha},\bm{\kappa}) = \textbf{c}+ \textbf{V}\left\lbrace\omega_{0}(\bm{\alpha})+ \mathrm{log}(\bm{\kappa})\right\rbrace\\
& \mathrm{cov}(\textbf{q}\vert \bm{\alpha},\bm{\kappa}) = \textbf{V}\hspace{2pt}\mathrm{diag}\left\lbrace\omega_{1}(\bm{\alpha})\right\rbrace\textbf{V}^{\prime},
\end{align}
\noindent
where for a generic $m$-dimensional real-valued vector $\textbf{k} = (k_{1},\ldots,k_{m})^{\prime}$, $\mathrm{diag}\left(\textbf{k}\right)$ denote an $m\times m$ dimensional diagonal matrix with main diagonal equal to $\textbf{k}$. The function $\omega_{j}(\textbf{k})$, for non-negative integer $j$, is a vector-valued polygamma function, where the $i$-th element of $\omega_{j}(\textbf{k})$ is defined to be $\frac{d^{j+1}}{dk_{i}^{j+1}} \mathrm{log}\left(\Gamma(k_{i})\right)$ for $i = 1,\ldots,m$.\\
\end{enumerate}
}

\noindent
The proof of Theorem~1($i$) can be found in the Appendix, and in general, let $ \mathrm{MLG}(\textbf{c},\textbf{V},\bm{\alpha},\bm{\kappa})$ be shorthand for the pdf in (\ref{mlg_pdf}). When comparing (\ref{univ_LG}), (\ref{propPois}), and (\ref{mlg_pdf}) we see that the univariate log-gamma pdf, the Poisson pdf, and the multivariate log-gamma pdf share a basic structure. Specifically, all three pdfs have an exponential term and a double exponential term. This pattern is the main reason why conjugacy exists between the Poisson distribution and the log gamma distribution, which we take advantage of in subsequent sections.

An important difference in the parameterization of MLG random vectors and multivariate normal random vectors, is that MLG random vectors have additional shape and scale parameters. This can be seen in Theorem~1($ii$), which shows that the mean of $\textbf{q}$ depends on $\textbf{c}$, $\bm{\alpha}$, and $\bm{\kappa}$; similarly, the covariance matrix of $\textbf{q}$ depends on both $\textbf{V}$ and $\bm{\alpha}$. This could potentially be problematic when performing posterior statistical inference due to identifiability considerations. Thus, in Section~2.2, we provide a specification of $\bm{\alpha}$ and $\bm{\kappa}$ so that the mean and variance of the MLG random vector $\textbf{q}$ depends only on the parameters $\textbf{c}$ and $\textbf{V}$.
 
 \subsection{Shape and Scale Parameter Specification} We now provide guidance on how to specify the shape and scale parameters $\bm{\alpha}$ and $\bm{\kappa}$ to define a two parameter (i.e., $\textbf{c}$ and $\textbf{V}$) MLG distribution. Our first consideration is to remove the functional dependence of the mean and covariance of a MLG random vector on $\bm{\alpha}$ and $\bm{\kappa}$ (see Equation (\ref{meancovar})). To do this, we standardize the vector of independent log-gamma random variables $\textbf{w}$ in (\ref{linear_comb}). Specifically, define $\alpha^{*}$ as the value such that $\omega_{1}(\alpha^{*}) = 1$, which is given by $\alpha^{*}\approx 1.4263$ (using the Newton-Raphson method). Then, consider $\textbf{q} \sim \mathrm{MLG}(\textbf{c},\textbf{V},\alpha^{*}\bm{1}_{m},\mathrm{exp}(-\omega_{0}(\alpha^{*}))\bm{1}_{m})$, where $\bm{1}_{m}$ is an $m$-dimensional vector of 1s. From Theorem~1($ii$) it follows that this specification of $\textbf{q}$ has mean $\textbf{c}$ and covariance $\textbf{V}\textbf{V}^{\prime}$. We refer to this two-parameter multivariate log-gamma distribution as the standardized multivariate log-gamma (sMLG) distribution, and is denoted as MLG$(\textbf{c},\textbf{V},\alpha^{*}\bm{1}_{m},\mathrm{exp}(-\omega_{0}(\alpha^{*}))\bm{1}_{m})$.
 
Both the MLG and sMLG distributions are non-standard; that is, it is more common to assume a multivariate normal distribution to incorporate dependence for count-valued data \citep{diggle}. Hence, we investigate a connection between the multivariate log-gamma distribution and the multivariate normal distribution.\\
 
 \noindent
  \textit{Proposition 1: Let $\textbf{q} \sim \mathrm{MLG}(\textbf{c},\alpha^{1/2}\textbf{V},\alpha\bm{1},\frac{1}{\alpha}\bm{1})$. Then $\textbf{q}$ converges in distribution to a multivariate normal random vector with mean $\textbf{c}$ and covariance matrix $\textbf{V}\textbf{V}^{\prime}$ as $\alpha$ goes to infinity.\\
  }
  
  \noindent
Figure 2 displays an approximation of a normal distribution using a log-gamma random variable based on Proposition 1. Here, we see a situation where the log-gamma distribution can provide an excellent approximation to a normal distribution - even for values that occur far out in the tails of the standard normal distribution. 
  
  The asymptotic result in Proposition 1 is on the shape parameter, which can be specified as any value that one would like. Thus, Proposition 1 provides motivation for the use of the multivariate log-gamma distribution, since one can always specify a MLG distribution to be ``close'' to the commonly used multivariate normal distribution. In practice, we have found that $\alpha = 1000$ to be sufficiently large; however, one must verify an appropriate value of $\alpha$ for their specific setting. Refer to this two-parameter MLG distribution as the normal approximation of the multivariate log-gamma (nMLG) distribution, and define it as MLG($\textbf{c},\alpha_{\mathrm{G}}^{1/2}\textbf{V},\alpha_{\mathrm{G}}\bm{1},\frac{1}{\alpha_{\mathrm{G}}}\bm{1})$ for $\alpha_{\mathrm{G}}$ very large; note that the ``n'' in ``nMLG'' stands for ``normal'' and the ``G'' in $\alpha_{\mathrm{G}}$ stands for ``Gaussian.'' In practice, we take $\alpha_{\mathrm{G}}= 1000$.

\subsection{Conditional and Marginal Distributions for Multivariate Log-Gamma Random Vectors} In this section, technical results are presented on conditional and marginal distributions for an $m$-dimensional MLG random vector $\textbf{q}$. Bayesian inference of count-valued data will require simulating from conditional distributions of $\textbf{q}$. Thus, we provide the technical results needed to simulate from these conditional distributions.\\

\noindent
\textit{Proposition 2: Let $\textbf{q} \sim \mathrm{MLG}(\textbf{c},\textbf{V},\bm{\alpha},\bm{\kappa})$, and let $\textbf{q} = (\textbf{q}_{1}^{\prime}, \textbf{q}_{2}^{\prime})^{\prime}$, so that $\textbf{q}_{1}$ is $g$-dimensional and $\textbf{q}_{2}$ is $(m-g)$-dimensional. In a similar manner, partition $\textbf{V}^{-1} = [\textbf{H}\hspace{5pt}\textbf{B}]$ into an $m\times g$ matrix $\textbf{H}$ and an $m\times (m-g)$ matrix $\textbf{B}$.
\begin{enumerate}[label=\roman*.]
\item Then, the conditional pdf of $\textbf{q}_{1}\vert \textbf{q}_{2} = \textbf{d},\textbf{c},\textbf{V},\bm{\alpha},\bm{\kappa}$ is given by
\begin{align}
\label{cond_pdf}
f(\textbf{q}_{1}\vert \textbf{q}_{2} = \textbf{d},\textbf{c},\textbf{V},\bm{\alpha},\bm{\kappa})&=f(\textbf{q}_{1}\vert \textbf{H},\bm{\alpha},\bm{\kappa}_{1.2})=M\hspace{5pt}\mathrm{exp}\left\lbrace\bm{\alpha}^{\prime}\textbf{H}\textbf{q}_{1} - \bm{\kappa}_{1.2}^{(-1)\prime}\mathrm{exp}(\textbf{H}\textbf{q}_{1})\right\rbrace,
\end{align}
where $\bm{\kappa}_{1.2}^{(-1)}\equiv \mathrm{exp}\left\lbrace\textbf{B}\textbf{d} - \textbf{V}^{-1}\textbf{c} + \mathrm{log}(\bm{\kappa}^{(-1)})\right\rbrace$ and the normalizing constant $M$ is
\begin{equation*}
M =  \frac{1}{\mathrm{det}(\textbf{V}\textbf{V}^{\prime})^{1/2}}\left(\prod_{i = 1}^{m}\frac{1}{\Gamma(\alpha_{i})\kappa_{i}^{\alpha_{i}}}\right)\frac{\mathrm{exp}\left(\bm{\alpha}^{\prime}\textbf{B}\textbf{d} - \bm{\alpha}^{\prime}\textbf{V}^{-1}\textbf{c}\right)}{\left[\int f(\textbf{q}\vert \textbf{c},\textbf{V},\bm{\alpha},\bm{\kappa})d\textbf{q}_{1}\right]_{\textbf{q}_{2} = \textbf{d}}}.
\end{equation*}
\item The conditional random vector $\textbf{q}_{1}\vert \textbf{q}_{2} = \bm{0}_{m-g},\textbf{c}=\bm{0}_{m},\textbf{V},\bm{\alpha},\bm{\kappa}_{1.2}$ is equal in distribution to $\textbf{q}_{1}\vert \textbf{q}_{2} = \textbf{d},\textbf{c},\textbf{V},\bm{\alpha},\bm{\kappa}$.\\
\end{enumerate}
}

%

\noindent
\textit{Remark 1:} Let $\mathrm{cMLG}(\textbf{H},\bm{\alpha},\bm{\kappa}_{1.2})$ be a shorthand for the pdf in (\ref{cond_pdf}), where ``cMLG'' stands for ``conditional multivariate log-gamma.'' Proposition 1($i$) shows that cMLG does not fall within the same class of pdfs as the joint distribution given in (\ref{mlg_pdf}). This is primarily due to the fact that the $m\times g$ real-valued matrix $\textbf{H}$, within the expression of cMLG in (\ref{cond_pdf}), is not square. This property is different from the multivariate normal distribution, where both marginal and conditional distributions obtained from a multivariate normal random vector, are multivariate normal \citep[e.g., see][among others]{anderson,Johnson}. The fact that cMLG in (\ref{cond_pdf}) is not MLG is especially important for posterior inference because we will need to simulate from cMLG, and we cannot use (\ref{linear_comb}) to do this. Thus, we require an additional result that allows us to simulate from cMLG.\\
 

\noindent
 \textit{Theorem 2: Let $\textbf{q} \sim \mathrm{MLG}(\bm{0}_{m},\textbf{V},\bm{\alpha},\bm{\kappa})$, and partition this $m$-dimensional random vector so that $\textbf{q} = (\textbf{q}_{1}^{\prime}, \textbf{q}_{2}^{\prime})^{\prime}$, where $\textbf{q}_{1}$ is $g$-dimensional and $\textbf{q}_{2}$ is $(m-g)$-dimensional. Additionally, consider the class of MLG random vectors that satisfy the following:
 \begin{equation}\label{special_precision}
 \textbf{V}^{-1}=\left[
 \begin{array}{cc}
 \textbf{Q}_{1} & \textbf{Q}_{2}
 \end{array}\right]
 \left[
  \begin{array}{cc}
  \textbf{R}_{1} & \bm{0}_{g,m-g} \\ 
  \bm{0}_{m-g,g} & \frac{1}{\sigma_{2}}\textbf{I}_{m-g},
  \end{array}\right]
 \end{equation}
 where in general $\bm{0}_{k,b}$ is a $k\times b$ matrix of zeros; $\textbf{I}_{m-g}$ is a $(m-g)\times (m-g)$ identity matrix;
  \begin{equation*}
  \textbf{H}=\left[
  \begin{array}{cc}
  \textbf{Q}_{1} & \textbf{Q}_{2}
  \end{array}\right]
  \left[
   \begin{array}{c}
   \textbf{R}_{1} \\ 
   \bm{0}_{m-g,g} ,
   \end{array}\right]
  \end{equation*}
 is the QR decomposition of the $m\times g$ matrix $\textbf{H}$; the $m\times g$ matrix $\textbf{Q}_{1}$ satisfies $\textbf{Q}_{1}^{\prime}\textbf{Q}_{1} = \textbf{I}_{g}$, the $m\times (m-g)$ matrix $\textbf{Q}_{2}$ satisfies $\textbf{Q}_{2}^{\prime}\textbf{Q}_{2} = \textbf{I}_{m-g}$ and $\textbf{Q}_{2}^{\prime}\textbf{Q}_{1} = \bm{0}_{m-g,g}$; $\textbf{R}_{1}$ is a $g\times g$ upper triangular matrix; and $\sigma_{2}>0$. Then, the following statements hold.
 \begin{enumerate}[label=\roman*.]
 \item The marginal distribution of the $g$-dimensional random vector $\textbf{q}_{1}$ is given by
 \begin{align}
 \label{marg_pdf}
 f(\textbf{q}_{1}\vert \textbf{H},\bm{\alpha},\bm{\kappa})&=M_{1}\hspace{5pt}\mathrm{exp}\left\lbrace\bm{\alpha}^{\prime}\textbf{H}\textbf{q}_{1} - \bm{\kappa}^{(-1)\prime}\mathrm{exp}(\textbf{H}\textbf{q}_{1})\right\rbrace,
 \end{align}
 where the normalizing constant $M_{1}$ is
 \begin{equation*}
 M_{1} =  \mathrm{det}\left([\textbf{H}\hspace{5pt}\textbf{Q}_{2}]\right)\left(\prod_{i = 1}^{m}\frac{1}{\Gamma(\alpha_{i})\kappa_{i}^{\alpha_{i}}}\right)\frac{1}{\left[\int f(\textbf{q}\vert \bm{0}_{m},\textbf{V} = [\textbf{H}\hspace{5pt}\textbf{Q}_{2}]^{-1},\bm{\alpha},\bm{\kappa})d\textbf{q}_{1}\right]_{\textbf{q}_{2} = \bm{0}_{m-g}}}.
 \end{equation*}
  \item The $g$-dimensional random vector $\textbf{q}_{1}$ is equal in distribution to $(\textbf{H}^{\prime}\textbf{H})^{-1}\textbf{H}^{\prime}\textbf{w}$, where the $m$-dimensional random vector $\textbf{w} \sim \mathrm{MLG}(\bm{0}_{m},\textbf{I}_{m},\bm{\alpha},\bm{\kappa})$.\\
 \end{enumerate}
 }

\noindent
 \textit{Remark 2:} In subsequent sections, let $\mathrm{mMLG}(\textbf{H},\bm{\alpha},\bm{\kappa})$ be a shorthand for the pdf in (\ref{marg_pdf}), where ``mMLG'' stands for ``marginal multivariate log-gamma.'' From Proposition 2($i$) and Theorem~2($i$) it is evident that \textit{this particular class} of marginal distributions falls into the \textit{same} class of distributions as the conditional distribution of $\textbf{q}_{1}$ given $\textbf{q}_{2}$. That is, from Proposition 2($i$) and Theorem~2($i$) $\mathrm{mMLG}(\textbf{H},\bm{\alpha},\bm{\kappa}) = \mathrm{cMLG}(\textbf{H},\bm{\alpha},\bm{\kappa})$. This equality is important because Theorem~2($ii$) provides a way to simulate from mMLG, and hence cMLG. \\

\noindent
 \textit{Remark 3:} Theorem~2($ii$) shows that it is (computationally) easy to simulate from mMLG (and equivalently cMLG) provided that $g \ll m$. Recall that $\textbf{H}$ is $m\times g$, which implies that computing the $g\times g$ matrix $(\textbf{H}^{\prime}\textbf{H})^{-1}$ is computationally feasible when $g$ is ``small.''\\
 
\noindent
 \textit{Remark 4:} If $\textbf{H}$ is square then from (\ref{linear_comb}) we have that $(\textbf{H}^{\prime}\textbf{H})^{-1}\textbf{H}^{\prime}\textbf{w} = \textbf{H}^{-1}\textbf{w}$, which follows a MLG distribution with precision matrix $\textbf{H}$ (if using the sMLG specification). Thus, the class of marginal distributions in Theorem~2($i$) can be seen as more general than the MLG random vector in (\ref{linear_comb}), since Theorem~2($ii$) allows for a non-square ``precision parameter matrix.''
 
 \subsection{Example Full-Conditional Distribution} We are now ready to apply the MLG distribution theory for posterior statistical inference. In this section, our goal is to provide a straightforward example of the use of the MLG distribution to model dependent count-valued data. There are many simplifying assumptions made within this section; specifically, that both $\textbf{c}$ and $\textbf{V}$ are known. This, of course, is for illustrative purposes, and more complex modeling strategies based on the MLG distribution will be introduced in Section 3.
 
 For $i = 1,..,m$, let $Z_{i}\vert q_{i}\ind \mathrm{Pois}\{\mathrm{exp}(q_{i})\}$, and let $\textbf{q} \equiv (q_{1},...,q_{m})^{\prime} \sim \mathrm{MLG}\left(\textbf{c}, \textbf{W}, \alpha_{k}\bm{1}_{m}, \kappa_{k}\bm{1}_{m}\right)$, where $\textbf{W} \equiv \alpha_{\mathrm{G}}^{k/2}\textbf{V}$, $\textbf{V} \in \mathbb{R}^{m}\times\mathbb{R}^{m}$, ${\alpha}_{k} \equiv \alpha_{\mathrm{G}}^{k}(\alpha^{*})^{1-k}$, and ${\kappa}_{k} \equiv \left(\frac{1}{\alpha_{\mathrm{G}}}\right)^{k}\left[\mathrm{exp}\left\lbrace-\omega_{0}(\alpha^{*})\right\rbrace\right]^{1-k}$; $ k = 0,1$. Notice, that if $k = 0$ then $\textbf{q}$ is based on the sMLG specification and if $k = 1$ then $\textbf{q}$ is based on the nMLG specification. It follows from (\ref{propPois}) that
 \begin{align}\label{simple_poisson}
 f(\bz\vert \textbf{q}) &= \prod_{i = 1}^{m} f(Z_{i}\vert q_{i})\propto \mathrm{exp}\left\lbrace\sum_{i = 1}^{m}Z_{i}q_{i}-\sum_{i = 1}^{m}\mathrm{exp}(q_{i})\right\rbrace
 = \mathrm{exp}\left\lbrace\bz^{\prime}\textbf{q}-\bm{1}_{m}^{\prime}\mathrm{exp}(\textbf{q})\right\rbrace,
 \end{align}
\noindent
where $\textbf{Z}\equiv (Z_{1},...,Z_{m})^{\prime}$. From Theorem 1($i$) we have that 
 \begin{align}\label{marginal}
 f(\textbf{q})
 &\propto \mathrm{exp}\left[\alpha_{k}\bm{1}_{m}^{\prime}\textbf{W}^{-1}\textbf{q}-\frac{1}{\kappa_{k}}\bm{1}_{m}^{\prime}\mathrm{exp}\left\lbrace\textbf{W}^{-1}(\textbf{q}-\textbf{c})\right\rbrace\right]=\mathrm{exp}\left[\alpha_{k}\bm{1}_{m}^{\prime}\textbf{W}^{-1}\textbf{q}-\frac{1}{\kappa_{k}}\mathrm{exp}\left\lbrace-\textbf{W}^{-1}\textbf{c}\right\rbrace^{\prime}\mathrm{exp}\left\lbrace\textbf{W}^{-1}\textbf{q}\right\rbrace\right].
 \end{align}
 \noindent
Then, define the $2m \times m$ matrix
\begin{equation*}
\textbf{H}\equiv\left[\begin{array}{c}
      \textbf{I}_{m} \\ 
      \textbf{W}^{-1}
      \end{array}\right].
\end{equation*}
\noindent
 Using (\ref{simple_poisson}) and (\ref{marginal}) we have that
  \begin{align}\label{conjugate}
  \nonumber
  f(\textbf{q}\vert \bz) &\propto f(\bz\vert \textbf{q})f(\textbf{q})\propto \mathrm{exp}\left\lbrace\bz^{\prime}\textbf{q}+ \alpha_{k}\bm{1}_{m}^{\prime}\textbf{W}^{-1}\textbf{q}-\bm{1}_{m}^{\prime}\mathrm{exp}(\textbf{q})-\frac{1}{\kappa_{k}}\mathrm{exp}(-\textbf{W}^{-1}\textbf{c})^{\prime}\mathrm{exp}(\textbf{W}^{-1}\textbf{q})\right\rbrace\\
  \nonumber
  &= \mathrm{exp}\left\lbrace \bz^{\prime}\textbf{q}+d\bm{1}_{m}^{\prime}\textbf{q}+ \alpha_{k}\bm{1}_{m}^{\prime}\textbf{W}^{-1}\textbf{q}-d\bm{1}_{m}^{\prime}\textbf{W}\textbf{W}^{-1}\textbf{q}-\bm{1}_{m}^{\prime}\mathrm{exp}(\textbf{q})-\frac{1}{\kappa_{k}}\mathrm{exp}(-\textbf{W}^{-1}\textbf{c})^{\prime}\mathrm{exp}(\textbf{W}^{-1}\textbf{q})\right\rbrace\\
  \nonumber
&= \mathrm{exp}\left\lbrace(\bz^{\prime}+d\bm{1}_{m}^{\prime}, \alpha_{k}\bm{1}_{m}^{\prime}-d\bm{1}_{m}^{\prime}\textbf{W}) \textbf{H}\textbf{q}-(\bm{1}_{m}^{\prime}, \frac{1}{\kappa_{k}}\mathrm{exp}(-\textbf{W}^{-1}\textbf{c})^{\prime})\mathrm{exp}\left(   \textbf{H}\textbf{q}\right)\right\rbrace\\
      &\propto \mathrm{mMLG}\left\lbrace\textbf{H}, (\bz^{\prime}+d\bm{1}_{m}^{\prime}, \alpha_{k}\bm{1}_{m}^{\prime}-d\bm{1}_{m}^{\prime}\textbf{W})^{\prime}, \bm{\kappa}_{q}\right\rbrace,
  \end{align}
where $\bm{\kappa}_{q}^{(-1)}\equiv \left\lbrace\bm{1}_{m}^{\prime}, \frac{1}{\kappa_{k}}\mathrm{exp}(-\textbf{W}^{-1}\textbf{c})^{\prime}\right\rbrace^{\prime}$,
\begin{equation*}
d \equiv \frac{\alpha_{k}}{1+\mathrm{max}\left\lbrace \mathrm{abs}\left(\bm{1}_{m}^{\prime}\textbf{W}\right) \right\rbrace },
\end{equation*}
max($\textbf{k}$) gives the maximum value of the elements in the $m$-dimensional real-valued vector $\textbf{k}$, and ``abs'' is the vector-valued absolute value function. The value $d$ accommodates possible zero-values within the data-vector $\textbf{Z}$. Since the absolute value of the elements in $d\bm{1}_{m}^{\prime}\textbf{W}$ are strictly between zero and 1, we have that the elements of $\alpha_{k}\bm{1}_{m}^{\prime}-d\bm{1}_{m}^{\prime}$ are strictly positive. Also, it is clear that $\bz^{\prime}+d\bm{1}_{m}^{\prime}$ consists of strictly positive elements. In the case that each element of $\textbf{Z}$ is non-zero we set $d=0$.

Notice that (\ref{simple_poisson}), (\ref{marginal}), and (\ref{conjugate}) are the multivariate analogs to (\ref{propPois}), (\ref{univ_LG}), and (\ref{univ_congj}), respectively. Additionally, from Theorem 2($ii$) it is straightforward to simulate from the posterior distribution in (\ref{conjugate}). Specifically, simulate
\begin{equation*}
\textbf{w} \sim \mathrm{MLG}\left\lbrace\bm{0}_{2m},\textbf{I}_{2m},(\bz^{\prime}+d\bm{1}_{m}^{\prime}, \alpha_{k}\bm{1}_{m}^{\prime}-d\bm{1}_{m}^{\prime})^{\prime},\bm{\kappa}_{q}\right\rbrace
\end{equation*}
\noindent
using (\ref{linear_comb}). Then a realization from the posterior distribution in (\ref{conjugate}) is given by $(\textbf{H}^{\prime}\textbf{H})^{-1}\textbf{H}^{\prime}\textbf{w}$, which is computationally feasible provided that $m$ is ``small'' (see Remark 3). 

\section{The Poisson Multivariate Spatio-Temporal Mixed Effects \\ Model} The simplicity of the example in Section 2.4 does not translate immediately into practice; recall that the example in Section 2.4 required both $\textbf{c}$ and $\textbf{V}$ to be known, which is almost never the case. Furthermore, modern datasets (count-valued and otherwise) are becoming increasingly more complex, \textit{and high-dimensional} \citep[e.g., see][for a review]{bradley2014_comp}. For example, a clear majority of federal and environmental datasets display dependencies over different variables, spatial locations, times, and spatial/temporal scales \citep{bradleyMSTM2,bradleySTCOS}. It is not immediately clear how one can incorporate the MLG distribution into these common data settings.

Thus, to facilitate the use of the MLG distribution among the applied statistics community we incorporate it into an extremely general setting: high-dimensional multivariate space-time data. Specifically, the novel distribution theory from Section 2 is used to define a Bayesian hierarchical model, which we call the \textit{Poisson multivariate spatio-temporal mixed effects model} (P-MSTM). The dependence between count data (observations), latent processes, and unknown parameters are intricate, and hence, we choose to present the conditional and marginal distributions that makeup the P-MSTM in several subsections. In particular, Section~3.1 describes the conditional distribution of the data given latent processes and unknown parameters; Section~3.2 defines the conditional distribution of the latent process given the unknown parameters; Section~3.3 describes the process model's temporal dynamics; and Section~3.4 gives the marginal distribution of unknown parameters. Then, in Section~3.5, we provide the full-conditional distributions needed for conducting Bayesian inference using a Gibbs sampler.

\subsection{The Data Model} Consider data that are recorded over $\ell = 1,\ldots,L$ different variables, $t = T_{\mathrm{L}}^{(\ell)},\ldots,T_{\mathrm{U}}^{(\ell)}$ different time points, and $N_{t}^{(\ell)}$ areas from the set $D_{t,\mathrm{P}}^{(\ell)} \equiv\{A_{i}: i = 1,\ldots,N_{t}^{(\ell)}\}$, where $A_{i}$ is a subset of the domain of interest $D \subset \mathbb{R}^{d}$ and the subscript ``P'' stands for ``prediction regions.'' Let $D_{t,\mathrm{P}}^{(\ell)}$ consist of disjoint areal units; that is, $\cup_{i = 1}^{n}A_{i} \subset D$ and $A_{i}\cap A_{j} = \emptyset$ ($i \ne j$). In practice, all possible prediction regions are not observed, and hence, we denote the set of $n_{t}^{(\ell)}$ areal units that are associated with the observed data by $D_{t,\mathrm{O}}^{(\ell)} \subset D_{t,\mathrm{P}}^{(\ell)}$, where the subscript ``O'' stands for ``observed regions.'' Let $\textbf{Z}_{t} \equiv (Z_{t}^{(\ell)}(A): \ell = 1,\ldots,L, \hspace{5pt}A \in D_{t,\mathrm{O}}^{(\ell)})^{\prime}$ be a generic $n_{t}$-dimensional data vector consisting of count-valued data, where $n_{t} = \sum_{\ell = 1}^{L}n_{t}^{(\ell)}$; $t = 1,\ldots,T$. In what follows, a general class of models to analyze the $n$-dimensional vector $\bz \equiv (\textbf{Z}_{t}^{\prime}: t = 1,\ldots,T)^{\prime}$ is introduced, where we let $\underset{\ell}{\mathrm{min}}(T_{\mathrm{L}}^{(\ell)}) = 1$, $T \equiv \underset{\ell}{\mathrm{max}}(T_{\mathrm{U}}^{(\ell)})$, and $n \equiv \sum_{t = 1}^{T}n_{t}$.

An element of the generic $n$-dimensional count-valued data vector $\bz$ is assumed to follow the conditional distribution
	\begin{equation}\label{data:model}
	Z_{t}^{(\ell)}(A)\vert Y_{t}^{(\ell)}(A) \ind \mathrm{Pois}\left(\mathrm{exp}\left\lbrace Y_{t}^{(\ell)}(A)\right\rbrace\right);\hspace{5pt}\ell = 1,\ldots,L, \hspace{5pt}t = T_{\mathrm{L}}^{(\ell)}, \ldots, T_{\mathrm{U}}^{(\ell)}, \hspace{5pt}A \in D_{t,\mathrm{P}}^{(\ell)},
	\end{equation}
\noindent
where we note that the canonical log-link is used. The Poisson distribution is not the only choice for modeling $Z_{t}^{(\ell)}(A)\vert Y_{t}^{(\ell)}(A)$; $t = T_{\mathrm{L}}^{(\ell)}, \ldots, T_{\mathrm{U}}^{(\ell)}$ and $\ell = 1,\ldots,L$. For example, one might consider a negative binomial distribution with a process defined as the integration of an inhomogeneous point process. However, the Poisson distribution has been the de facto model for modeling high-dimensional spatially referenced count data \citep[e.g., see][]{aritrajsm}. Furthermore, the Poisson distribution with the canonical log-link is a common choice for modeling count data within a generalized linear mixed effects model (GLMM) framework \citep[e.g., see][]{glm-nelder}.

\subsection{The Process Models}We use the following specification of the conditional distribution of $\{Y_{t}^{(\ell)}(A)\}$ given unknown parameters:
\begin{equation}\label{process:model:truth}
Y_{t}^{(\ell)}(A)= \textbf{x}_{t}^{(\ell)}(A)^{\prime}\bm{\beta}+\bm{\psi}_{t}^{(\ell)}(A)^{\prime}\bm{\eta}_{t} + \xi_{t}^{(\ell)}(A);\hspace{10pt} \ell = 1,\ldots,L, \hspace{10pt}t = 1, \ldots, T, \hspace{10pt}A \in D_{t,\mathrm{P}}^{(\ell)},
\end{equation}
\noindent
where $\textbf{x}_{t}^{(\ell)}(A)^{\prime}\bm{\beta}$ represents the ``large-scale'' variability of $Y_{t}^{(\ell)}(A)$, $\textbf{x}_{t}^{(\ell)}$ is a $p$-dimensional vector of known multivariate spatio-temporal covariates, and $\bm{\beta}\in \mathbb{R}^{p}$ is an unknown vector-valued parameter. Here, let the $n_{t}\times p$ matrix $\textbf{X}_{t}\equiv (\textbf{x}_{t}^{(\ell)}(A)^{\prime}: \ell = 1,\ldots,L, \hspace{5pt}A \in D_{t,\mathrm{O}}^{(\ell)})^{\prime}$. For our purposes it is reasonable to take $\bm{\beta}$ as constant over time, however, the results in this manuscript can easily be extended to the setting where $\bm{\beta}$ is time-varying. The $r$-dimensional random vector $\bm{\eta}_{t}$ is assumed to be mean-zero and have an unknown covariance matrix, and the set $\{\xi_{t}^{(\ell)}(A)\}$ consists of independent log-gamma random variables with mean zero and unknown variance $\sigma_{\xi}^{2}>0$. We consider both sMLG and nMLG specifications for the distribution of $\{\xi_{t}^{(\ell)}(A)\}$. Notice that the distributional assumptions governing $\{\bm{\eta}_{t}\}$ have not been specified $\--$ these definitions are provided in Section~3.3.

The $r$-dimensional real vectors $\{\bm{\psi}_{t}^{(\ell)}(A)\}$ can belong to any class of areal basis functions \citep[see][for different choices of areal basis functions]{bradleyCAGE}. Here, let the $n_{t}\times r$ matrix $\bm{\Psi}_{t}\equiv (\bm{\psi}_{t}^{(\ell)}(A)^{\prime}: \ell = 1,\ldots,L, \hspace{5pt}A \in D_{t,\mathrm{O}}^{(\ell)})^{\prime}$. In general, we only require that $\{\textbf{x}_{t}^{(\ell)}(A): \ell = 1,\ldots,L, \hspace{5pt}A \in D_{t,\mathrm{P}}^{(\ell)}\}$ and $\{\bm{\psi}_{t}^{(\ell)}(A): \ell = 1,\ldots,L, \hspace{5pt}A \in D_{t,\mathrm{P}}^{(\ell)}\}$ are not confounded for each $t = 1,\ldots,T$. That is, the $N_{t}\times p$ matrix $\textbf{X}_{t}^{\mathrm{P}}\equiv(\textbf{x}_{t}^{(\ell)}(A)^{\prime}: \ell = 1,\ldots,L, \hspace{5pt}A \in D_{t,\mathrm{P}}^{(\ell)})^{\prime}$ and the $N_{t}\times r$ matrix $\bm{\Psi}_{t}^{\mathrm{P}}\equiv(\bm{\psi}_{t}^{(\ell)}(A)^{\prime}: \ell = 1,\ldots,L, \hspace{5pt}A\in D_{t,\mathrm{P}}^{(\ell)})^{\prime}$ have columns that are linearly independent, where $N_{t} = \sum_{\ell = 1}^{L}N_{t}^{(\ell)}$. This specification allows us to perform inference on $\bm{\beta}$, since $\bm{\beta}$ and each random vector in $\{\bm{\eta}_{t}\}$ are not confounded \citep[see,][and the references therein]{griffith2000, griffith2002, griffith2004, griffith2007, hughes,aaronp,bradleyMSTM}. In this manuscript, we set $\{\psi_{t}^{(\ell)}(A)\}$ equal to the MI basis functions from \citet{bradleyMSTM}, which is defined a priori so that $\textbf{X}_{t}^{\mathrm{P}}$ and $\bm{\Psi}_{t}^{\mathrm{P}}$ are not confounded. For convenience of exposition the formal definition of the MI basis functions are left to Supplemental Materials.

\subsection{Process Model Dynamics} Process dynamics are modeled using a first-order vector autoregressive (VAR(1)) model
\begin{align}
\label{var1}
\bm{\eta}_{t} &= \textbf{M}_{t}\bm{\eta}_{t-1} + \textbf{b}_{t};\hspace{5pt} t = 2,\ldots,T,
\end{align}
\noindent
where $\bm{\eta}_{1}\sim \mathrm{MLG}(\bm{0},\textbf{W}_{1,k}^{1/2}, {\alpha}_{k}\bm{1}_{r},\bm{\kappa}_{k}\bm{1}_{r})$, $\{\textbf{M}_{t}\}$ are known $r\times r$ propagator matrices, and for $t\ge 2$ the $r$-dimensional random vector $\textbf{b}_{t}\sim \mathrm{MLG}(\bm{0},\textbf{W}_{t,k}^{1/2}, {\alpha}_{k}\bm{1}_{r},\bm{\kappa}_{k}\bm{1}_{r})$ and is independent of $\bm{\eta}_{t-1}$. Additionally, let $\textbf{W}_{t,k} \equiv \alpha_{\mathrm{G}}^{k}\textbf{K}_{t}$, $\textbf{K}_{t}$ be an unknown $r\times r$ positive definite real-valued matrix, ${\alpha}_{k} \equiv \alpha_{\mathrm{G}}^{k}(\alpha^{*})^{1-k}$, and let ${\kappa}_{k} \equiv \left(\frac{1}{\alpha_{\mathrm{G}}}\right)^{k}\left\lbrace\mathrm{exp}\left[-\omega_{0}(\alpha^{*})\right]\right\rbrace^{1-k}$; $ k = 0,1$ and $t = 1,\ldots,T$. Notice, that if $k = 0$ then (\ref{var1}) is based on the sMLG specification and if $k = 1$ then (\ref{var1}) is based on the nMLG specification. 

From (\ref{process:model:truth}) and (\ref{var1}) it is immediate that the VAR(1) model allows for nonseparable asymmetric nonstationary multivariate spatio-temporal dependencies within the latent process of interest. To the best of our knowledge, we are the first to consider the MLG specification for a VAR(1) model. Additionally, if $T = 1$ we simply let $\bm{\eta}_{1}\sim \mathrm{MLG}(\bm{0},\textbf{W}_{1,k}^{1/2}, {\alpha}_{k}\bm{1}_{r},\bm{\kappa}_{k}\bm{1}_{r})$, which is used for multivariate spatial, multivariate-only, and spatial-only datasets.

\citet{bradleyMSTM} have provided a class of propagator matrices (for the Gaussian setting) informed by the covariates $\{\textbf{x}_{t}^{(\ell)}\}$, which they call the MI propagator matrix. For high-dimensional data it is advantageous to avoid MCMC sampling of $\{\textbf{M}_{t}\}$ by assuming that it is known. This class of propagator matrices is motivated by removing confounding over time, which is similar to the motivation of the MI basis functions. Thus, we use this class of propagator matrices to model the dynamics of count-valued data. For convenience of exposition, the details of the MI propagator matrix are provided in Supplemental Materials.

\subsection{Parameter Models} An important step for modeling high-dimensional data is reducing the dimension of the parameter space of interest. This general theme is consistent throughout the ``big spatial data literature'' (and the ``big data'' literature in general) and is done in a variety of ways (e.g., see the rejoinder in \citet{bradleyTEST}). We use the same perspective when specifying the distributions for ${\textbf{K}_{t}}$; $t = 1,\ldots,T$. In particular, place the following assumption on $\textbf{K}_{t}$,
\begin{equation}\label{mi}
\textbf{K}_{t} = \sigma_{\mathrm{K}}^{2}\textbf{K}_{t}^{*};\hspace{5pt}t = 1,\ldots,T,
\end{equation}
\noindent
where $\sigma_{\mathrm{K}}^{2}>0$ is unknown,
\begin{equation}\label{Kstar}
 \textbf{K}_{t}^{*-1}=\underset{\textbf{C}}{\mathrm{arg\hspace{5pt}min}}\left\lbrace ||\textbf{Q}_{t} - \bm{\Psi}_{t}^{\mathrm{P}}\textbf{C}^{-1}\bm{\Psi}_{t}^{\mathrm{P}\prime}||_{\mathrm{F}}^{2}\right\rbrace;\hspace{5pt} t = 1,\ldots,T,
 \end{equation} 
 \noindent
 $\textbf{Q}_{t} = \textbf{I}_{N_{t}} - \textbf{A}_{t}$, the minimization in (\ref{Kstar}) is among $r\times r$ positive semi-definite matrices $\textbf{C}$, and for a generic real-valued square matrix $\textbf{H}$ the Frobenius norm is defined as $||\textbf{H}||_{F}^{2} = \mathrm{trace}(\textbf{H}^{\prime}\textbf{H})$. In the setting where $|D_{t}^{(\ell)}|>1$ for at least one $t$ and $\ell$, let $\textbf{A}_{t}$ be the adjacency matrix corresponding to the edges formed by $\{D_{t,\mathrm{P}}^{(\ell)}:\ell = 1,\ldots,L\}$; hence $\textbf{Q}_{t}$ is the precision matrix from an intrinsic conditional autoregressive (ICAR) model in this case. However, in practice $\textbf{A}_{t}$ is allowed to be any generic $N_{t}\times N_{t}$ real-valued matrix. The functional form of $\textbf{K}_{t}^{*}$ can be found in \citet{bradleyMSTM} and \citet{bradleyCAGE}.
 
 The parameterization in (\ref{mi}) is called the MI parameter model, and has recently become a popular parameterization. In fact, the MI parameter model, has been successfully used to model covariances for large-dimensional spatially referenced data \citep[e.g., see][among others]{hughes,aaronp,bradleyMSTM,bradleyCAGE}. The main motivating feature of this approach is that it calibrates the covariance (or precision) of the random effect towards a covariance (or precision) matrix that is not simplified due to computational considerations. Moreover, it reduces the parameter space significantly, where a single parameter is defined (i.e., $\sigma_{\mathrm{K}}^{2}$) as opposed to $r^{2}$ parameters.

Hence, parameter models are required for $\sigma_{\mathrm{K}}^{2}$, $\sigma_{\xi}^{2}$, and $\bm{\beta}$. For simplicity, consider simple discrete uniform priors for $\sigma_{\mathrm{K}}^{2}$ and $\sigma_{\xi}^{2}$. That is, it is assumed that
\begin{align}
\nonumber
f(\sigma_{\mathrm{K}}) = \frac{1}{U_{\mathrm{K}}};\hspace{10pt}\sigma_{\mathrm{K}} = a_{1}^{(\mathrm{K})},\ldots,a_{U_{\mathrm{K}}}^{(\mathrm{K})}\\
\label{sigma_models}
f(\sigma_{\xi})  = \frac{1}{U_{\mathrm{\xi}}};\hspace{10pt}\sigma_{\xi} = a_{1}^{(\xi)},\ldots,a_{U_{\mathrm{\xi}}}^{(\xi)},
\end{align}
\noindent
where for our application in Section~5 many different choices for $a_{1}^{(\mathrm{K})},\ldots,a_{U_{\mathrm{K}}}^{(\mathrm{K})}$ and $a_{1}^{(\xi)},\ldots,a_{U_{\mathrm{\xi}}}^{(\xi)}$ were considered, and we found that $a_{1}^{(\mathrm{K})} = a_{1}^{(\mathrm{\xi})} = 0.01, a_{2}^{(\mathrm{K})} = a_{2}^{(\mathrm{\xi})} = 0.02,\ldots,a_{U_{\mathrm{K}}}^{(\mathrm{K})} = a_{U_{\mathrm{\xi}}}^{(\xi)} = 2$ is appropriate for that application. Any number of different parameter models may be considered, and we suggest that one seriously considers alternatives to what we use in (\ref{sigma_models}). However, for our purpose of prediction, the simple discrete uniform prior is appropriate. Additionally, the $p$-dimensional vector $\bm{\beta}$ is assumed to have the following parameter model:
\begin{align}
\label{beta_dist}
\bm{\beta}&\sim \mathrm{MLG}\left(\bm{0},\sigma_{\beta}\textbf{I},\bm{\alpha}_{\beta}\bm{1}_{p},\bm{\kappa}_{\beta}\bm{1}_{p}\right).
\end{align}
\noindent
Set $\sigma_{\beta}$ equal to a fixed large number (we use $\sigma_{\beta} = 10$) so that the prior on $\bm{\beta}$ is flat. In general, we have found that prediction is robust to this specification.

\subsection{Full Conditional Distributions for Markov Chain Monte Carlo} The joint distribution of the data, processes, and parameters is written as the product of the following conditional and marginal distributions:
\begin{align}\label{summary}
\nonumber
&\mathrm{Data\hspace{5pt}Model:}\hspace{5pt} Z_{t}^{(\ell)}(A)\vert \bfbeta,\bm{\eta}_{t}, \xi_{t}^{(\ell)}(A) \ind \mathrm{Pois}\left[\mathrm{exp}\left\lbrace\textbf{x}_{t}^{(\ell)}(A)^{\prime}\beta + \bm{\psi}_{t}^{(\ell)}(A)^{\prime}\bm{\eta}_{t} + \xi_{t}^{(\ell)}(A)\right\rbrace\right];\\
\nonumber
&\mathrm{Process\hspace{5pt}Model\hspace{5pt}1:}\hspace{5pt} \bm{\eta}_{t}\vert \bm{\eta}_{t-1},\sigma_{\mathrm{K}}\sim \mathrm{MLG}\left(\textbf{M}_{t}\bm{\eta}_{t-1}, \textbf{W}_{t,k}^{1/2},{\alpha}_{k}\bm{1}_{r},{\kappa}_{k}\bm{1}_{r}\right);2\le t \le T, \hspace{15pt} (provided \hspace{2pt} T>1)\\
\nonumber
&\mathrm{Process\hspace{5pt}Model\hspace{5pt}2:}\hspace{5pt} \bm{\eta}_{1}\vert \sigma_{\mathrm{K}} \sim \mathrm{MLG}\left(\bm{0}, \textbf{W}_{1,k}^{1/2},{\alpha}_{k}\bm{1}_{r},{\kappa}_{k}\bm{1}_{r}\right);\\
\nonumber
&\mathrm{Process\hspace{5pt}Model\hspace{5pt}3:}\hspace{5pt} \bm{\xi}_{t}\vert \sigma_{\xi} \sim \mathrm{MLG}\left(\bm{0}, \alpha_{k}^{1/2}\sigma_{\xi}\textbf{I}_{n_{t}},{\alpha}_{k}\bm{1}_{n_{t}},{\kappa}_{k}\bm{1}_{n_{t}}\right);\\
\nonumber
&\mathrm{Parameter\hspace{5pt}Model\hspace{5pt}1:}\hspace{5pt} \bm{\beta}\sim \mathrm{MLG}\left(\bm{0}_{p,1}, \alpha_{k}^{1/2}\sigma_{\beta}\textbf{I}_{p},\alpha_{k}\bm{1}_{p},\kappa_{k}\bm{1}_{p}\right);\\
\nonumber
&\mathrm{Parameter\hspace{5pt}Model\hspace{5pt}2:}\hspace{5pt} f(\sigma_{\mathrm{K}}) = \frac{1}{U_{\mathrm{K}}};\\
&\mathrm{Parameter\hspace{5pt}Model\hspace{5pt}3:}\hspace{5pt} f(\sigma_{\xi})  = \frac{1}{U_{\mathrm{\xi}}},
\end{align}
\noindent
where recall $\textbf{W}_{t,k} \equiv \alpha_{\mathrm{G}}^{k}\sigma_{\mathrm{K}}^{2}\textbf{K}_{t}^{*}$, ${\alpha}_{k} \equiv \alpha_{\mathrm{G}}^{k}(\alpha^{*})^{1-k}$, and ${\kappa}_{k} \equiv \left(\frac{1}{\alpha_{\mathrm{G}}}\right)^{k}\left[ \mathrm{exp}\left\lbrace-\omega_{0}(\alpha^{*})\right\rbrace\right]^{1-k}$; $ k = 0,1, \hspace{2pt}\ell = 1,\ldots,L, \hspace{5pt}t = T_{\mathrm{L}}^{(\ell)},\ldots,T_{\mathrm{U}}^{(\ell)}, \hspace{4pt}A \in D_{t,\mathrm{O}}^{(\ell)}$, $\sigma_{\mathrm{K}}= a_{1}^{(\mathrm{K})},\ldots,a_{U_{\mathrm{K}}}^{(\mathrm{K})}$, and $\sigma_{\xi} = a_{1}^{(\xi)},\ldots,a_{U_{\mathrm{\xi}}}^{(\xi)}$. As with the VAR(1) model in (\ref{var1}), if $k = 0$ then (\ref{summary}) is based on the sMLG specification, and if $k = 1$ then (\ref{summary}) is based on the nMLG specification. In practice, we use the deviance information criterion (DIC) \citep{spiegel-2002} to select between sMLG and nMLG. 

The P-MSTM presented in (\ref{summary}) is extremely general, and can be adapted in variety of ways. For example, one can easily cast the P-MSTM within paradigms outside Bayesian statistics and areal spatial data. That is, one can set unknown parameters equal to estimates to produce empirical Bayesian versions of the P-MSTM; see, for example, \citet{aritrajsm}. Likewise, the P-MSTM is flexible enough to handle different basis functions, propagator matrices, and parameter models that may be more suitable in the context of different problems. This includes point referenced basis functions that are often used in environmental and ecological settings \citep[e.g., see][among others]{wikle2001,johan}.

Additionally, the model in (\ref{summary}) is defined for the case when $L=1$, $T = 1$, and/or $|D_{t,\mathrm{P}}^{(\ell)}| = 1$ for each $t$ and $\ell$. This implies that our modeling framework can be readily applied to multivariate-only, spatial-only, times series, multivariate spatial, multivariate time series, and spatio-temporal datasets (in addition to multivariate spatio-temporal data). The generality of (\ref{summary}) is especially notable because it is rather straightforward to simulate from the full-conditional distributions implied by the model in (\ref{summary}).\\

\noindent
 \textit{Proposition 3: Suppose the $n$-dimensional data vector $\textbf{Z}$ follows the P-MSTM distribution given in (\ref{summary}). Then, we have the following full conditional distribution for the unknown latent random vectors, and unknown parameters. For  $k = 0$ and 1 we have
  \begin{align}\label{full_cond}
  \nonumber
   & f(\bm{\beta}\vert \cdot)= \mathrm{mMLG}(\textbf{H}_{\beta,k},\bm{\alpha}_{\beta,k},\bm{\kappa}_{\beta,k})\\
  \nonumber
  & f(\bm{\eta}_{t}\vert \cdot)= \mathrm{mMLG}(\textbf{H}_{\eta,t,k},\bm{\alpha}_{\eta,t,k},\bm{\kappa}_{\eta,t,k});\hspace{5pt} 2\le t \le T-1\hspace{15pt} (provided \hspace{2pt} T>1)\\
  \nonumber
  & f(\bm{\eta}_{1}\vert \cdot)= \mathrm{mMLG}(\textbf{H}_{\eta,1,k},\bm{\alpha}_{\eta,1,k},\bm{\kappa}_{\eta,1,k})\hspace{85pt} (provided \hspace{2pt} T>1)\\
  \nonumber
  & f(\bm{\eta}_{T}\vert \cdot)= \mathrm{mMLG}(\textbf{H}_{\eta,T,k},\bm{\alpha}_{\eta,T,k},\bm{\kappa}_{\eta,T,k})\\
  \nonumber
    & f(\bm{\xi}_{t}\vert \cdot)= \mathrm{mMLG}(\textbf{H}_{\xi,t,k},\bm{\alpha}_{\xi,t,k},\bm{\kappa}_{\xi,t,k});\hspace{5pt}1\le t \le T,\\
    \nonumber
& f(\sigma_{\mathrm{K}}\vert \cdot) = p_{\mathrm{K}}(\sigma_{\mathrm{K}});\hspace{5pt} \sigma_{\mathrm{K}}=a_{1}^{(\mathrm{K})},\ldots,a_{U_{\mathrm{K}}}^{(\mathrm{K})},\\
& f(\sigma_{\xi}\vert \cdot)= p_{\xi}(\sigma_{\xi});\hspace{5pt} \sigma_{\xi}=a_{1}^{(\xi)},\ldots,a_{U_{\mathrm{\xi}}}^{(\xi)},
  \end{align}
  where $f(\bm{\beta}\vert \cdot)$ represents the pdf of $\bm{\beta}$ given all other process variables, parameters, and the data. For each $t$, we define $f(\bm{\eta}_{t}\vert \cdot)$, $f(\bm{\xi}_{t}\vert \cdot)$, $f(\sigma_{\mathrm{K}}\vert \cdot)$, and $f(\sigma_{\xi}\vert \cdot)$ in a similar manner. For ease of exposition, in Table 1 we provide the definitions of the remaining unknown quantities in (\ref{full_cond}).\\
 }
 
\noindent
The proof of Proposition 3 is given in the Appendix. Additionally, the step-by-step instructions outlining the implementation of the Gibbs sampler based on (\ref{full_cond}) is given in {Algorithm 1}. Notice that it is relatively easy to simulate from (\ref{full_cond}) using Theorem~2($ii$) provided that $r \ll n$ and $p \ll n$; that is, from Theorem~2($ii$) simulating from the full-conditionals in (\ref{full_cond}) involves computing the inverse of $p\times p$ and $r\times r$ matrices, which involves computations on the order of $p^{3}$ and $r^{3}$, respectively.

Proposition 3 can be applied to the aforementioned special cases of multivariate spatio-temporal data (i.e., spatial-only, times series, multivariate-only, spatio-temporal, multivariate spatial, and multivariate times series datasets). Thus, the implications of Proposition 3 are enormous, as it provides a way to \textit{efficiently model} a wide range of less general but interesting special cases not considered in this manuscript. As an example, the full conditional distributions for the spatial-only setting (i.e., when $L = 1$, $T = 1$, and when $|D_{1}^{(1)}| > 1$) is presented in Supplemental Materials, along with a demonstration using ACS period estimates.

\section{Results: LEHD Simulations and Analysis} We now return to the QWI dataset presented in Figure 1(a). The QWIs have become a staple data source for understanding US economics and have lead to many scientific insights across several areas (e.g., for the economics setting see \citet{supermarket}, \citet{thompson}, \citet{workpaper}, and \citet{irle}, among others; for the social-economic setting see \citet{Glaeser1}, \citet{Glaeser2}, and \citet{ford}, among others; and for the statistics setting see \citet{abowd}, \citet{abowdlmm}, and \citet{bradleyMSTM}). From the statistical perspective, the MSTM from \citet{bradleyMSTM} is exceedingly important, as it leads to an enhancement of the QWIs. Specifically, the MSTM can be used to produce estimates of continuous QWIs (e.g., average quarterly income) that have complete spatio-temporal coverage, which is otherwise not available \citep[][Section 5.5.1 and 5.6]{abowd}. However, the MSTM (based on the Gaussian assumption) has a limited utility on QWIs. This is because approximately 70$\%$ of the LEHD QWIs are count-valued (e.g., county-level beginning of quarter employment), which implies that the MSTM is applicable to a small portion of the entire scope of the QWIs. Thus, applying the P-MSTM to count-valued QWIs will be extremely useful, and is likely to have a large impact on the QWI user community. 

Note that all computations were computed using Matlab (Version 8.0) on a dual 10 core 2.8 GHz Intel Xeon E5-2680 v2 processor, with 256 GB of RAM.

\subsection{A Simulation Study} We choose to calibrate our simulation model towards QWIs. That is, we set the mean of a Poisson random variable equal to a count-valued QWI and use this distribution to generate a ``pseudo data-value,'' which are treated as unknown. Then, the pseudo data and the P-MSTM are used to predict the QWIs. This empirical simulation study design is similar to what is done in \citet{bradleyMSTM}, and is motivated as a way to produce simulated data that is similar to what one might observe in practice.

Let $Z_{t}^{(\ell)}(A)$ represent the number of individuals employed at the beginning of the quarter (i.e., the count-valued QWIs considered in Section~1), for industry $\ell$, Minnesota county $A$, and quarter $t$. Then, simulate pseudo-data as follows,
\begin{equation}\label{pseudo}
R_{t}^{(\ell)}\sim \mathrm{Pois}(Z_{t}^{(\ell)}(A)+1);\hspace{5pt} \ell = 1,2, \hspace{5pt}t = 76,\ldots,96, \hspace{5pt}A \in D_{\mathrm{MN},t}^{(\ell)},
\end{equation}
\noindent
where $D_{\mathrm{MN},t}^{(\ell)}$ represents the set of counties in Minnesota (MN) that have available quarterly average monthly income estimates, $\ell = 1$ denotes the information industry, and $\ell = 2$ represents the professional, scientific and technical services industry. These two industries were chosen for this simulation study since they are highly correlated. Notice that we add 1 in (\ref{pseudo}) so that the mean of the Poisson random variables are also greater than 0. 

Randomly select 65$\%$ of the areal units in $D_{\mathrm{MN},t}^{(\ell)}$ to be ``observed,'' and denote this new  set with $D_{\mathrm{MN, O}, t}^{(\ell)}$. For illustration, we use the following covariates $\textbf{x}_{t}^{(\ell)}(A) = (1,I(\ell = 1),\ldots,I(\ell = 19), |A|, I(t = 1),\ldots,I(t = 1,\ldots,95),\mathrm{population}(A))^{\prime}$, where $\mathrm{population}(A)$ is the 2010 decennial Census value of the population of county $A$ and $I(\cdot)$ is the indicator function. Following \citet{hughes}'s rule of thumb for specifying $r$, we set $r = 42$ (see the Supplemental Material). Both the sMLG and nMLG specifications are considered in this section, and we investigate whether the specification leads to noticeable out-of-sample properties.

In Figure 3, we present the QWIs (panel a), the pseudo data (panel b), and the predictor (panels c and d) given by
\begin{equation*}
E\left[ Z_{t}^{(\ell)}(A)\vert \{R_{t}^{(\ell)}(A): \ell = 1,2, t = 76,\ldots,96, A \in D_{\mathrm{MN, O}, t}^{(\ell)}\}\right];\hspace{5pt} \ell = 1,2,\hspace{5pt} t = 96, \hspace{5pt}A \in D_{\mathrm{MN},t}^{(\ell)},
\end{equation*}
\noindent
where the expectation is obtained using the P-MSTM and Algorithm 1. In general, the predictions based on both the sMLG assumption (panel c) and the nMLG assumption (panel d) lead to predictions that reflect the overall pattern of the data. This is further supported in Figure 4(a,b), where we plot the log QWIs and the log predictions over an arbitrary ordering of the regions. Again we see that both methods tend to track the truth fairly closely. In Figure 4(c,d) we provide a scatterplot of the log QWIs versus the log predictions. Here, the predictions are similar to the truth (the correlations are 0.8333 and 0.8708 for panels c and d, respectively), however, there is more error in the predictions for small and moderate values of the log QWIs.

Now, consider 100 independent replications of the set $\{R_{t,j}^{(\ell)}(A): \ell = 1,2,\hspace{5pt} t = 76,\ldots,96, \hspace{5pt}A \in D_{\mathrm{MN, O}, t}^{(\ell)}\}$, where $j = 1,\ldots,100$ and for each $j$ we have that $R_{t,j}^{(\ell)}(A)$ is simulated according to (\ref{pseudo}). To evaluate the predictors we compute the following average absolute error diagnostic
\begin{equation}\label{abs_error}
\mathrm{average}\left\lbrace \mathrm{abs}\left(Z_{t}^{(\ell)}(A)-E\left[Z_{t}^{(\ell)}(A)\vert \{R_{t,j}^{(\ell)}(A)\}\right]\right)\right\rbrace;\hspace{5pt}j = 1,\ldots,100,
\end{equation}
\noindent
where ``average'' is the sample average function taken over $\ell = 1,2, t = 76,\ldots,96,$ and $A \in D_{\mathrm{MN},t}^{(\ell)}$, and ``abs'' is the absolute value function. In Figure 5, we present boxplots of the average absolute error diagnostic over the 100 replications of $\{R_{t,j}^{(\ell)}(A)\}$ by the MLG specification. Here, we see that when one assumes sMLG marginal improvements are obtained, in the average absolute error diagnostic, over when one assumes nMLG. The $p$-value of a sign-test, testing for a difference between the true mean of (\ref{abs_error}) when using sMLG and nMLG, is 9.8$\times 10^{-4}$, which suggests that sMLG outperforms nMLG. However, it is clear that the practical difference in the predictors is extremely small.

\subsection{Predicting the Mean Beginning of the Quarter Employment} We show that one can obtain reasonable predictions of the mean number of individuals employed at the beginning of a quarter, over all 3,145 US counties, 20 NAICS sectors, and 96 quarters, using the high-dimensional QWI dataset of size 4,089,755 (partially presented in Figure 1(a)). To use the P-MSTM from Section~3.1 through 3.4 we need to specify multivariate spatio-temporal covariates $\{\textbf{x}_{t}^{(\ell)}(A)\}$, the rank of the MI basis functions $r$, and the assumptions on the MLG distribution (i.e., sMLG or nMLG). For illustration, we again use the following covariates $\textbf{x}_{t}^{(\ell)}(A) = (1,I(\ell = 1), |A|, I(t = 76),\ldots,I(t = 1,\ldots,95),\mathrm{population}(A))^{\prime}$. As in Section~4.1 we set $r = 42$ (see the Supplemental Material). Both the sMLG and nMLG specifications were considered, and the DIC was considerably larger when using nMLG; that is, sMLG produced a DIC of $1.5585\times 10^{14}$, and nMLG resulted in a DIC of $1.2317\times 10^{35}$. Thus, we present the results using the sMLG specification.

In Figure 6(a), we display a histogram of the CPU time (in seconds) to compute one replicate from the P-MSTM fitted using the 4,089,755 QWIs. Notice that, it consistently takes approximately 6 seconds to obtain 1 MCMC replicate from the P-MSTM. Furthermore, the entire chain (of 10,000 iterations) took approximately 17 hours to compute. Compare this to the 6 or more days to compute each diverging chain in Figure 1(b,c,d). As with the discussion surrounding Figure 1, one should note that there are many confounding factors involved with citing computation times, including the software, computer, and code used to compute the algorithm. Here, we simply state the CPU times based on our computer and codes. In Figure 6(b), we give one trace plot for the sample chain, and, through visual inspection, see that convergence appears to be obtained for this parameter. Additionally, we randomly selected other latent random variables and parameters and the trace plots of the sample chains gave similar results. Moreover, we check batch means estimates of Monte Carlo error (with batch size 50) \citep[e.g., see][]{batch1,batch2}, and compute Gelman-Rubin diagnostics based on three independent chains \citep[e.g., see][]{gr1}, which were consistently less than 1.02. These diagnostics all provide evidence to suggest that there is no lack of convergence of the MCMC algorithm.

In Figures 7(b) and 7(c), we plot the predictions and the associated posterior standard deviation for the mean beginning of quarter employment during the 4-th quarter of 2013 and for the education industry. It should be emphasized that predictions have been made over all 3,145 US counties, 20 NAICS sectors, and 96 quarters. Upon comparison of Figure 1(a) to Figure 7(b) we see that the predictions reflect the general patterns of the data. Furthermore, the posterior standard deviations in Figure 7(c) are very small (at most around 400) considering that Poisson random variables have their mean equal to their variance. Additionally, we display a boxplot of the residuals (i.e., log QWI minus log prediction) in Figure 7(d). Notice that the median is close to 0, and the range of the residuals are extremely small (between -0.3 and 0.3). Thus, we see that the in-sample error of the predictors based on the P-MSTM tends to be small and have relatively little bias.

\section{Discussion} We have introduced a comprehensive framework for jointly modeling Poisson data that could possibly be referenced over different variables, regions, and times. This methodology is rooted in the development of new distribution theory that makes Bayesian inference for correlated count-valued data computationally feasible. Specifically, we propose a multivariate log-gamma distribution that incorporates nonseparable asymmetric nonstationary dependencies, and leads to computationally efficient sampling of full conditional distributions within a Gibbs sampler.

Several theoretical results were required to create this multivariate log-gamma paradigm. In particular, we show that full conditional distributions are of the same form of a marginal distribution of a multivariate log-gamma random vector. Also, for those wary of dropping the conventional multivariate normal assumption, we provide a result that shows that the multivariate log-gamma distribution can be specified to be arbitrarily close to a multivariate normal distribution.

This multivariate log-gamma distribution is used within the multivariate spatio-temporal mixed effects model (MSTM) framework of \citet{bradleyMSTM}, leading to what we call the Poisson multivariate spatio-temporal mixed effects model (P-MSTM). The implications of a general (easy to fit) model for multivariate spatio-temporal count data are vast. First, it is well-known that it is more difficult to fit correlated Poisson data than correlated Gaussian data, since Poisson generalized linear mixed models require computational expensive Metropolis-Hasting updates within a Gibbs sampler (e.g., see Figure 1(b,c,d)). However, this is no longer the case as Proposition 3 shows that the multivariate log-gamma distribution leads to full-conditional distributions that are easy to simulate from, and by Proposition 2, one can incorporate the same types of dependencies as the LGP approach. Another important implication of the P-MSTM is that it can be used in a wide range of special cases including: spatial-only, times series, multivariate-only, spatio-temporal, multivariate spatial, and multivariate times series datasets.

The generality of the P-MSTM is especially notable considering that the P-MSTM can be applied to ``big datasets.'' It is absolutely crucial that modern statistical methodology be computationally feasible, since ``big data'' has become the norm with sizes that are ever-increasing. Thus, in this article we demonstrated that the P-MSTM is computationally feasible for a big dataset (of 4,089,755 observations) consisting of count-valued QWIs obtained from US Census Bureau's LEHD program. Furthermore, the P-MSTM was shown to give small in-sample errors. Using an empirically motivated simulation study, we also show that the P-MSTM leads to small out-of-sample errors.

The P-MSTM is flexible enough to allow for many different specifications. For example, one could use a different class of areal basis functions, point referenced basis functions, a different class of propagator matrices, and different parameter models (or even estimates) for covariances. Thus, there are many exciting open research directions, that build on this new distributional framework for count-data.

\section*{Acknowledgments} This research was partially supported by the U.S. National Science Foundation (NSF) and the U.S. Census Bureau under NSF grant SES-1132031, funded through the NSF-Census Research Network (NCRN) program.

\section*{Appendix: Proofs}
\renewcommand{\theequation}{A.\arabic{equation}}
\setcounter{equation}{0}
In this appendix we provide {proofs for the} technical results stated in the paper. 

\noindent
\paragraph{\large{Proof of Theorem 1:}}
\normalsize
Theorem 1($i$): From (\ref{univ_LG}) we see that the distribution of the random vector \textbf{w} in (\ref{linear_comb}) is given by,\\
\begin{equation*}
\left(\prod_{i = 1}^{m}\frac{1}{\Gamma(\alpha_{i})\kappa_{i}^{\alpha_{i}}}\right)\mathrm{exp}\left\lbrace\bm{\alpha}^{\prime}\textbf{w} - \bm{\kappa}^{(-1)\prime}\mathrm{exp}(\textbf{w})\right\rbrace;\hspace{10pt} \textbf{w} \in \mathbb{R}^{m}.
\end{equation*}
The inverse of the transform of (\ref{linear_comb}) is given by $\textbf{w} = \textbf{V}^{-1}(\textbf{q} - \textbf{c})$, and the Jacobian is given by $|\mathrm{det}(\textbf{V}^{-1})| = \mathrm{det}(\textbf{V}^{-1\prime}\textbf{V}^{-1})^{1/2} = \mathrm{det}(\textbf{V}\textbf{V}^{\prime})^{-1/2}$. Then, by a change-of-variables \citep[e.g., see][]{casellaBerger}, we have that the pdf of $\textbf{q}$ is given by,
\begin{align}
\nonumber
&\frac{1}{\mathrm{det}(\textbf{V}\textbf{V}^{\prime})^{1/2}}\left(\prod_{i = 1}^{m}\frac{1}{\Gamma(\alpha_{i})\kappa_{i}^{\alpha_{i}}}\right)\mathrm{exp}\left[\bm{\alpha}^{\prime}\textbf{V}^{-1}(\textbf{q} - \textbf{c}) - \bm{\kappa}^{(-1)\prime}\mathrm{exp}\left\lbrace\textbf{V}^{-1}(\textbf{q} - \textbf{c})\right\rbrace\right];\hspace{5pt} \textbf{q} \in \mathbb{R}^{m};\hspace{10pt}\textbf{q} \in \mathbb{R}^{m}.
\end{align}
\noindent
This completes the proof of Theorem~1($i$). 

Theorem 1($ii$): Take the mean and covariance of $\textbf{q}$ in (\ref{linear_comb}) to obtain,
\begin{align}
\nonumber
& E(\textbf{q}\vert \bm{\alpha},\bm{\kappa}) = \textbf{c}+\textbf{V}E(\textbf{w}\vert \bm{\alpha},\bm{\kappa})\\
\nonumber
& \mathrm{cov}(\textbf{q}\vert \bm{\alpha},\bm{\kappa}) = \textbf{V}\hspace{2pt}\mathrm{diag}\left\lbrace\mathrm{cov}(\textbf{w}\vert \bm{\alpha},\bm{\kappa})\right\rbrace\textbf{V}^{\prime}.
\end{align}
Because the elements of $\textbf{w}$ are independent, we only require the mean and variance of of the LG random variable ${w}_{i}$ for $i = 1,\ldots,m$. This result is well known \citep[e.g., see][among others]{Prentice} and given by
\begin{align*}
& E(w_{i}\vert \alpha_{i},\kappa_{i}) =\omega_{0}(\alpha_{i}) +  \mathrm{log}(\kappa_{i})\\
& \mathrm{Var(w_{i}}\vert \alpha_{i},\kappa_{i}) = \omega_{1}(\alpha_{i});\hspace{5pt} i = 1,\ldots,m,
\end{align*}
\noindent
where the function $\omega_{k}$, for non-negative integer $k$, is the polygamma function, and for a real value $z$ we have that $\omega_{k}(z)\equiv \frac{d^{k+1}}{dz^{k+1}} \mathrm{log}\left\lbrace\Gamma(z)\right\rbrace$. This proves Theorem~1($ii$).\\

\noindent
\paragraph{\large{Proof of Proposition 1:}}
\normalsize
 Consider the transformation $Q = \alpha^{1/2}W$, where $W\sim \mathrm{LG}\left(\alpha,\frac{1}{\alpha}\right)$. Then we have that
\begin{equation*}
f(Q\vert \alpha,\kappa)\propto \mathrm{exp}\left\lbrace\frac{\alpha}{\alpha^{1/2}}Q - \alpha\hspace{2pt}\mathrm{exp}\left(\frac{1}{\alpha^{1/2}}Q\right)\right\rbrace,
\end{equation*}
\noindent
and using the Taylor Series expansion of $\mathrm{exp}(x)$ we have
\begin{align*}
& f(Q\vert \alpha,\kappa)\propto \mathrm{exp}\left[\alpha^{1/2}Q - \alpha\left\lbrace\frac{1}{\alpha^{1/2}}Q + \frac{1}{2\alpha}Q^{2} + O\left(\frac{Q^{3}}{\alpha^{3/2}}\right)\right\rbrace\right],
\end{align*}
\noindent
where ``$O(\cdot)$'' is the ``Big-O'' notation \citep[e.g., see][among others]{lehman}. Then, letting $\alpha$ go to infinity yields,
\begin{align*}
& \underset{\alpha \rightarrow \infty}{\mathrm{lim}}f(Q\vert \alpha,\kappa)\propto \mathrm{exp}\left(-\frac{Q^{2}}{2}\right) \propto \mathrm{Normal}(0,1).
\end{align*}
\noindent
Thus, $Q$ converges in distribution to a standard normal distribution as $\alpha$ goes to infinity. Now suppose $\textbf{w} = (w_{1},\ldots.,w_{m})^{\prime}\sim \mathrm{MLG}\left(\bm{0}_{m},\alpha^{1/2}\textbf{I}_{m}, \alpha, \frac{1}{\alpha}\right)$. Then it follows from the result above that $\alpha^{1/2}\textbf{w}$ converges to a standard Gaussian distribution. By ``standard Gaussian distribution'' we mean a multivariate normal distribution with mean zero and $m\times m$ identity covariance matrix. Now define the transformation,
\begin{equation*}
\textbf{q} = \textbf{c}+\textbf{V}(\alpha^{1/2}\textbf{w}),
\end{equation*}
which follows a $\mathrm{MLG}(\textbf{c},\alpha^{1/2}\textbf{V},\alpha\bm{1},\frac{1}{\alpha}\bm{1})$. It follows from Theorem~5.1.8 of \citet{lehman}, and the fact that  $\alpha^{1/2}\textbf{w}$ converges to a standard Gaussian distribution, that $\textbf{q}$ converges in distribution to a multivariate normal distribution with mean $\textbf{c}$ and covariance matrix $\textbf{V}\textbf{V}^{\prime}$.\\

\noindent
\paragraph{\large{Proof of Proposition 2:}}
\normalsize
Proposition 2($i$): It follows from Theorem~1($i$) that the conditional distribution is given by\\
\begin{align*}
f(\textbf{q}_{1}\vert \textbf{q}_{2},\textbf{c},\textbf{V},\bm{\alpha},\bm{\kappa}) &= \frac{\left[f(\textbf{q}\vert \textbf{c},\textbf{V},\bm{\alpha},\bm{\kappa})\right]_{\textbf{q}_{2} = \textbf{d}}}{\left[\int f(\textbf{q}\vert \textbf{c},\textbf{V},\bm{\alpha},\bm{\kappa})d\textbf{q}_{1}\right]_{\textbf{q}_{2} = \textbf{d}}}\\
&=M\hspace{5pt}\mathrm{exp}\left\lbrace\bm{\alpha}^{\prime}\textbf{H}\textbf{q}_{1} - \bm{\kappa}_{1.2}^{(-1)\prime}\mathrm{exp}(\textbf{H}\textbf{q}_{1})\right\rbrace;\hspace{5pt}\textbf{q} \in \mathbb{R}^{m},
\end{align*}
\noindent
which proves the result. To prove Proposition 2($ii$), we see that from Proposition 2($i$) that
\begin{align*}
f(\textbf{q}_{1}\vert \textbf{q}_{2} = \bm{0}_{m-g},\textbf{c}=\bm{0}_{m},\textbf{V},\bm{\alpha},\bm{\kappa}_{1.2})&\propto \mathrm{exp}\left\lbrace\bm{\alpha}^{\prime}\textbf{H}\textbf{q}_{1} - \bm{\kappa}_{1.2}^{(-1)\prime}\mathrm{exp}(\textbf{H}\textbf{q}_{1})\right\rbrace\\
&\propto f(\textbf{q}_{1}\vert \textbf{q}_{2} = \textbf{d},\textbf{c},\textbf{V},\bm{\alpha},\bm{\kappa}).
\end{align*}
This proves Proposition 2($ii$).\\

\noindent
\paragraph{\large{Proof of Theorem 2:}}
\normalsize
Theorem~2($ii$) is used within the proof of Theorem~2($i$). Thus, we shall start by proving Theorem~2($ii$): Notice that 
  \begin{equation*}
  \textbf{V}=
  \left[
   \begin{array}{c}
   (\textbf{H}^{\prime}\textbf{H})^{-1}\textbf{H}^{\prime} \\ 
   \sigma_{2}\textbf{Q}_{2}^{\prime}
   \end{array}\right].
  \end{equation*}
\noindent
From (\ref{linear_comb}) we see that
  \begin{equation}\label{margstep1}
  \left[
   \begin{array}{c}
   \textbf{q}_{1} \\ 
   \textbf{q}_{2}
   \end{array}\right] = \left[
      \begin{array}{c}
      (\textbf{H}^{\prime}\textbf{H})^{-1}\textbf{H}^{\prime}\textbf{w} \\ 
      \sigma_{2}\textbf{Q}_{2}^{\prime}\textbf{w}
      \end{array}\right],
  \end{equation}
\noindent
where the $m$-dimensional random vector $\textbf{w}\sim \mathrm{MLG}(\bm{0}_{m},\textbf{I}_{m},\bm{\alpha},\bm{\kappa})$. Multiplying both sides of (\ref{margstep1}) by $[\textbf{I}_{g},\bm{0}_{g,m-g}]$ we have
\begin{equation*}
\textbf{q}_{1} = (\textbf{H}^{\prime}\textbf{H})^{-1}\textbf{H}^{\prime}\textbf{w},
\end{equation*}
\noindent
which is the desired result.

Theorem 2($i$): From Proposition 2($ii$) the conditional distribution of $\textbf{q}_{1}\vert \textbf{q}_{2},\textbf{c} = \bm{0}_{m},\textbf{V},\bm{\alpha},\bm{\kappa}$ has the following pdf:
\begin{align*}
f(\textbf{q}_{1}\vert \textbf{q}_{2},\textbf{c} = \bm{0}_{m},\textbf{V},\bm{\alpha},\bm{\kappa})&=f(\textbf{q}_{1}\vert \textbf{q}_{2} = \bm{0}_{m-g},\textbf{c} = \bm{0}_{m},\textbf{V},\bm{\alpha},\bm{\kappa}_{1.2})\\
& =M\hspace{5pt}\mathrm{exp}\left\lbrace\bm{\alpha}^{\prime}\textbf{H}\textbf{q}_{1} - \bm{\kappa}_{1.2}^{(-1)\prime}\mathrm{exp}(\textbf{H}\textbf{q}_{1})\right\rbrace,
\end{align*}
where, recall $\bm{\kappa}_{1.2}^{(-1)}\equiv \mathrm{exp}\left\lbrace\frac{1}{\sigma_{2}}\textbf{Q}_{2}\textbf{q}_{2} + \mathrm{log}(\bm{\kappa}^{(-1)})\right\rbrace$ and the normalizing constant $M$ is
\begin{align*}
M &= |\mathrm{det}([\textbf{H}\hspace{5pt}\frac{1}{\sigma_{2}}\textbf{Q}_{2}])|\left(\prod_{i = 1}^{m}\frac{1}{\Gamma(\alpha_{i})\kappa_{1.2,i}^{\alpha_{i}}}\right)\frac{1}{\left[\int f(\textbf{q}\vert \textbf{c} = \bm{0}_{m},\textbf{V} = [\textbf{H}\hspace{5pt}\frac{1}{\sigma_{2}}\textbf{Q}_{2}]^{-1},\bm{\alpha},\bm{\kappa}_{1.2})d\textbf{q}_{1}\right]_{\textbf{q}_{2} = \bm{0}_{m-g}}}\\
&= |\mathrm{det}([\textbf{H}\hspace{5pt}\textbf{Q}_{2}])|\left(\prod_{i = 1}^{m}\frac{1}{\Gamma(\alpha_{i})\kappa_{1.2,i}^{\alpha_{i}}}\right)\frac{1}{\left[\int f(\textbf{q}\vert \textbf{c} = \bm{0}_{m},\textbf{V} = [\textbf{H}\hspace{5pt}\textbf{Q}_{2}]^{-1},\bm{\alpha},\bm{\kappa}_{1.2})d\textbf{q}_{1}\right]_{\textbf{q}_{2} = \bm{0}_{m-g}}},
\end{align*}
where $1/\kappa_{1.2,i}$ is the $i$-th element of $\bm{\kappa}_{1.2}^{(-1)}$. Thus, the only quantity in the conditional pdf $f(\textbf{q}_{1}\vert \textbf{q}_{2},\textbf{c} = \bm{0}_{m},\textbf{V},\bm{\alpha},\bm{\kappa})$ that depends on $\textbf{q}_{2}$ and $\sigma_{2}$ is $\bm{\kappa}_{1.2}$. This gives us that, 
\begin{equation}\label{limcond}
\underset{\sigma_{2}\rightarrow \infty}{\mathrm{lim}}f(\textbf{q}_{1}\vert \textbf{q}_{2} = \bm{0}_{m-g},\textbf{c} = \bm{0}_{m},\textbf{V},\bm{\alpha},\bm{\kappa}_{1.2}) = M_{1}\hspace{5pt}\mathrm{exp}\left\lbrace\bm{\alpha}^{\prime}\textbf{H}\textbf{q}_{1} - \bm{\kappa}^{(-1)\prime}\mathrm{exp}(\textbf{H}\textbf{q}_{1})\right\rbrace,
\end{equation}
\noindent
where $M_{1}$ is defined in Theorem~2($i$). Notice that the limit in (\ref{limcond}) does not depend on $\textbf{q}_{2}$. Let
\begin{equation*}
\rho(\textbf{q}_{1},\textbf{H},\bm{\alpha},\bm{\kappa}) \equiv M_{1}\hspace{5pt}\mathrm{exp}\left\lbrace\bm{\alpha}^{\prime}\textbf{H}\textbf{q}_{1} - \bm{\kappa}^{(-1)\prime}\mathrm{exp}(\textbf{H}\textbf{q}_{1})\right\rbrace,
\end{equation*}
\noindent
so that $\rho(\textbf{q}_{1},\textbf{H},\bm{\alpha},\bm{\kappa}) = \underset{\sigma_{2}\rightarrow \infty}{\mathrm{lim}}f(\textbf{q}_{1}\vert \textbf{q}_{2} ,\textbf{c} = \bm{0}_{m},\textbf{V},\bm{\alpha},\bm{\kappa}_{1.2})$.

Now, it follows from Theorem~2($ii$) that $f(\textbf{q}_{1}\vert \textbf{c}= \bm{0}_{m},\textbf{V},\bm{\alpha},\bm{\kappa})$ does not depend on $\sigma_{2}$. This implies that
\begin{align*}
f(\textbf{q}_{1}\vert \textbf{c} = \bm{0}_{m},\textbf{V},\bm{\alpha},\bm{\kappa})&=\underset{\sigma_{2}\rightarrow \infty}{\mathrm{lim}}f(\textbf{q}_{1}\vert \textbf{c} = \bm{0}_{m},\textbf{V},\bm{\alpha},\bm{\kappa})\\
&=\underset{\sigma_{2}\rightarrow \infty}{\mathrm{lim}}\int f(\textbf{q}_{1}\vert \textbf{q}_{2},\textbf{c} = \bm{0}_{m},\textbf{V},\bm{\alpha},\bm{\kappa})f(\textbf{q}_{2}\vert \textbf{c} = \bm{0}_{m},\textbf{V},\bm{\alpha},\bm{\kappa})d\textbf{q}_{2}\\
&=\underset{\sigma_{2}\rightarrow \infty}{\mathrm{lim}}\int f(\textbf{q}_{1}\vert \textbf{q}_{2} = \bm{0}_{m-g},\textbf{c} = \bm{0}_{m},\textbf{V},\bm{\alpha},\bm{\kappa}_{1.2})f(\textbf{q}_{2}\vert \textbf{c} = \bm{0}_{m},\textbf{V},\bm{\alpha},\bm{\kappa})d\textbf{q}_{2}\\
&=\int \underset{\sigma_{2}\rightarrow \infty}{\mathrm{lim}}f(\textbf{q}_{1}\vert \textbf{q}_{2} = \bm{0}_{m-g},\textbf{c} = \bm{0}_{m},\textbf{V},\bm{\alpha},\bm{\kappa}_{1.2})\underset{\sigma_{2}\rightarrow \infty}{\mathrm{lim}}f(\textbf{q}_{2}\vert \textbf{c} = \bm{0}_{m},\textbf{V},\bm{\alpha},\bm{\kappa})d\textbf{q}_{2}\\
&=\int \rho(\textbf{q}_{1},\textbf{H},\bm{\alpha},\bm{\kappa})\underset{\sigma_{2}\rightarrow \infty}{\mathrm{lim}}f(\textbf{q}_{2}\vert \textbf{c} = \bm{0}_{m},\textbf{V},\bm{\alpha},\bm{\kappa})d\textbf{q}_{2}\\
&=\rho(\textbf{q}_{1},\textbf{H},\bm{\alpha},\bm{\kappa})\underset{\sigma_{2}\rightarrow \infty}{\mathrm{lim}}\int f(\textbf{q}_{2}\vert \textbf{c} = \bm{0}_{m},\textbf{V},\bm{\alpha},\bm{\kappa})d\textbf{q}_{2}\\
&=\rho(\textbf{q}_{1},\textbf{H},\bm{\alpha},\bm{\kappa}).
\end{align*}
\noindent
Thus, $\rho(\textbf{q}_{1},\textbf{H},\bm{\alpha},\bm{\kappa})$ is the marginal pdf of $\textbf{q}_{1}$, which is the desired result.\\

\noindent
\paragraph{\large{Proof of Proposition 3:}} Finding the full-conditional distributions associated with the P-MSTM in (\ref{summary}) is a matter of algebra, however, there is a considerable amount of bookkeeping that is required. To aide the reader, we have included Table 1, which organizes all the bookkeeping that is required. Additionally, we organize the proof of each full conditional distribution systematically in the list below. 
\begin{enumerate}
\item From (\ref{summary}), the full conditional distribution for $\bm{\beta}$ satisfies
      \begin{align}
      \nonumber
      f(\bm{\beta}\vert \cdot)& \propto f(\bm{\beta})\prod_{t = 1}^{T}f(\textbf{Z}_{t}\vert \bm{\beta},\bm{\eta}_{t}, \bm{\xi}_{t})\\
       \nonumber
      &\propto \hspace{5pt}\mathrm{exp}\left[\sum_{t = 1}^{T}\bm{Z}_{t}^{\prime}\textbf{X}_{t}\bm{\beta} - \sum_{t = 1}^{T}\underset{A \in D_{\mathrm{O},T}^{(\ell)}}{\sum}\sum_{\ell=1}^{L}\mathrm{exp}\left\lbrace\bm{\psi}_{t}^{(\ell)}(A)^{\prime}\bm{\eta} + \xi_{t}^{(\ell)}(A)\right\rbrace^{\prime}\mathrm{exp}\left\lbrace\textbf{x}_{t}(A)^{\prime}\bm{\beta}\right\rbrace\right]\\
      \nonumber
      &\hspace{5pt} \times \mathrm{exp}\left\lbrace\alpha_{k}^{1-1/2}\sigma_{\beta}^{-1}\bm{1}_{p}^{\prime}\bm{\beta}-\frac{1}{\kappa_{k}}\bm{1}_{p}^{\prime}\mathrm{exp}\left(\alpha_{k}^{-1/2}\sigma_{\beta}^{-1}\bm{\beta}\right)\right\rbrace.
       \end{align}
       Rearranging terms we have
        \begin{equation*}
       f(\bm{\beta}\vert \cdot) \propto \mathrm{exp}\left\lbrace \bm{\alpha}_{\beta,k}^{\prime}\textbf{H}_{\beta,k}\bm{\beta}-\bm{\kappa}_{\beta,k}^{(-1)\prime}\mathrm{exp}\left(\textbf{H}_{\beta,k}\bm{\beta}\right)\right\rbrace,
        \end{equation*}
        \noindent
        which implies that $f(\bm{\beta}\vert \cdot)$ is equal to $\mathrm{mMLG}(\textbf{H}_{\beta,k},\bm{\alpha}_{\beta,k},\bm{\kappa}_{\beta,k})$. 
\item If $T>1$, then it follows from (\ref{summary}) that for $1<t<T$ the full conditional distribution for $\bm{\eta}_{t}$ satisfies
\begin{align}
\nonumber
f(\bm{\eta}_{t}\vert \cdot)& \propto f(\textbf{Z}_{t}\vert \bm{\beta},\bm{\eta}_{t}, \bm{\xi}_{t})f(\bm{\eta}_{t}\vert \bm{\eta}_{t-1},\sigma_{\mathrm{K}})f(\bm{\eta}_{t+1}\vert \bm{\eta}_{t},\sigma_{\mathrm{K}})\\
 \nonumber
&\propto \hspace{5pt}\mathrm{exp}\left[\bm{Z}_{t}^{\prime}\bm{\Psi}_{t}\bm{\eta}_{t} - \underset{A \in D_{\mathrm{O},t}^{(\ell)}}{\sum}\sum_{\ell=1}^{L}\mathrm{exp}\left\lbrace\textbf{x}_{t}(A)^{\prime}\bm{\beta} + \xi_{t}^{(\ell)}(A)\right\rbrace^{\prime}\mathrm{exp}\left\lbrace\bm{\psi}_{t}^{(\ell)}(A)^{\prime}\bm{\eta}_{t}\right\rbrace\right]\\
\nonumber
&\hspace{5pt} \times \mathrm{exp}\left\lbrace\alpha_{k}\bm{1}_{r}^{\prime}\textbf{W}_{t,k}^{-1/2}\bm{\eta}_{t}-\frac{1}{\kappa_{k}}\mathrm{exp}\left(-\textbf{W}_{t,k}^{-1/2}\textbf{M}_{t}\bm{\eta}_{t-1}\right)^{\prime}\mathrm{exp}\left(\textbf{W}_{t,k}^{-1/2}\bm{\eta}_{t}\right)\right\rbrace\\
 \nonumber
 &\hspace{5pt} \times \mathrm{exp}\left\lbrace -\alpha_{k}\bm{1}_{r}^{\prime}\textbf{W}_{t+1,k}^{-1/2}\textbf{M}_{t+1}\bm{\eta}_{t}-\frac{1}{\kappa_{k}}\mathrm{exp}\left(\textbf{W}_{t+1,k}^{-1/2}\bm{\eta}_{t+1}\right)^{\prime}\mathrm{exp}\left(-\textbf{W}_{t+1,k}^{-1/2}\textbf{M}_{t+1}\bm{\eta}_{t}\right)\right\rbrace.
\end{align}
 Again, rearranging terms we have
 \begin{equation*}
f(\bm{\eta}_{t}\vert \cdot) \propto \mathrm{exp}\left\lbrace \bm{\alpha}_{\eta,t,k}^{\prime}\textbf{H}_{\eta,t,k}\bm{\eta}_{t}-\bm{\kappa}_{\eta,t,k}^{(-1)\prime}\mathrm{exp}\left(\textbf{H}_{\eta,t,k}\bm{\eta}_{t}\right)\right\rbrace,
 \end{equation*}
 \noindent
 which implies that $f(\bm{\eta}_{t}\vert \cdot)$ is equal to $\mathrm{mMLG}(\textbf{H}_{\eta,t,k},\bm{\alpha}_{\eta,t,k},\bm{\kappa}_{\eta,t,k})$. 
 
\item If $T>1$, then it follows from (\ref{summary}), that the full conditional distribution for $\bm{\eta}_{1}$ satisfies
\begin{align}
\nonumber
f(\bm{\eta}_{1}\vert \cdot)& \propto f(\textbf{Z}_{1}\vert \bm{\beta},\bm{\eta}_{1}, \bm{\xi}_{t})f(\bm{\eta}_{1}\vert \sigma_{\mathrm{K}})f(\bm{\eta}_{2}\vert \bm{\eta}_{1},\sigma_{\mathrm{K}})\\
 \nonumber
&\propto \hspace{5pt}\mathrm{exp}\left[\bm{Z}_{1}^{\prime}\bm{\Psi}_{1}\bm{\eta}_{1} - \underset{A \in D_{\mathrm{O},1}^{(\ell)}}{\sum}\sum_{\ell=1}^{L}\mathrm{exp}\left\lbrace\textbf{x}_{1}(A)^{\prime}\bm{\beta} + \xi_{1}^{(\ell)}(A)\right\rbrace^{\prime}\mathrm{exp}\left\lbrace\bm{\psi}_{1}^{(\ell)}(A)^{\prime}\bm{\eta}\right\rbrace\right]\\
\nonumber
&\hspace{5pt} \times \mathrm{exp}\left\lbrace\alpha_{k}\bm{1}_{r}^{\prime}\textbf{W}_{1,k}^{-1/2}\bm{\eta}_{1}-\frac{1}{\kappa_{k}}\bm{1}_{r}^{\prime}\mathrm{exp}\left(\textbf{W}_{1,k}^{-1/2}\bm{\eta}_{1}\right)\right\rbrace\\
 \nonumber
 &\hspace{5pt} \times \mathrm{exp}\left\lbrace -\alpha_{k}\bm{1}_{r}^{\prime}\textbf{W}_{2,k}^{-1/2}\textbf{M}_{2}\bm{\eta}_{1}-\frac{1}{\kappa_{k}}\mathrm{exp}\left(\textbf{W}_{2,k}^{-1/2}\bm{\eta}_{2}\right)^{\prime}\mathrm{exp}\left(-\textbf{W}_{2,k}^{-1/2}\textbf{M}_{2}\bm{\eta}_{1}\right)\right\rbrace.
 \end{align}
 Again, rearranging terms we have
  \begin{equation*}
 f(\bm{\eta}_{1}\vert \cdot) \propto \mathrm{exp}\left\lbrace \bm{\alpha}_{\eta,1,k}^{\prime}\textbf{H}_{\eta,1,k}\bm{\eta}_{t}-\bm{\kappa}_{\eta,1,k}^{(-1)\prime}\mathrm{exp}\left(\textbf{H}_{\eta,1,k}\bm{\eta}_{1}\right)\right\rbrace,
  \end{equation*}
  \noindent
  which implies that $f(\bm{\eta}_{1}\vert \cdot)$ is equal to $\mathrm{mMLG}(\textbf{H}_{\eta,1,k},\bm{\alpha}_{\eta,1,k},\bm{\kappa}_{\eta,1,k})$. 
\item From (\ref{summary}), the full conditional distribution for $\bm{\eta}_{T}$ satisfies
  \begin{align}\label{etaTfull}
  \nonumber
  f(\bm{\eta}_{T}\vert \cdot)& \propto f(\textbf{Z}_{T}\vert \bm{\beta},\bm{\eta}_{T}, \bm{\xi}_{T})f(\bm{\eta}_{T}\vert \bm{\eta}_{T-1},\sigma_{\mathrm{K}})\\
 \nonumber
  &\propto \hspace{5pt}\mathrm{exp}\left[\bm{Z}_{T}^{\prime}\bm{\Psi}_{T}\bm{\eta}_{T} - \underset{A \in D_{\mathrm{O},T}^{(\ell)}}{\sum}\sum_{\ell}^{L}\mathrm{exp}\left\lbrace\textbf{x}_{T}(A)^{\prime}\bm{\beta} + \xi_{T}^{(\ell)}(A)\right\rbrace^{\prime}\mathrm{exp}\left\lbrace\bm{\psi}_{T}^{(\ell)}(A)^{\prime}\bm{\eta}\right\rbrace\right]\\
  &\hspace{5pt} \times \mathrm{exp}\left\lbrace\alpha_{k}\bm{1}_{r}^{\prime}\textbf{W}_{T,k}^{-1/2}\bm{\eta}_{T}-\frac{1}{\kappa_{k}}\mathrm{exp}\left(-\textbf{W}_{T,k}^{-1/2}\textbf{M}_{T}\bm{\eta}_{T-1}\right)^{\prime}\mathrm{exp}\left(\textbf{W}_{T,k}^{-1/2}\bm{\eta}_{T}\right)\right\rbrace.
   \end{align}
   Again, rearranging terms we have
    \begin{equation*}
   f(\bm{\eta}_{T}\vert \cdot) \propto \mathrm{exp}\left\lbrace \bm{\alpha}_{\eta,T,k}^{\prime}\textbf{H}_{\eta,T,k}\bm{\eta}_{T}-\bm{\kappa}_{\eta,T,k}^{(-1)\prime}\mathrm{exp}\left(\textbf{H}_{\eta,T,k}\bm{\eta}_{T}\right)\right\rbrace,
    \end{equation*}
    \noindent
    which implies that $f(\bm{\eta}_{T}\vert \cdot)$ is equal to $\mathrm{mMLG}(\textbf{H}_{\eta,T,k},\bm{\alpha}_{\eta,T,k},\bm{\kappa}_{\eta,T,k})$. If $T=1$ then replace $\textbf{M}_{T}$ and $\bm{\eta}_{T-1}$ with $\bm{0}_{r,r}$ and $\bm{0}_{r}$ within the expression of (\ref{etaTfull}).
\item From (\ref{summary}), we have that for each $t$ the full conditional distribution for $\bm{\xi}_{t}$ satisfies
        \begin{align}
        \nonumber
        f(\bm{\xi}_{t}\vert \cdot)& \propto f(\textbf{Z}_{t}\vert \bm{\beta},\bm{\eta}_{t}, \bm{\xi}_{t})f(\bm{\xi}_{t}\vert \sigma_{\xi})\\
         \nonumber
        &\propto \hspace{5pt}\mathrm{exp}\left[\bm{Z}_{t}^{\prime}\bm{\xi}_{t}- \underset{A \in D_{\mathrm{O},t}^{(\ell)}}{\sum}\sum_{\ell}^{L}\mathrm{exp}\left\lbrace\textbf{x}_{t}(A)^{\prime}\bm{\beta} + \bm{\psi}_{t}^{(\ell)}(A)^{\prime}\bm{\eta} \right\rbrace^{\prime}\mathrm{exp}\left\lbrace\xi_{t}^{(\ell)}(A)\right\rbrace\right]\\
              \nonumber
              &\hspace{5pt} \times \mathrm{exp}\left\lbrace\alpha_{k}^{1-1/2}\sigma_{\xi}^{-1}\bm{1}_{n_{t}}^{\prime}\bm{\xi}_{t}-\frac{1}{\kappa_{k}}\bm{1}_{p}^{\prime}\mathrm{exp}\left(\alpha_{k}^{-1/2}\sigma_{\xi}^{-1}\bm{\xi}_{t}\right)\right\rbrace.
        \end{align}
         Rearranging terms we have
         \begin{equation*}
 f(\bm{\xi}_{t}\vert \cdot) \propto \mathrm{exp}\left\lbrace \bm{\alpha}_{\xi,t,k}^{\prime}\textbf{H}_{\xi,t,k}\bm{\xi}_{t}-\bm{\kappa}_{\xi,t,k}^{(-1)\prime}\mathrm{exp}\left(\textbf{H}_{\xi,t,k}\bm{\xi}_{t}\right)\right\rbrace,
         \end{equation*}
         \noindent
         which implies that $f(\bm{\xi}_{t}\vert \cdot)$ is equal to a $\mathrm{mMLG}(\textbf{H}_{\xi,t,k},\bm{\alpha}_{\xi,t,k},\bm{\kappa}_{\xi,t,k})$.
\item Notice that, since Process Models 1 and 2 are the only quantities that include $\sigma_{K}$ we have that
         \begin{equation}\label{discrete_fc}
         f(\sigma_{\mathrm{K}}\vert \cdot) = \frac{f(\bm{\eta}_{1}\vert \sigma_{K})\prod_{i = 2}^{T}f(\bm{\eta}_{t}\vert \bm{\eta}_{t-1},\sigma_{K})}{\sum_{\sigma_{\mathrm{K}}}f(\bm{\eta}_{1}\vert \sigma_{K})\prod_{i = 2}^{T}f(\bm{\eta}_{t}\vert \bm{\eta}_{t-1},\sigma_{K})}.
         \end{equation}
 Then substituting (\ref{mlg_pdf}) into (\ref{discrete_fc}), we obtain 
   \begin{align}\label{K_fc}
 \nonumber
 & p_{\mathrm{K}}(\sigma_{\mathrm{K}})=\\
 &\frac{\prod_{t=1}^{T}\mathrm{exp}\left\lbrace\frac{\alpha_{k}}{\sigma_{\mathrm{K}}}\bm{1}_{r}^{\prime}\textbf{W}_{t,k}^{-1/2}\bm{\eta}_{t} - \frac{\alpha_{k}}{\sigma_{\mathrm{K}}}\bm{1}_{r}^{\prime}\textbf{W}_{t,k}^{-1/2}\textbf{M}_{t}\bm{\eta}_{t-1} - \frac{1}{\kappa_{k}}\bm{1}_{r}^{\prime}\mathrm{exp}(\frac{1}{\sigma_{\mathrm{K}}}\textbf{W}_{t,k}^{-1/2}\bm{\eta}_{t} -\frac{1}{\sigma_{\mathrm{K}}}\textbf{W}_{t,k}^{-1/2}\textbf{M}_{t}\bm{\eta}_{t-1})\right\rbrace}{\sum_{\sigma_{\mathrm{K}}}\prod_{t=1}^{T}\mathrm{exp}\left\lbrace\frac{1}{\sigma_{\mathrm{K}}}\alpha_{k}\bm{1}_{r}^{\prime}\textbf{W}_{t,k}^{-1/2}\bm{\eta}_{t} - \frac{1}{\sigma_{\mathrm{K}}}\alpha_{k}\bm{1}_{r}^{\prime}\textbf{W}_{t,k}^{-1/2}\textbf{M}_{t}\bm{\eta}_{t-1} - \frac{1}{\kappa_{k}}\bm{1}_{r}^{\prime}\mathrm{exp}(\frac{1}{\sigma_{\mathrm{K}}}\textbf{W}_{t,k}^{-1/2}\bm{\eta}_{t} -\frac{1}{\sigma_{\mathrm{K}}}\textbf{W}_{t,k}^{-1/2}\textbf{M}_{t}\bm{\eta}_{t-1})\right\rbrace}.
   \end{align}
\nonumber
The derivation of $f(\sigma_{\xi}\vert \cdot)$, in Table 1,follows similarly.\\
 \end{enumerate}
 
 \noindent
 This completes the proof.
 \bibliographystyle{jasa}  
 \bibliography{myref}
 \section*{Figures and Tables}
 
 \begin{table}[H]
 \centering
 \noindent\adjustbox{max width=\textwidth}{%
 \begin{tabular}{ |p{0.85\textwidth}|p{0.5\textwidth}|  }
 \hline
 	Definition & Additional Notes\\ \hline
 $\textbf{H}_{\beta,k} = (\textbf{X}_{1}^{\prime},\ldots,\textbf{X}_{T}^{\prime},\alpha_{k}^{-1/2}\sigma_{\beta}^{-1}\textbf{I}_{p})^{\prime}$	&	  \\ \hline
 $\textbf{H}_{\eta,t,k} = (\bm{\Psi}_{t}^{\prime}, \textbf{W}_{t,k}^{-1/2},-\textbf{W}_{t+1,k}^{-1/2}\textbf{M}_{t+1})^{\prime}$	&	  \\ \hline
 $\textbf{H}_{\eta,T,k} = (\bm{\Psi}_{T}^{\prime}, \textbf{W}_{T,k}^{-1/2})^{\prime}$	&	  \\ \hline
 $\textbf{H}_{\xi,t,k} = (\textbf{I}_{n_{t}}, \alpha_{k}^{-1/2}\sigma_{\xi}^{-1}\textbf{I}_{n_{t}})^{\prime}$	&	  \\ \hline
 $\bm{\kappa}_{\beta,k}^{(-1)} = \left\lbrace\mathrm{exp}(\bm{\Psi}_{1}\bm{\eta}_{1}+ \bm{\xi}_{1})^{\prime}, \ldots, \mathrm{exp}(\bm{\Psi}_{T}\bm{\eta}_{T}+ \bm{\xi}_{T})^{\prime},\frac{1}{\kappa_{k}}\bm{1}_{p}^{(1)\prime}\right\rbrace^{\prime}$	&	  \\ \hline
 $\bm{\kappa}_{\eta,t,k}^{(-1)} = \left\lbrace\mathrm{exp}(\textbf{X}_{t}\bm{\beta} + \bm{\xi}_{t})^{\prime}, \frac{1}{\kappa_{k}}\mathrm{exp}(-\textbf{W}_{t,k}^{-1/2}\textbf{M}_{t}\bm{\eta}_{t-1})^{\prime}, \frac{1}{\kappa_{k}}\mathrm{exp}(\textbf{W}_{t+1,k}^{-1/2}\bm{\eta}_{t+1})^{\prime}\right\rbrace^{\prime}$	&	$1 <t<T$  \\ \hline
 $\bm{\kappa}_{\eta,1,k}^{(-1)} = \left\lbrace\mathrm{exp}(\textbf{X}_{1}\bm{\beta} + \bm{\xi}_{1})^{\prime}, \frac{1}{\kappa_{k}}\bm{1}_{r}^{\prime}, \frac{1}{\kappa_{k}}\mathrm{exp}(\textbf{W}_{2,k}^{-1/2}\bm{\eta}_{2})^{\prime}\right\rbrace^{\prime}$	&	Provided $T>1$  \\ \hline
 $\bm{\kappa}_{\eta,T,k}^{(-1)} = \left\lbrace\mathrm{exp}(\textbf{X}_{T}\bm{\beta} + \bm{\xi}_{T})^{\prime}, \frac{1}{\kappa_{k}}\mathrm{exp}(-\textbf{W}_{T,k}^{-1/2}\textbf{M}_{T}\bm{\eta}_{T-1})^{\prime}\right\rbrace^{\prime}$	&	If $T=1$ then replace $\textbf{M}_{T}$ and $\bm{\eta}_{T-1}$ with $\bm{0}_{r,r}$ and $\bm{0}_{r}$ within the expression of $\bm{\kappa}_{\eta,T,k}^{(-1)}$ above.  \\ \hline
 $\bm{\kappa}_{\xi,t,k}^{(-1)} = \left\lbrace\mathrm{exp}(\textbf{X}_{t}\bm{\beta}+\bm{\Psi}_{t}\bm{\eta}_{t})^{\prime},\frac{1}{\kappa_{k}}\bm{1}_{n_{t}}^{\prime}\right\rbrace^{\prime}$	&	  \\ \hline
 $\bm{\alpha}_{\beta,k} = \left\lbrace \textbf{Z}_{1}^{\prime}+ d_{\beta,k}\bm{1}_{n_{1}}^{\prime},\ldots,Z_{T}^{\prime}+ d_{\beta,k}\bm{1}_{n_{T}}^{\prime},\alpha_{k}\bm{1}_{p}^{\prime} - d_{\beta,k}\alpha_{k}^{1/2}\sigma_{\beta}\sum_{t = 1}^{T}\bm{1}_{n_{t}}^{\prime}\textbf{X}_{t}\right\rbrace^{\prime}$	&	  \\ \hline
 $\bm{\alpha}_{\eta,t,k} = \left\lbrace \textbf{Z}_{t}^{\prime} + d_{\eta,t,k} \bm{1}_{n_{t}}^{\prime}, \alpha_{k}\bm{1}_{r}^{\prime} - \frac{d_{\eta,t,k}}{2} \bm{1}_{n_{t}}^{\prime}\bm{\Psi}_{t}\textbf{W}_{t,k}^{1/2},\alpha_{k}\bm{1}_{r}^{\prime}+ \frac{d_{\eta,t,k}}{2} \bm{1}_{n_{t}}^{\prime}\bm{\Psi}_{t}\textbf{M}_{t+1}^{\prime}\textbf{W}_{t+1,k}^{1/2} \right\rbrace^{\prime}$	&	$1<t<T$  \\ \hline
 $\bm{\alpha}_{\eta,1,k} = \left\lbrace \textbf{Z}_{1}^{\prime} + d_{\eta,1,k} \bm{1}_{n_{1}}^{\prime}, \alpha_{k}\bm{1}_{r}^{\prime} - \frac{d_{\eta,1,k}}{2} \bm{1}_{n_{1}}^{\prime}\bm{\Psi}_{1}\textbf{W}_{1,k}^{1/2},\alpha_{k}\bm{1}_{r}^{\prime}+ \frac{d_{\eta,1,k}}{2} \bm{1}_{n_{1}}^{\prime}\bm{\Psi}_{1}\textbf{M}_{2}^{\prime}\textbf{W}_{2,k}^{1/2} \right\rbrace^{\prime}$	&	Provided $T>1$  \\ \hline
 $\bm{\alpha}_{\eta,T,k} = \left\lbrace \textbf{Z}_{T}^{\prime} + d_{\eta,T,k} \bm{1}_{n_{T}}^{\prime}, \alpha_{k}\bm{1}_{r}^{\prime} - d_{\eta,1,k} \bm{1}_{n_{1}}^{\prime}\bm{\Psi}_{1}\textbf{W}_{1,k}^{1/2}\right\rbrace^{\prime}$	&	  \\ \hline
 $\bm{\alpha}_{\xi,t,k} = \left\lbrace \textbf{Z}_{t}^{\prime} + d_{\xi,t,k} \bm{1}_{n_{T}}^{\prime}, \alpha_{k}\bm{1}_{n_{t}}^{\prime} - d_{\xi,t,k}\alpha_{k}^{1/2}\sigma_{\xi}\bm{1}_{n_{t}}^{\prime}\right\rbrace^{\prime}$	&	  $1 \le t \le T$\\ \hline
 $d_{\beta,k} = {\alpha_{k}}\Bigg/{\left[1 + \mathrm{max}\left\lbrace \mathrm{abs}\left(\alpha_{k}^{1/2}\sigma_{\beta}\sum_{t = 1}^{T}\bm{1}_{n_{t}}^{\prime}\textbf{X}_{t}\right) \right\rbrace\right]}$	&	  \\ \hline
 $d_{\eta,t,k} = {\alpha_{k}}\Bigg/\left({1 + \mathrm{max}\left[ \mathrm{abs}\left\lbrace (\bm{1}_{n_{t}}^{\prime}\bm{\Psi}_{t}\textbf{W}_{t,k}^{1/2},-\bm{1}_{n_{t}}^{\prime}\bm{\Psi}_{t}\textbf{M}_{t+1}^{\prime}\textbf{W}_{t,k}^{1/2})\right\rbrace \right]}\right)$	&	$1 \le t < T$  \\ \hline
 $d_{\eta,T,k} = {\alpha_{k}}\Bigg/\left({1 + \mathrm{max}\left[ \mathrm{abs}\left\lbrace \bm{1}_{n_{t}}^{\prime}\bm{\Psi}_{t}\textbf{W}_{t,k}^{1/2}\right\rbrace \right]}\right)$	&	  \\ \hline
 $d_{\xi,t,k} = {\alpha_{k}}\Bigg/\left[{1 + \mathrm{max}\left\lbrace \mathrm{abs}\left(\alpha_{k}^{1/2}\sigma_{\xi}\bm{1}_{n_{t}}^{\prime}\right) \right\rbrace}\right]$	&	$1 \le t \le T$  \\ \hline
 $
          f(\sigma_{\mathrm{K}}\vert \cdot) = \frac{f(\bm{\eta}_{1}\vert \sigma_{K})\prod_{i = 2}^{T}f(\bm{\eta}_{t}\vert \bm{\eta}_{t-1},\sigma_{K})}{\sum_{\sigma_{\mathrm{K}}}f(\bm{\eta}_{1}\vert \sigma_{K})\prod_{i = 2}^{T}f(\bm{\eta}_{t}\vert \bm{\eta}_{t-1},\sigma_{K})}$	&	The exact expression can be found in (\ref{K_fc}).  \\ \hline
 $
       p_{\xi}(\sigma_{\xi})=\frac{\prod_{t=1}^{T}\mathrm{exp}\left\lbrace\frac{\alpha_{k}^{1/2}}{\sigma_{\xi}}\alpha_{k}\bm{1}_{n_{t}}^{\prime}\bm{\xi}_{t} - \frac{1}{\kappa_{k}}\bm{1}_{n_{t}}^{\prime}\mathrm{exp}(\frac{\alpha_{k}^{1/2}}{\sigma_{\xi}}\bm{\xi}_{t} )\right\rbrace}{\sum_{\sigma_{\xi}}\prod_{t=1}^{T}\mathrm{exp}\left\lbrace\frac{\alpha_{k}^{1/2}}{\sigma_{\xi}}\alpha_{k}\bm{1}_{n_{t}}^{\prime}\bm{\xi}_{t} - \frac{1}{\kappa_{k}}\bm{1}_{n_{t}}^{\prime}\mathrm{exp}(\frac{\alpha_{k}^{1/2}}{\sigma_{\xi}}\bm{\xi}_{t} )\right\rbrace}$	&	  \\ \hline
 \end{tabular}}
 \caption{A comprehensive list of functions, matrices, vectors, and constants used within the Proposition 3. If there are no zero counts within the dataset, then set $d_{\beta,k} = d_{\eta,1,k} = \ldots = d_{\eta,T,k} = d_{\xi,1,k} = \ldots = d_{\xi,T,k} = 0$.}
 \label{gprop}
 \end{table}
 
 \newpage
 \normalsize
 \begin{algorithm}[H]
 \caption{The Gibbs Sampler for the P-MSTM}
 \begin{enumerate}
 \item Choose a value of $k$ in (\ref{summary}), either, $k=0$ to specify sMLG distributions, or $k = 1$ to specify nMLG distributions.
 \item Initialize $\bm{\beta}$, $\sigma_{\mathrm{K}}$, $\sigma_{\xi}$, and $\bm{\xi}_{t}$ and $\bm{\eta}_{t}$ for each $t$. Denote these initializations with $\bm{\beta}^{[0]}$, $\sigma_{\mathrm{K}}^{[0]}$, $\sigma_{\xi}^{[0]}$, and $\bm{\xi}_{t}^{[0]}$ and $\bm{\eta}_{t}^{[0]}$ for each $t$.
 \item Set $b = 1$.
 \item Set $\bm{\beta}^{[b]}$ equal to a draw from $\mathrm{mMLG}(\textbf{H}_{\beta,k},\bm{\alpha}_{\beta,k},\bm{\kappa}_{\beta,k})$ using Theorem~2($ii$).
 \item If $t<T$, then set $\bm{\eta}_{t}^{[b]}$ equal to a draw from $\mathrm{mMLG}(\textbf{H}_{\eta,t,k},\bm{\alpha}_{\eta,t,k},\bm{\kappa}_{\eta,t,k})$ using Theorem~2($ii$).
 \item Set $\bm{\eta}_{T}^{[b]}$ equal to a draw from $\mathrm{mMLG}(\textbf{H}_{\eta,T,k},\bm{\alpha}_{\eta,T,k},\bm{\kappa}_{\eta,T,k})$ using Theorem~2($ii$).
 \item For each $t$ let $\bm{\xi}_{t}^{[b]}$ be a draw from $\mathrm{mMLG}(\textbf{H}_{\xi,t,k},\bm{\alpha}_{\xi,t,k},\bm{\kappa}_{\xi,t,k})$ using Theorem~2($ii$).
 \item Set $\sigma_{\mathrm{K}}^{[b]}$ equal to randomly selected value from $\{a_{1}^{(\mathrm{K})},\ldots,a_{U_{\mathrm{K}}}^{(\mathrm{K})}\}$, with respective probabilities determined by $p_{\mathrm{K}}(\cdot)$ in (\ref{full_cond}).
 \item Set $\sigma_{\xi}^{[b]}$ equal to randomly selected value from $\{a_{1}^{(\xi)},\ldots,a_{U_{\xi}}^{(\xi)}\}$, with respective probabilities determined by $p_{\xi}(\cdot)$ in (\ref{full_cond}).
 \item Set $b = b+1$.
 \item Repeat steps 4 through 10 until $b$ is equal to the desired value (i.e., convergence is achieved).
 \end{enumerate}
 \end{algorithm} 
 
       \newpage
         \begin{figure}[H]
         \begin{center}
         \begin{tabular}{cc}
            \includegraphics[width=9.5cm,height=9cm]{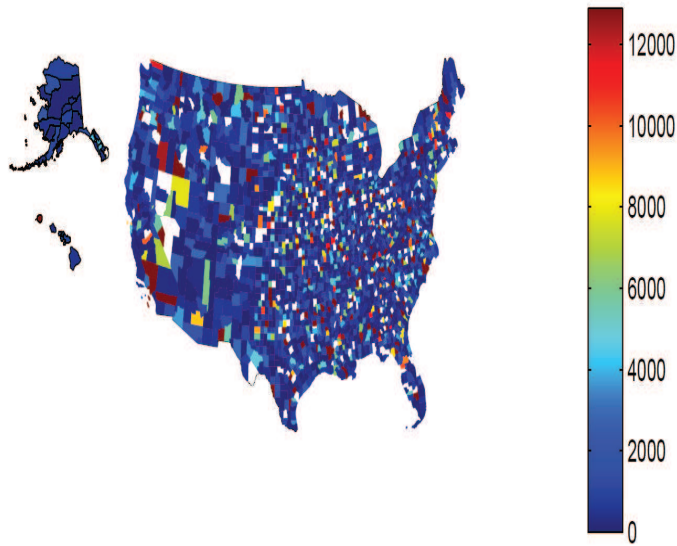}&
            \includegraphics[width=7.5cm,height=9cm]{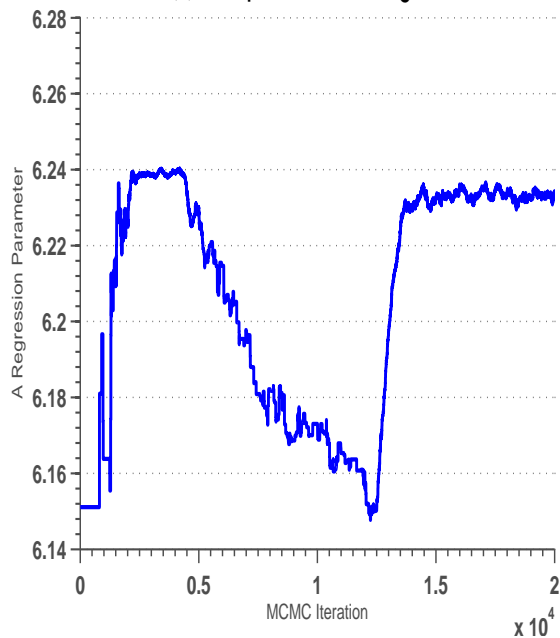}\\
            \includegraphics[width=7.5cm,height=9cm]{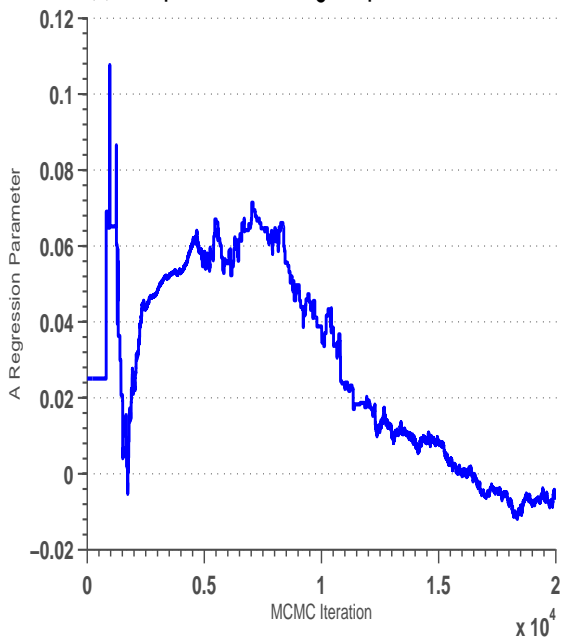}&
            \includegraphics[width=7.5cm,height=9cm]{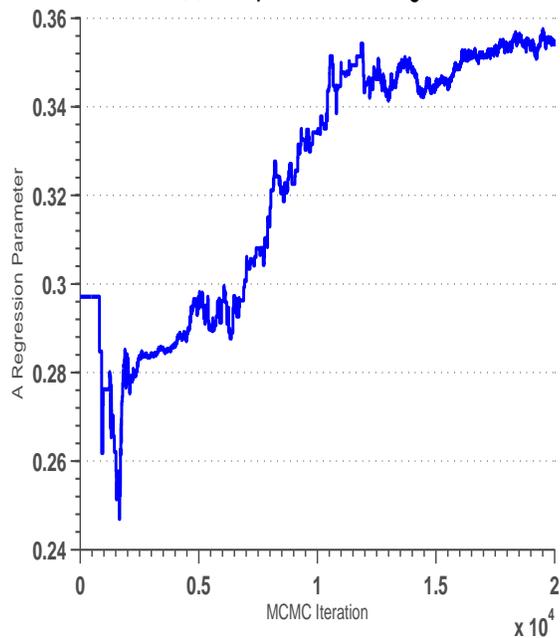}
         \end{tabular}
         \caption{\baselineskip=10pt{(a), Map of the LEHD estimated number of individuals employed in the beginning of the fourth quarter of 2013 within the information industry. White areas indicate ``suppressed'' QWIs. In panels (b), (c), and (d) we plot the trace plot from an MCMC using a Metroplis-within-Gibbs algorithm associated with a Gaussian process model from \citet{bradleyMSTM}. These trace plots are for an intercept parameter. Panels (b), (c), and (d) were computed using different methods for tuning a Metropolis-within-Gibbs algorithm. Specifically, (b), (c), and (d) were computed using MALA, an adaptive Robbins-Monroe process, and LAP, respectively. In these three panels we see that convergence is not achieved. }}
         \end{center}
         \end{figure}

       \newpage
         \begin{figure}[H]
         \begin{center}
         \begin{tabular}{cc}
            \includegraphics[width=16.5cm,height=18cm]{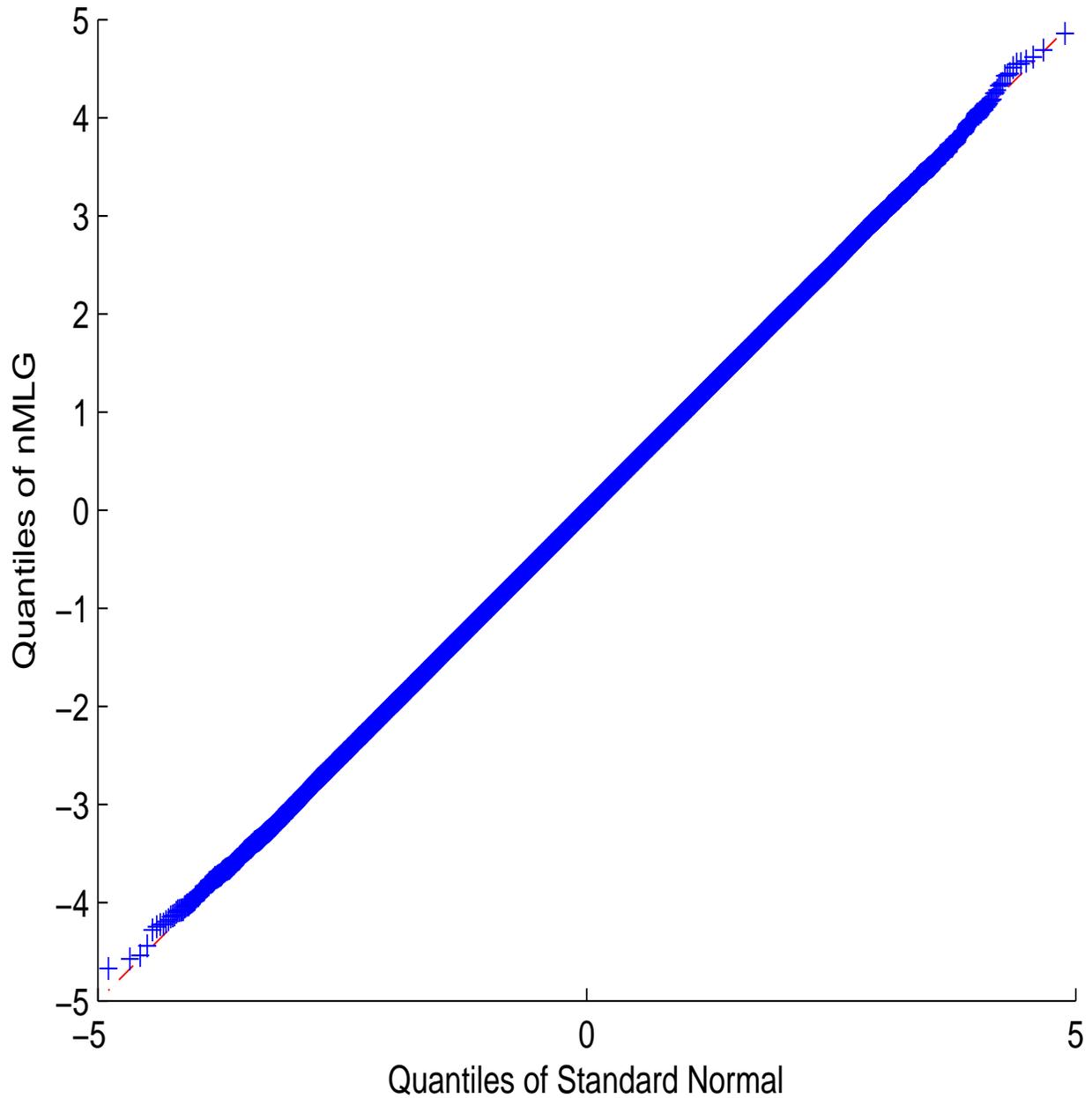}
         \end{tabular}
         \caption{\baselineskip=10pt{(a), In this plot we provide the normal QQ plot associated with a scaled log-gamma random variable. Specifically, we randomly select 10,000,000 values from a $Q\equiv \alpha_{\mathrm{G}}\hspace{5pt}\mathrm{LG}(\alpha_{\mathrm{G}},1/\alpha_{\mathrm{G}})$, and computed the normal QQ plot above. There are enough simulated values in Figure 2 to obtain values from -5 to 5 (recall that the standard normal distribution has 99.7$\%$ of it's mass between -3 and 3). This demonstrates that nMLG provides an excellent approximation of a normal random variable.}}
         \end{center}
         \end{figure}

       \newpage
         \begin{figure}[H]
         \begin{center}
         \begin{tabular}{cc}
            \includegraphics[width=8cm,height=9cm]{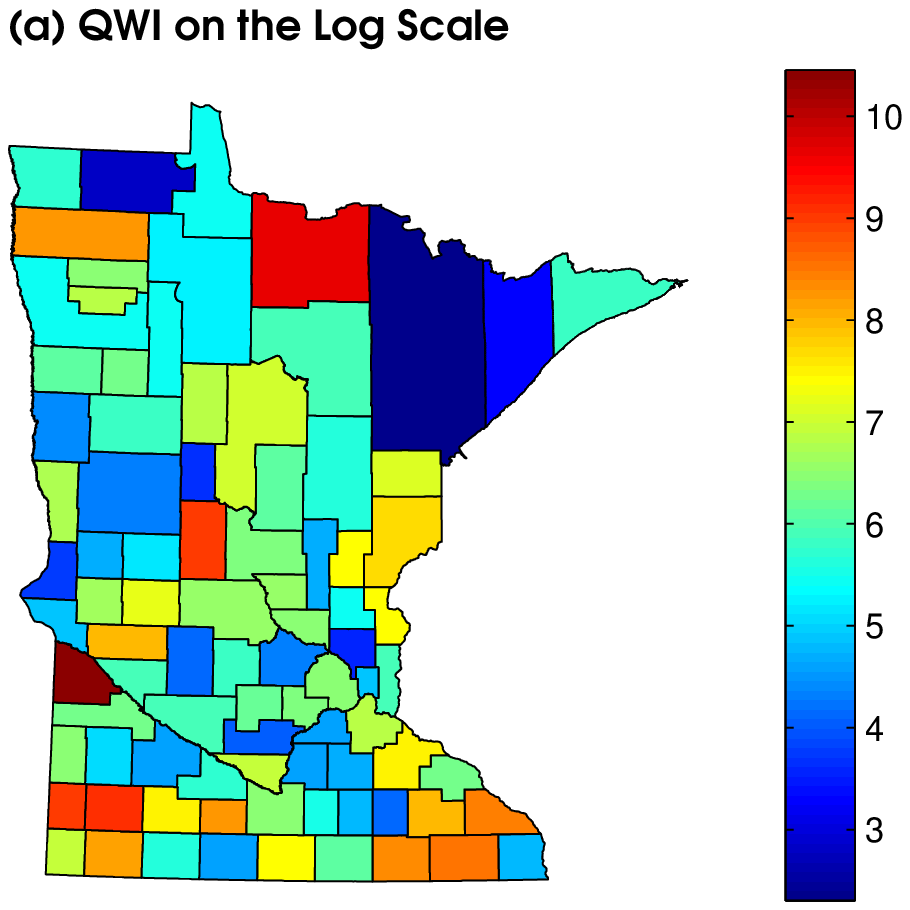}&
            \includegraphics[width=8cm,height=9cm]{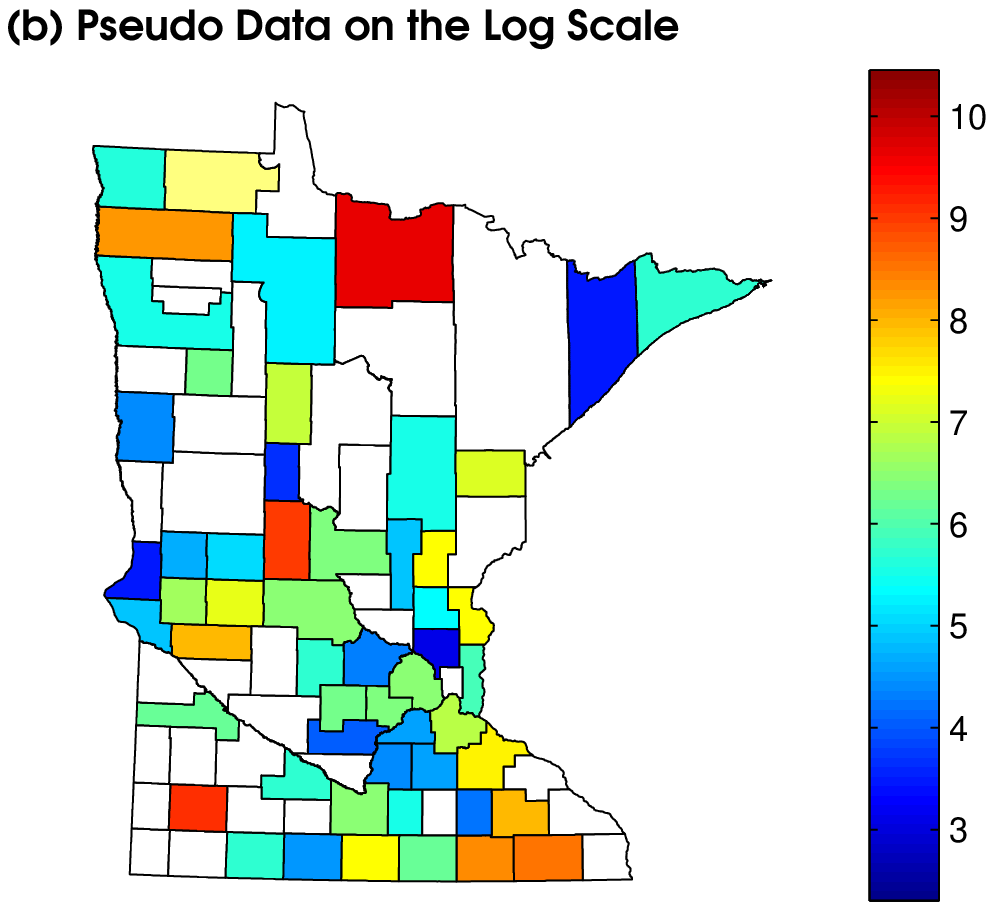}\\
            \includegraphics[width=8cm,height=9cm]{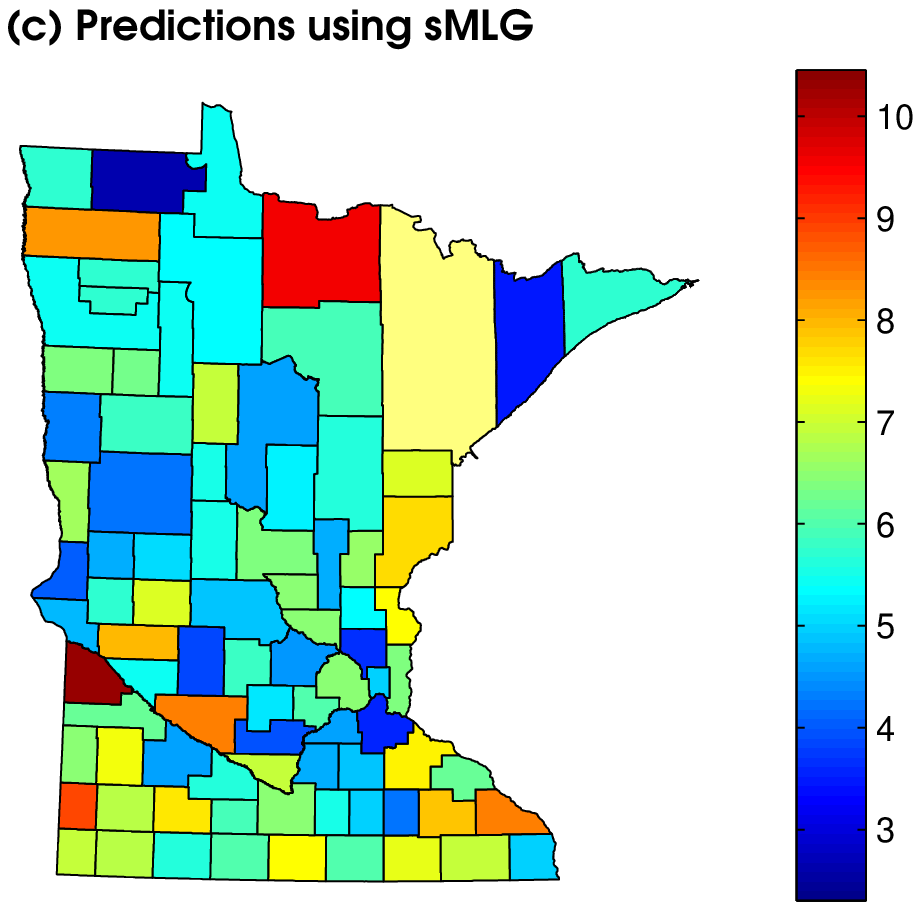}&
            \includegraphics[width=8cm,height=9cm]{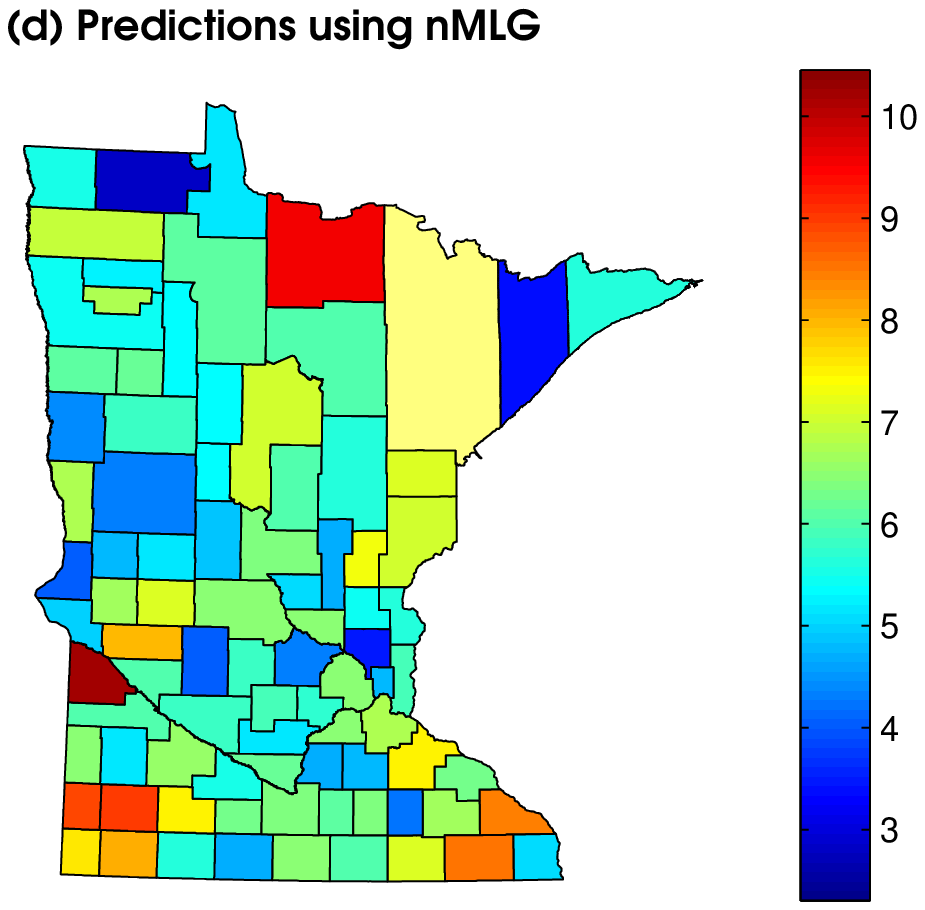}
         \end{tabular}
         \caption{\baselineskip=10pt{(a), The LEHD estimated number of individuals employed in the beginning of the fourth quarter of 2013 within the information industry (i.e., $\{Z_{96}^{(1)}(A)\}$) in Minnesota. For comparison, a map of the pseudo-data is $\{R_{96}^{(1)}(A)\}$ computed using (\ref{pseudo}) is given in (b). The white areas indicate ``suppressed'' QWIs. In (c) and (d), we provide the predictions of $\{\widehat{Z}_{96}^{(1)}(A)\}$ that are computed using P-MSTM and the pseudo-data $\{R_{t}^{(\ell)}(A)\}$ from Equation (\ref{pseudo}). In (c), the predictions are done using the sMLG specification and nMLG is computed using the nMLG specification.}}
         \end{center}
         \end{figure}
         
               \newpage
                 \begin{figure}[H]
                 \begin{center}
                 \begin{tabular}{cc}
                    \includegraphics[width=7.5cm,height=9cm]{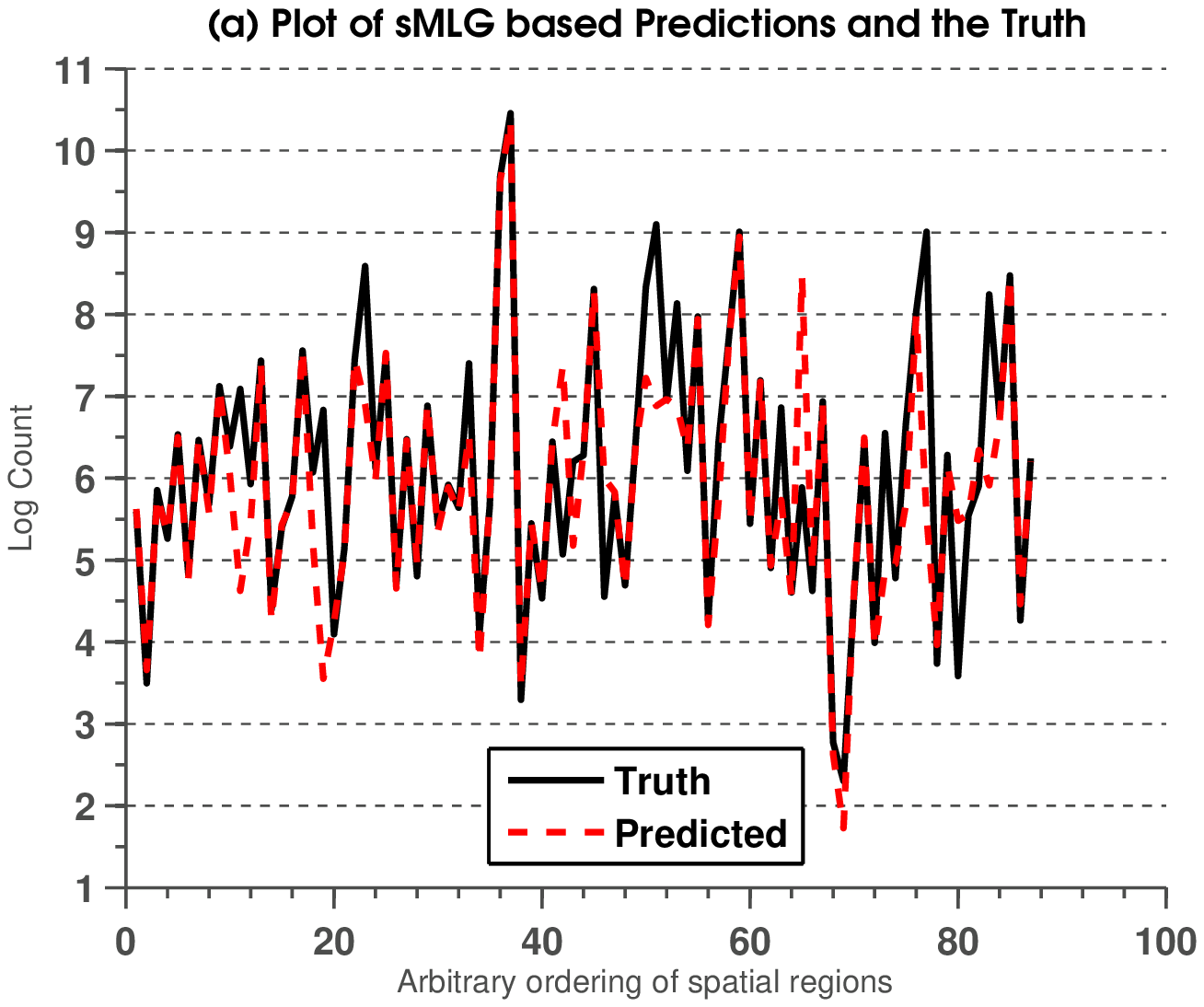}&
                    \includegraphics[width=7.5cm,height=9cm]{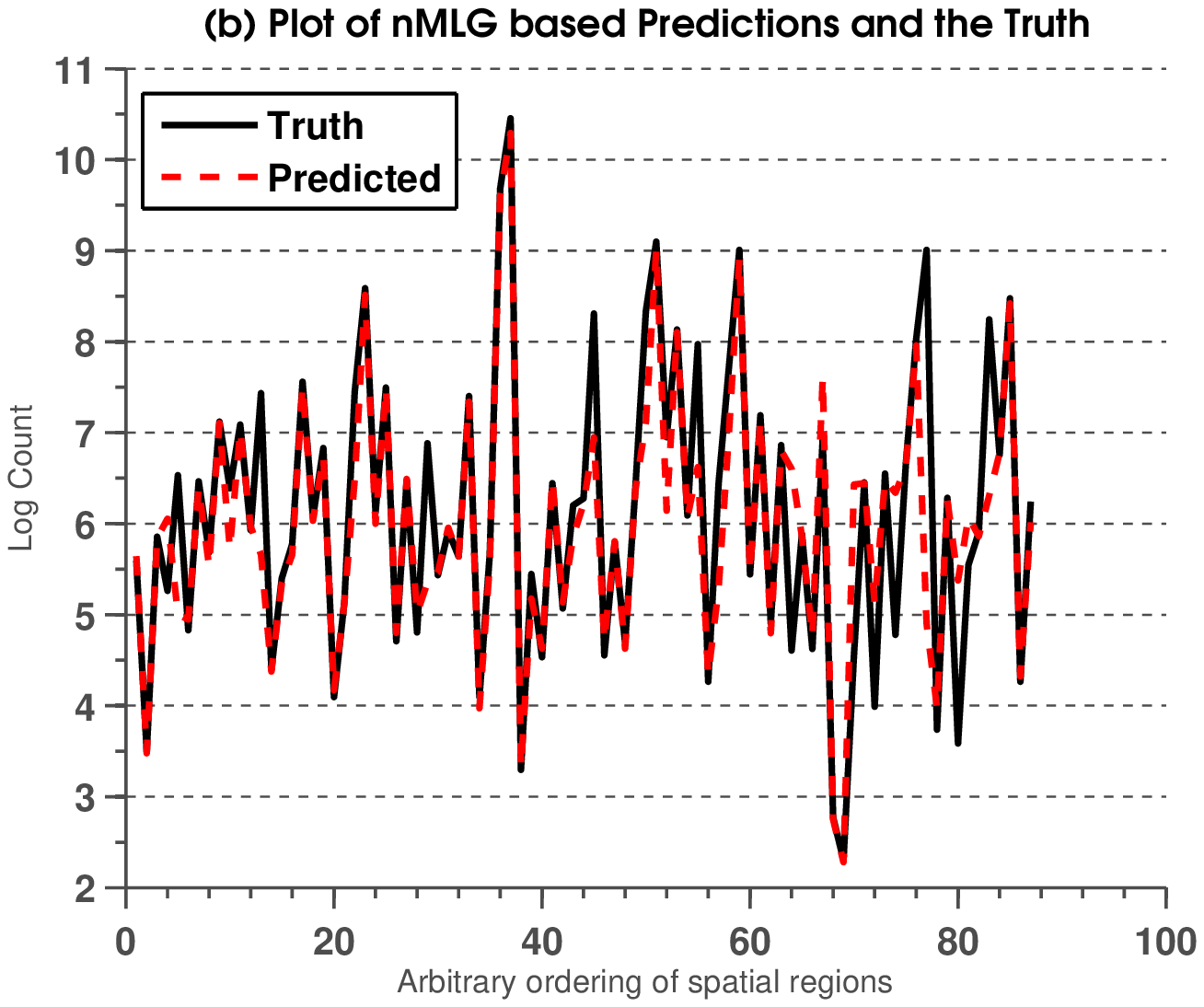}\\
                    \includegraphics[width=7.5cm,height=9cm]{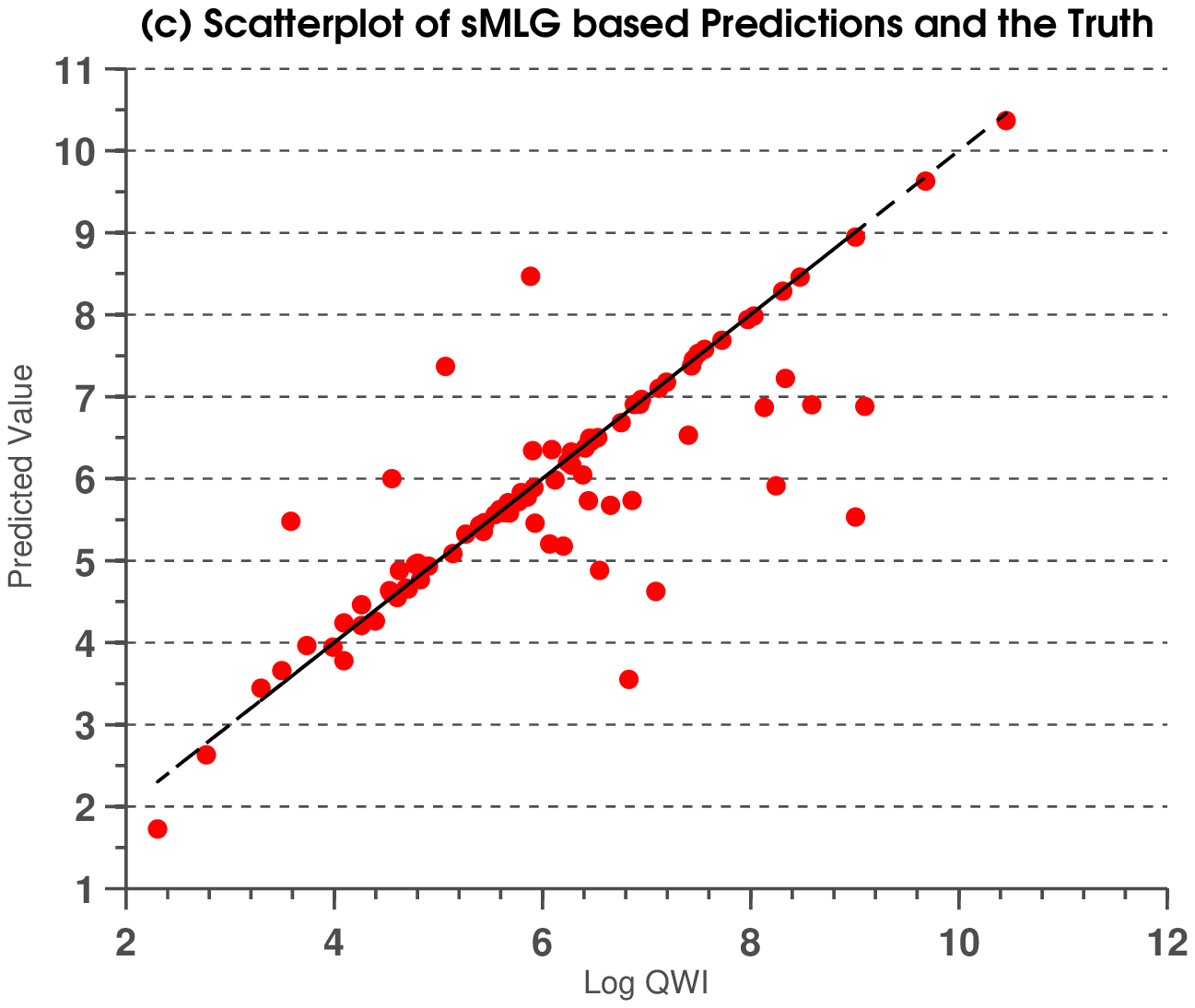}&
                    \includegraphics[width=7.5cm,height=9cm]{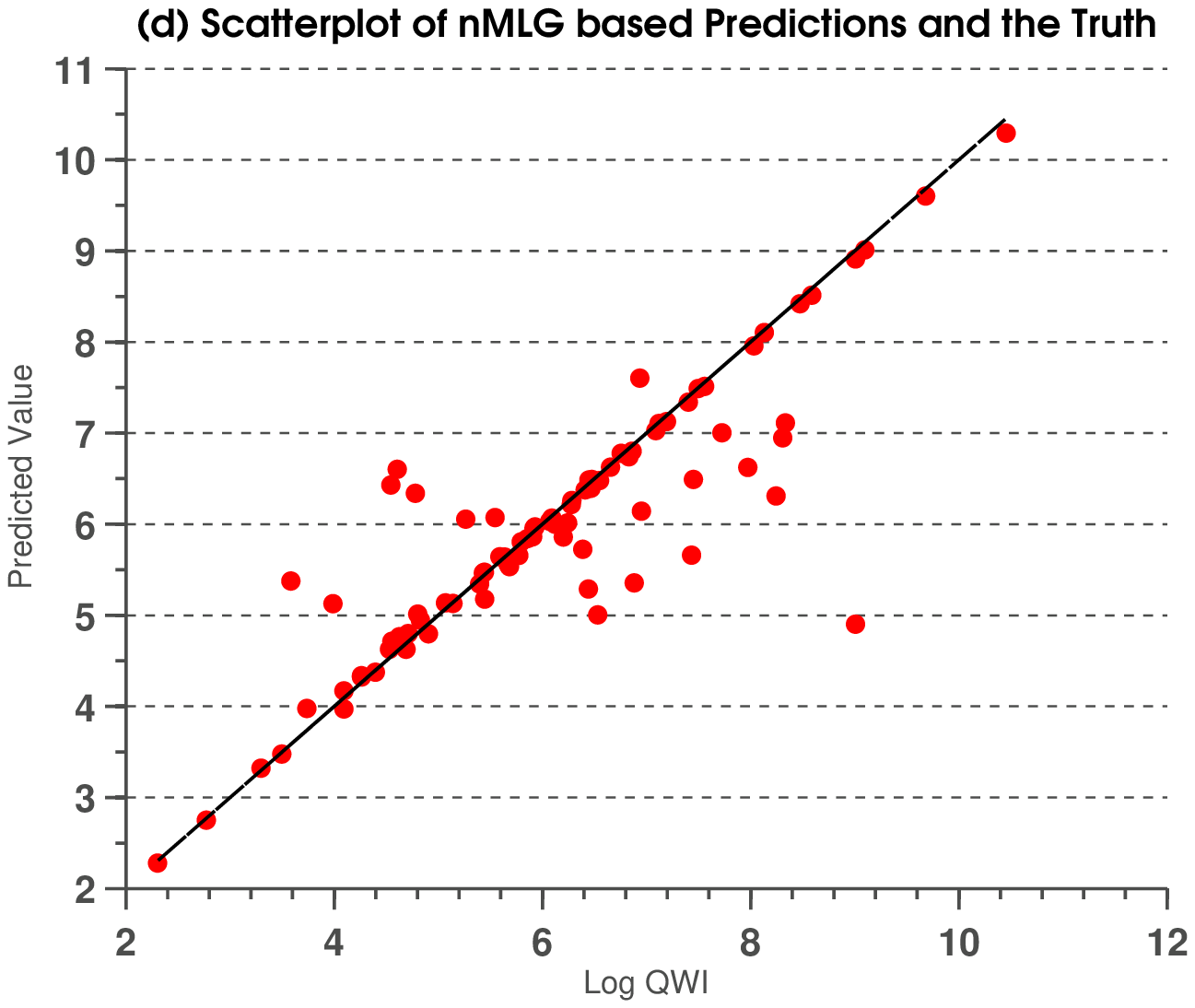}
                 \end{tabular}
                 \caption{\baselineskip=10pt{In (a) and (b), we plot the LEHD estimated number of individuals employed in the beginning of the fourth quarter of 2013 within the information industry in Minnesota, and the predicted values. In (a) the predictions are based on the sMLG specification, and in (b) the predictions are based on the nMLG specification. In (b) and (c), we present the predictions and standard deviations, respectively. In (c) and (d), we produce scatterplots of the LEHD estimated number of individuals employed in the beginning of the fourth quarter of 2013 within the information industry in Minnesota, versus the predicted values. In (c) the predictions are based on the sMLG specification, and in (d) the predictions are based on the nMLG specification.}}
                 \end{center}
                 \end{figure}
                 
               \newpage
                 \begin{figure}[H]
                 \begin{center}
                 \begin{tabular}{c}
                    \includegraphics[width=15.5cm,height=16cm]{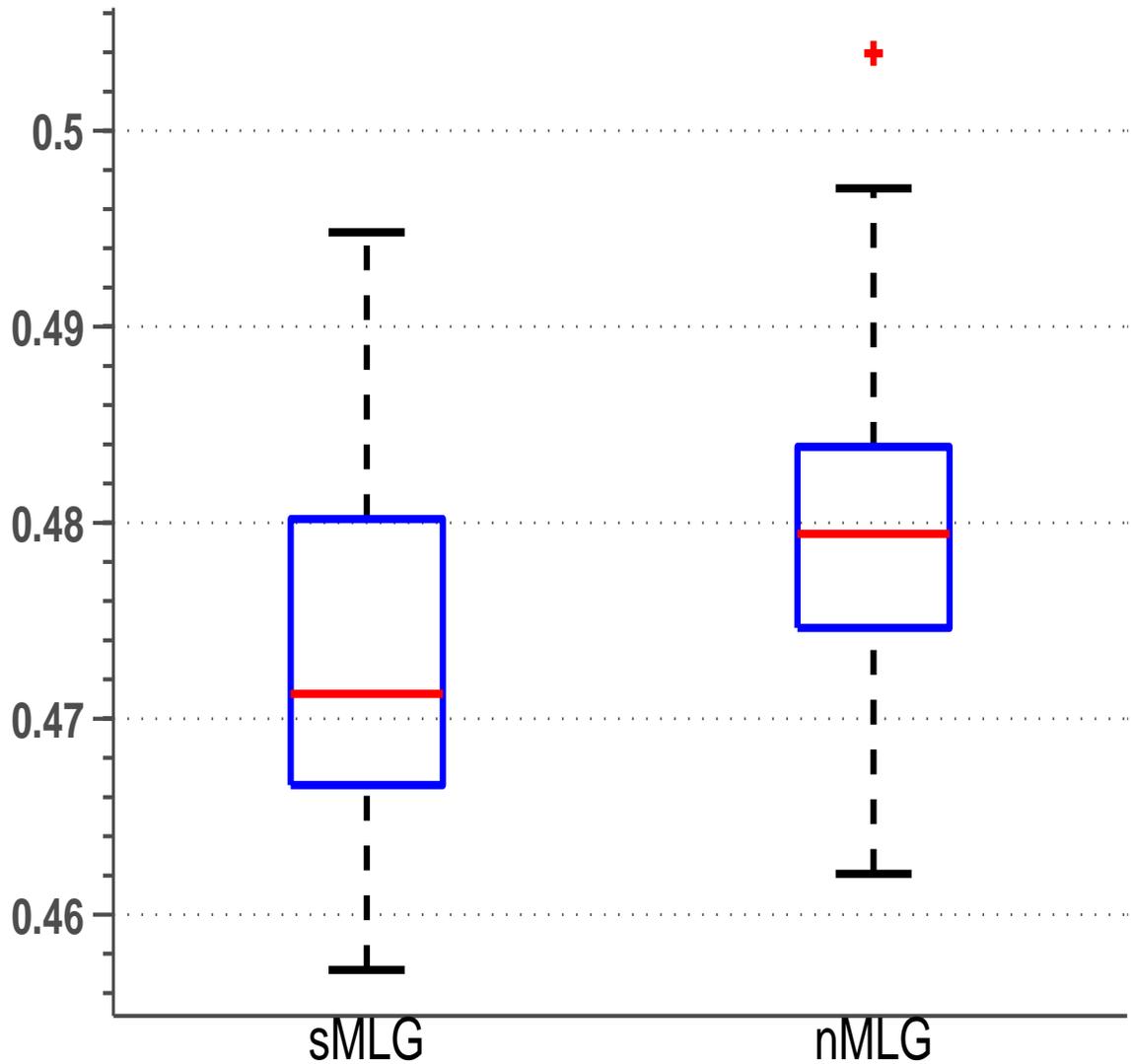}
                 \end{tabular}
                 \caption{\baselineskip=10pt{A boxplot of the average absolute error diagnostic in (\ref{abs_error}) over 100 replicates (i.e., $j = 1,\ldots,100$). The distributional assumption used to produce the predictions are given in on the $x$-axis.}}
                 \end{center}
                 \end{figure}
         
       \newpage
         \begin{figure}[H]
         \begin{center}
         \begin{tabular}{c}
            \includegraphics[width=16.5cm,height=9.5cm]{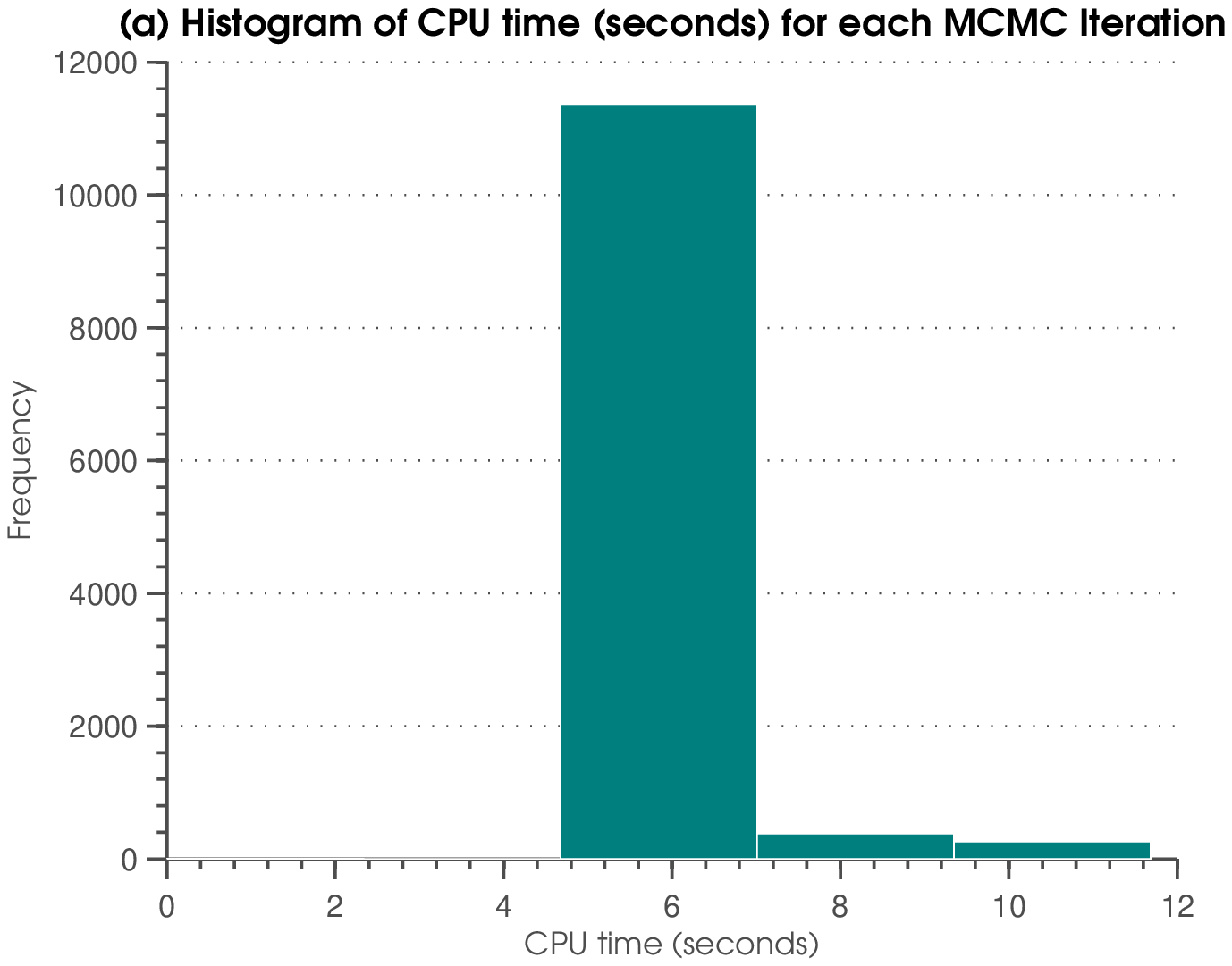}\\
            \includegraphics[width=16.5cm,height=9.5cm]{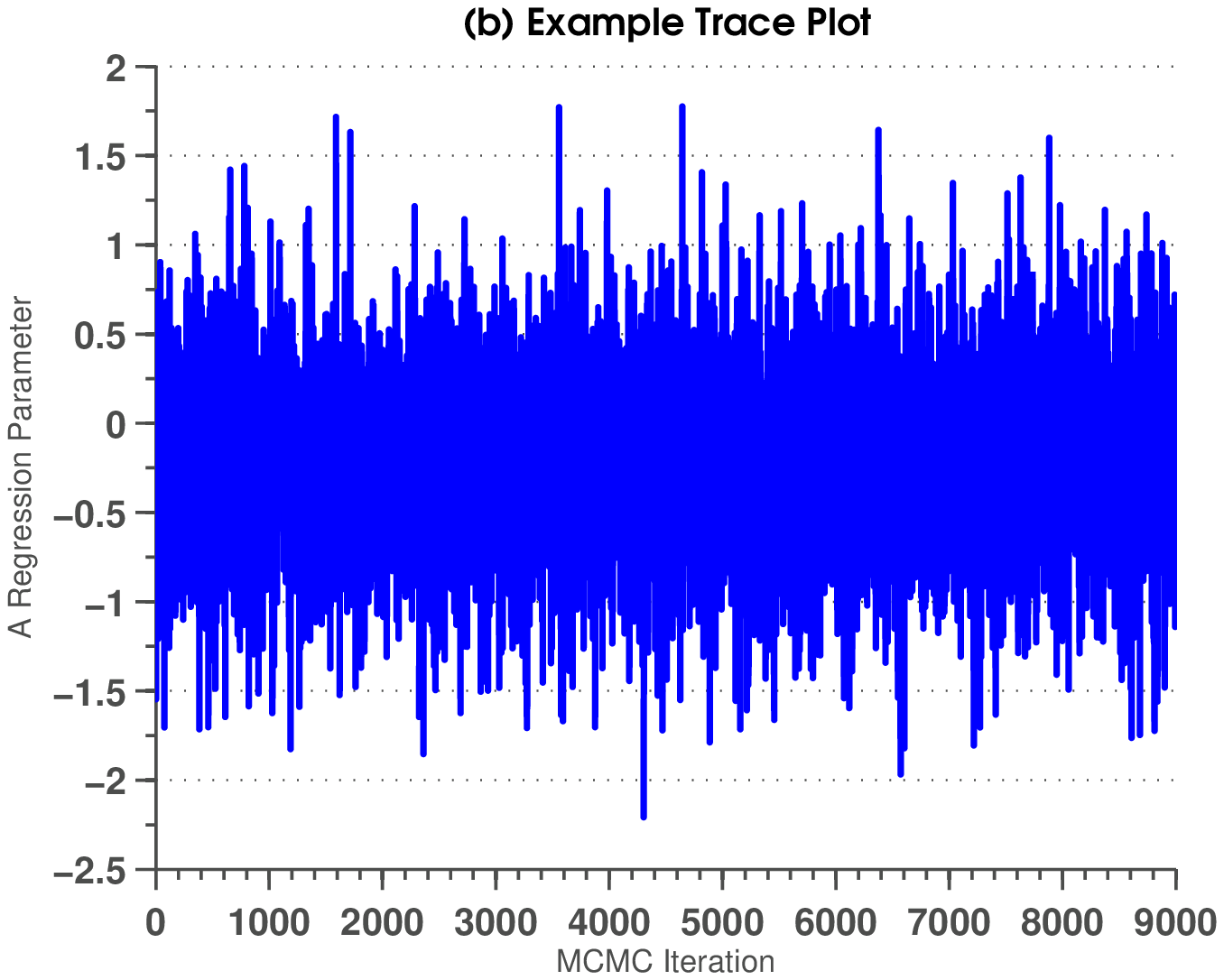}
         \end{tabular}
         \caption{\baselineskip=10pt{In (a), we plot the CPU time (in seconds) to produce each MCMC replicate from the P-MSTM applied to the 4,089,755 QWIs analyzed in Section~4. All computations were computed using Matlab (Version 8.0) on a dual 10 core 2.8 GHz Intel Xeon E5-2680 v2 processor, with 256 GB of RAM. In (b), we provide the trace plot associated with an intercept term (compare to Figure 1(b,c,d)).}}
         \end{center}
         \end{figure}
         
               \newpage
                 \begin{figure}[H]
                 \begin{center}
                 \begin{tabular}{cc}
                    \includegraphics[width=8.5cm,height=8.5cm]{LEHD_data.eps}&
                    \includegraphics[width=8.5cm,height=8.5cm]{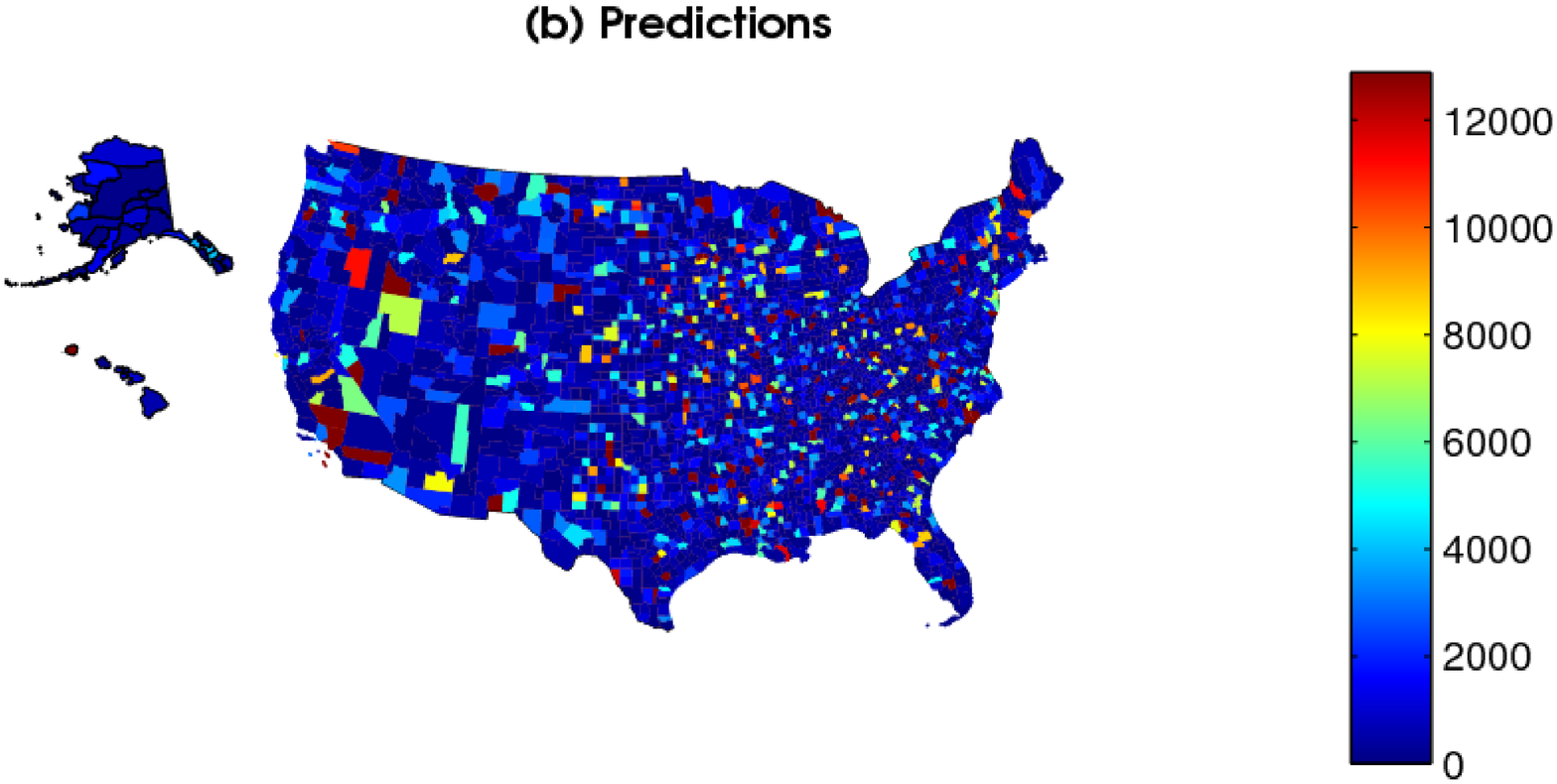}\\
                    \includegraphics[width=8.5cm,height=8.5cm]{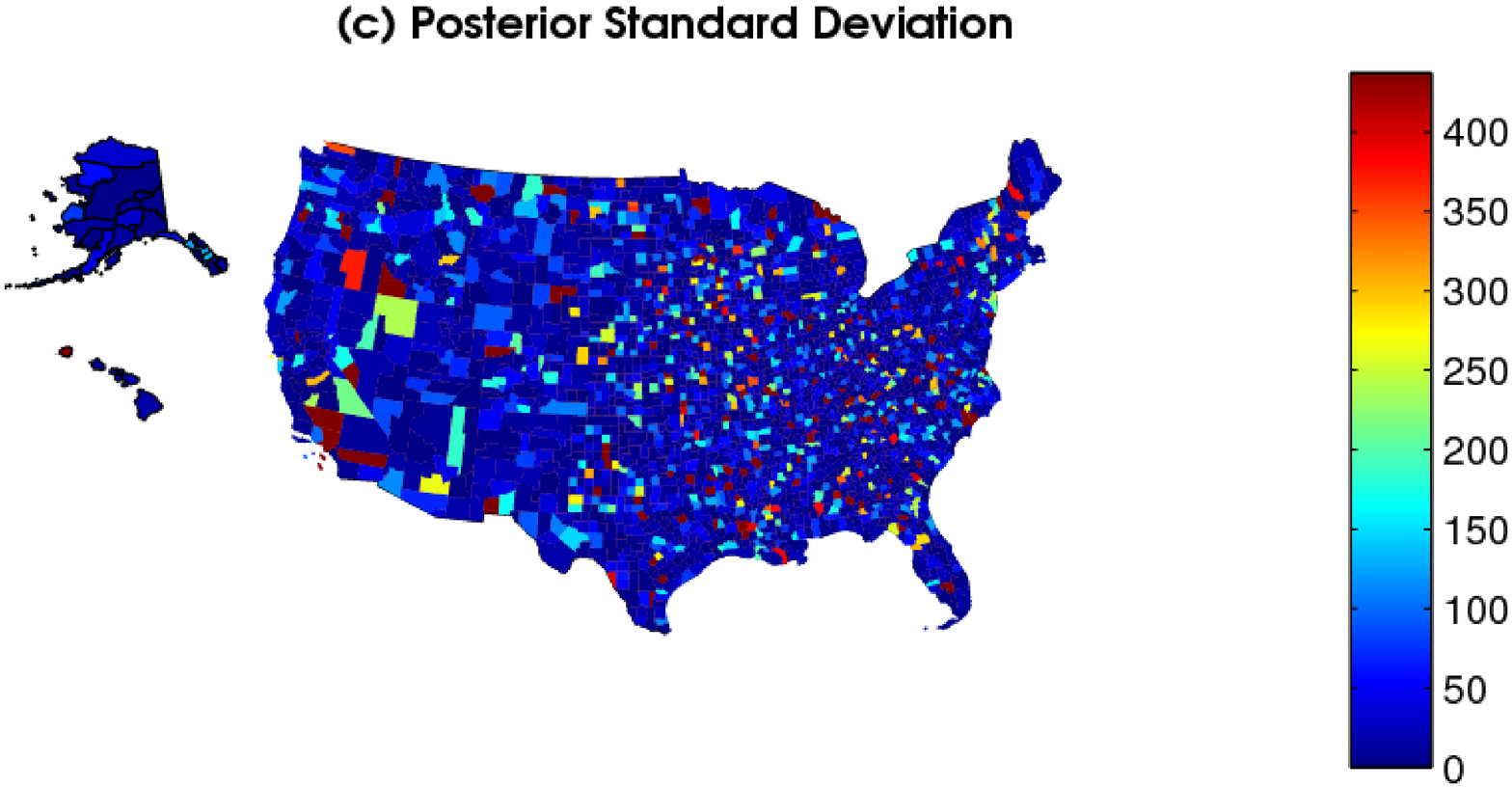}&
                    \includegraphics[width=8.5cm,height=8.5cm]{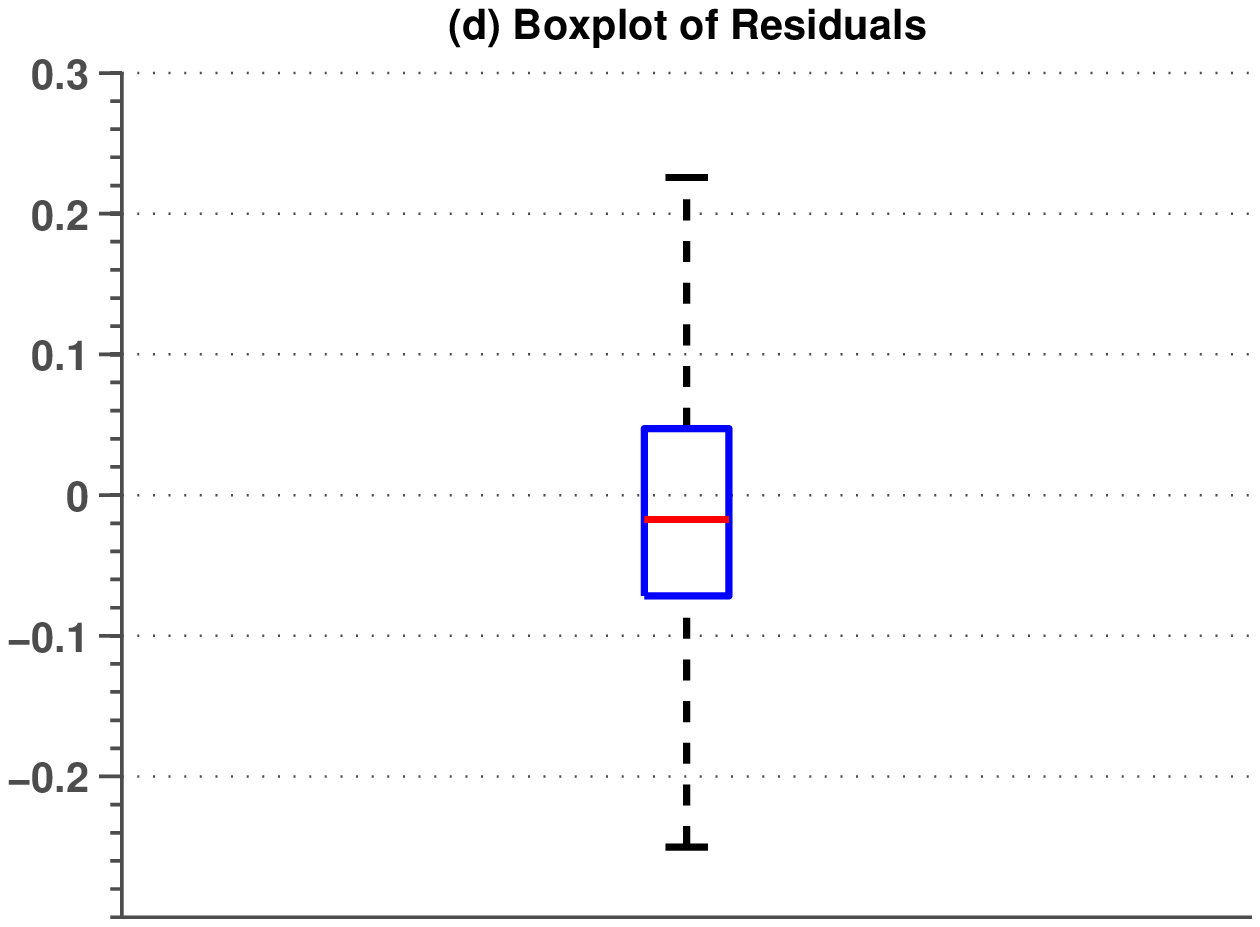}
                 \end{tabular}
                 \caption{\baselineskip=10pt{(a), Map of the LEHD estimated number of individuals employed in the beginning of the fourth quarter of 2013 within the information industry (i.e., $\{Z_{96}^{(1)}(A)\}$). In (b) and (c), we present the predictions and standard deviations, respectively. Note that (a,b,c) are only a subset of the avialable QWIs, predictions, and posterior standard deviations. Specifically, there are QWIs available over the 20 NAICS sectors, US counties, and 96 quarters. Additionally, the predictions and posterior standard deviations have complete coverage over all 20 sectors, 3,145 US counties, and 96 quarters. In (d), we plot the difference between the log predictions (plus 1) and the log QWI (plus 1), to demonstrate the in-sample error of the predictions.}}
                 \end{center}
                 \end{figure}
 
\newpage

\part*{}
\thispagestyle{empty} \baselineskip=28pt

\thispagestyle{empty} \baselineskip=28pt

\begin{center}
{\LARGE{\bf Supplemental Material: Computationally Efficient Distribution Theory for Bayesian Inference of High-Dimensional Correlated Count-Valued Data}}
\end{center}

\baselineskip=12pt

\vskip 2mm
\begin{center}
Jonathan R. Bradley$^{1}$,
Scott H. Holan$^{2}$,
Christopher K. Wikle$^{2}$  
\\
\end{center}

\pagenumbering{arabic}

\baselineskip=24pt

\section*{Appendix A: Details on Basis Functions and Propagator Matrices} 
\renewcommand{\theequation}{A.\arabic{equation}}
\setcounter{equation}{0}
In this section, we review the MI basis functions, and the MI propagator matrix from \citet{bradleyMSTM}. These quantities can be used to define $\{\bm{\psi}_{t}^{(\ell)}(A)\}$ and $\{\textbf{M}_{t}\}$ in the expression of the statistical model given in (21).\\

\noindent
\textit{A Review of the MI Basis Functions:} In this manuscript we choose to set $\{\psi_{t}^{(\ell)}(A)\}$ equal to the MI basis functions, which are a priori defined to be linearly independent of $\{\textbf{x}_{t}^{(\ell)}(A)\}$. That is, the $N_{t}\times r$ matrix $\bm{\Psi}_{t}^{\mathrm{P}}$ is specified to be contained within the orthogonal complement of the column space of $\textbf{X}_{t}^{\mathrm{P}}(\textbf{X}_{t}^{\mathrm{P}\prime}\textbf{X}_{t}^{\mathrm{P}})^{-1}\textbf{X}_{t}^{\mathrm{P}\prime}$. Specifically, define the MI operator as
\begin{equation}\label{mioperator}
\textbf{G}(\textbf{X}_{t}^{\mathrm{P}},\textbf{A}_{t}) \equiv\left(\textbf{I}_{N_{t}} - \textbf{X}_{t}^{\mathrm{P}}\left(\textbf{X}_{t}^{\mathrm{P}\prime}\textbf{X}_{t}^{\mathrm{P}}\right)^{-1}\textbf{X}_{t}^{\mathrm{P}\prime}\right)\textbf{A}_{t}\left(\textbf{I}_{N_{t}} - \textbf{X}_{t}^{\mathrm{P}}\left(\textbf{X}_{t}^{\mathrm{P}\prime}\textbf{X}_{t}^{\mathrm{P}}\right)^{-1}\textbf{X}_{t}^{\mathrm{P}\prime}\right);\hspace{5pt}t = 1,\ldots,T,
\end{equation}
where $\textbf{I}_{N_{t}}$ is an $N_{t}\times N_{t}$ identity matrix, and $\textbf{A}_{t}$ is a generic $N_{t}\times N_{t}$ weight matrix. In the setting where $|D_{t}^{(\ell)}|>1$ for at least one $(t,\ell)$ combination, we let $\textbf{A}_{t}$ be the adjacency matrix corresponding to the edges formed by $\{D_{t,\mathrm{P}}^{(\ell)}:\ell = 1,\ldots,L\}$. Notice that the MI operator in (\ref{mioperator}) defines a column space that is orthogonal to $\textbf{X}_{t}^{\mathrm{P}}$. Then, let the spectral representation $\textbf{G}(\textbf{X}_{t}^{\mathrm{P}},\textbf{A}_{t}) = \bm{\Phi}_{t}\bm{\Lambda}_{t}{\bm{\Phi}_{t}^{\prime}}$, and denote the $N_{t}\times r$ real matrix formed from the  first $r$ columns of $\bm{\Phi}_{t}$ as $\bm{\Psi}_{t}^{\mathrm{P}}$. As done in \citet{bradleyMSTM}, we set the row of $\bm{\Psi}_{t}^{\mathrm{P}}$ that corresponds to variable $\ell$ and areal unit $A$ equal to $\bm{\psi}_{t}^{(\ell)}(A)$. \citet{hughes} suggests setting $r$ equal to roughly 10$\%$ of the positive eigenvalues given on the diagonal of $\bm{\Lambda}_{t}$.\\

\noindent
\textit{A Review of the MI Propagator Matrix:} 
 Recall, we specify $\{\bm{\psi}_{t}^{(\ell)}\}$ such that it is not in the column space spanned by $\{\textbf{x}_{t}^{(\ell)}\}$; this allows one to perform inference on $\bm{\beta}$. We can use this thinking for the VAR(1) model. That is, substitute the VAR(1) expansion into (15) to obtain
\begin{equation}\label{process:model:truth2}
Y_{t}^{(\ell)}(A)= \textbf{x}_{t}^{(\ell)}(A)^{\prime}\bm{\beta}+\bm{\psi}_{t}^{(\ell)}(A)^{\prime}\textbf{M}_{t}\bm{\eta}_{t-1} + \bm{\psi}_{t}^{(\ell)}(A)^{\prime}\textbf{b}_{t}+\xi_{t}^{(\ell)}(A);\hspace{5pt} \ell = 1,\ldots,L, \hspace{5pt} t = T_{\mathrm{L}}^{(\ell)}, \ldots, T_{\mathrm{U}}^{(\ell)}, \hspace{5pt}A \in D_{t,\mathrm{P}}^{(\ell)}.
\end{equation}
\noindent
Stack the components of Equation (\ref{process:model:truth2}) to obtain
\begin{equation}\label{process:model:truth3}
\by_{t}^{\mathrm{P}} = \textbf{X}_{t}^{\mathrm{P}}\bm{\beta} + \bm{\Psi}_{t}^{\mathrm{P}}\textbf{M}_{t}\bm{\eta}_{t-1} + \bm{\Psi}_{t}^{\mathrm{P}}\textbf{b}_{t} + \bm{\xi}_{t}^{\mathrm{P}},
\end{equation}
\noindent
where $\by_{t}^{\mathrm{P}}\equiv (Y_{t}^{(\ell)}(A): \ell = 1,\ldots,L, \hspace{5pt}A \in D_{t,\mathrm{P}}^{(\ell)})^{\prime}$ and $\bm{\xi}_{t}^{\mathrm{P}}\equiv (\xi_{t}^{(\ell)}(A): \ell = 1,\ldots,L, \hspace{5pt}A \in D_{t,\mathrm{P}}^{(\ell)})^{\prime}$ are $N_{t}$-dimensional latent random vectors. Then, organize (\ref{process:model:truth3}) to get
\begin{equation*}
\bm{\Psi}_{t}^{\mathrm{P}\prime}(\by_{t}^{\mathrm{P}}-\bm{\xi}_{t}^{\mathrm{P}}) = \textbf{B}_{t}\bm{\zeta}_{t} + \textbf{M}_{t}\bm{\eta}_{t-1};\hspace{5pt} t = 2,\ldots, T,
\end{equation*}
where the $r\times(p+r)$ matrix $\textbf{B}_{t}\equiv (\bm{\Psi}_{t}^{\mathrm{P}\prime}\textbf{X}_{t}^{\mathrm{P}}, \textbf{I})$ and the $(p+r)$-dimensional random vector $\bm{\zeta}_{t} \equiv (\bm{\beta}^{\prime},\textbf{b}_{t}^{\prime})^{\prime}$. To ensure that $\textbf{M}_{t}$ is not confounded with $\textbf{B}_{t}$, for $t = 1,\ldots,T$, we let $\textbf{M}_{t}$ equal the $r$ eigenvectors of $\textbf{G}(\textbf{B}_{t},\textbf{U}_{t})$, where in general $\textbf{U}_{t}$ can be any $r\times r$ weight matrix. In Section~4, we let $\textbf{U}_{t}\equiv \textbf{I}_{r}$ as is done in \citet{bradleyMSTM}. This specification of $\{\textbf{M}_{t}\}$ is called the Moran's I (MI) propagator matrix \citep[see][for a discussion]{bradleyMSTM}.

\begin{figure}[H]
                \begin{center}
                \begin{tabular}{cc}
                   \includegraphics[width=8.5cm,height=9cm]{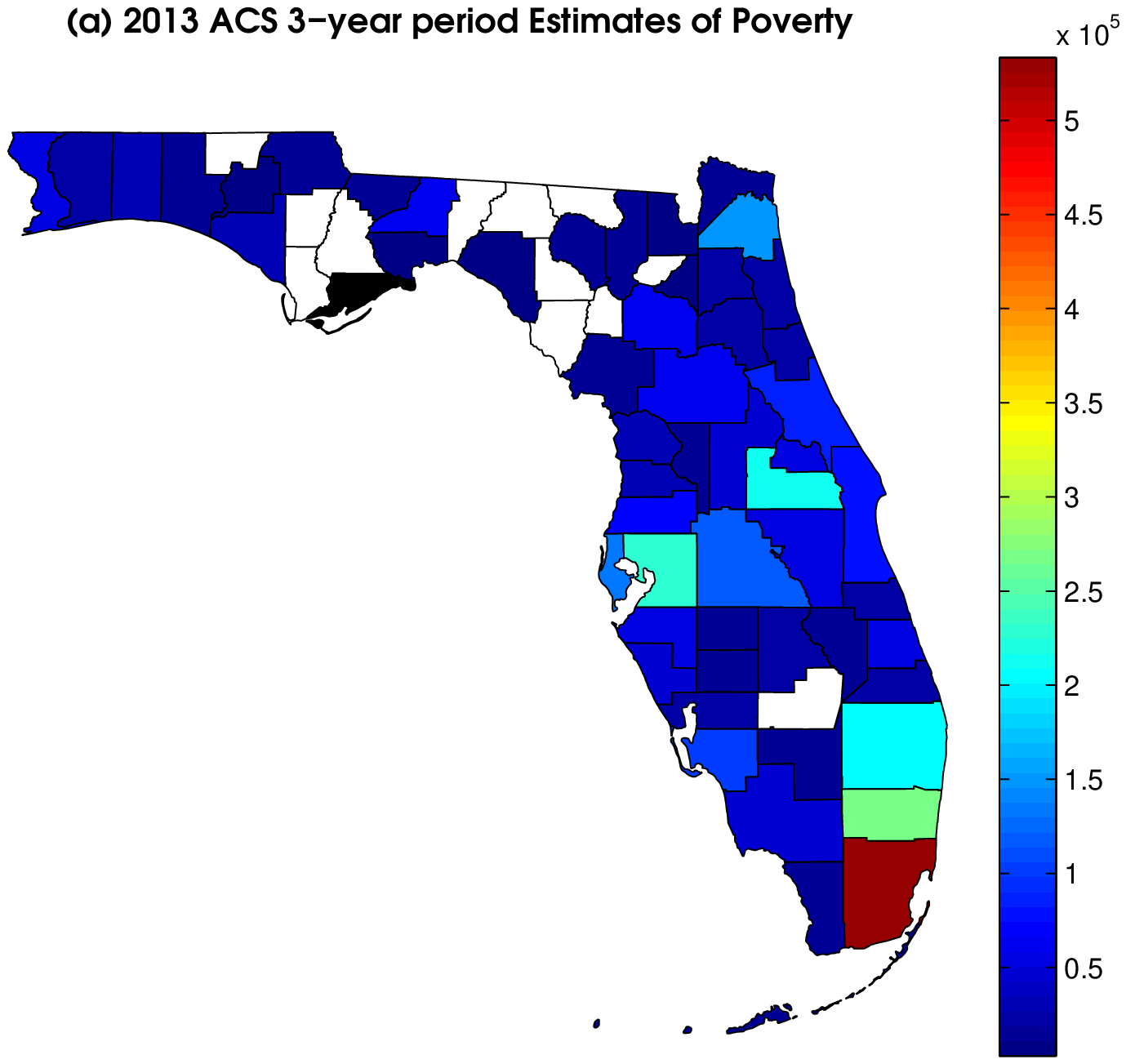}&
                   \includegraphics[width=8.5cm,height=9cm]{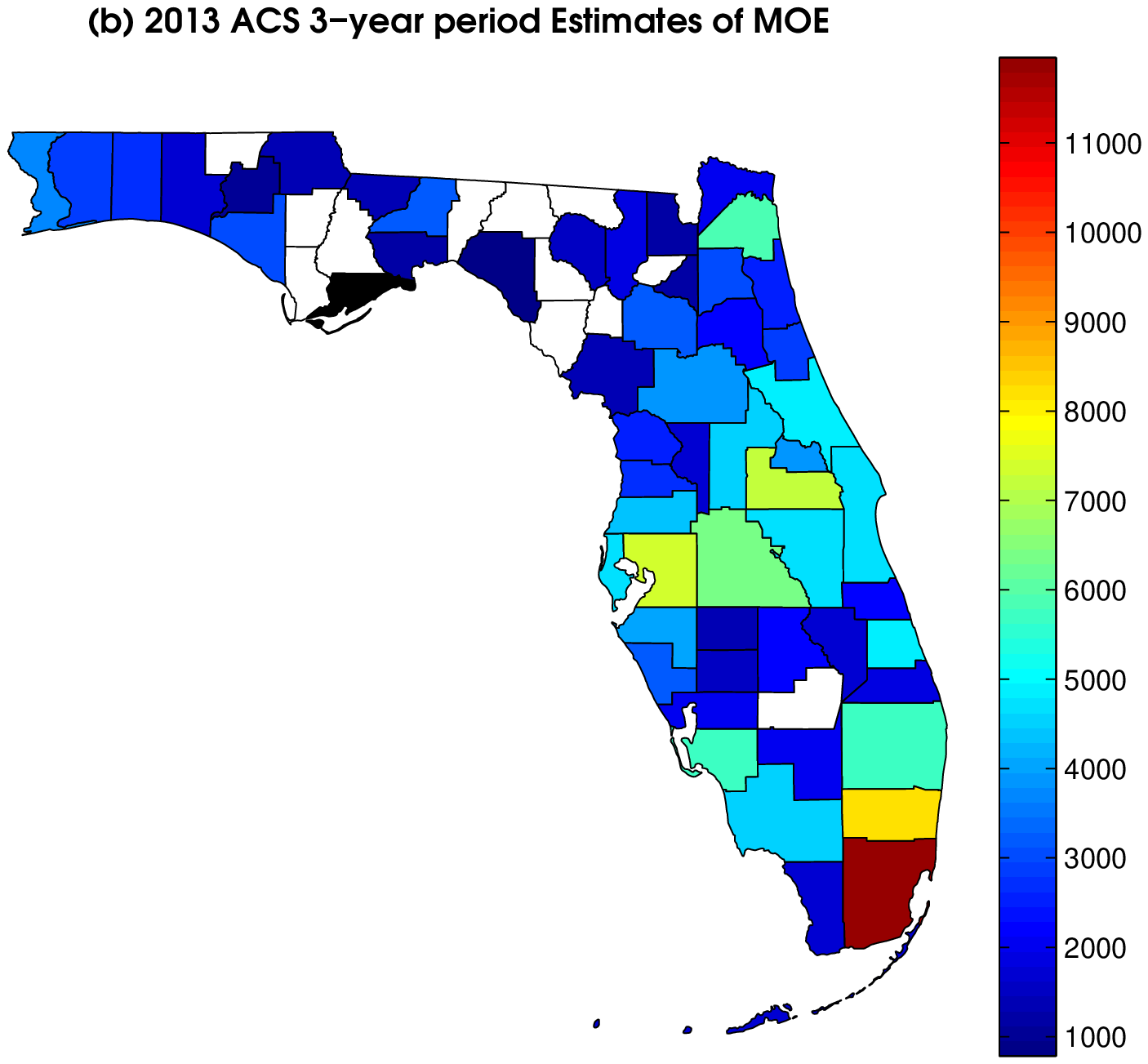}\\
                   \includegraphics[width=8.5cm,height=9cm]{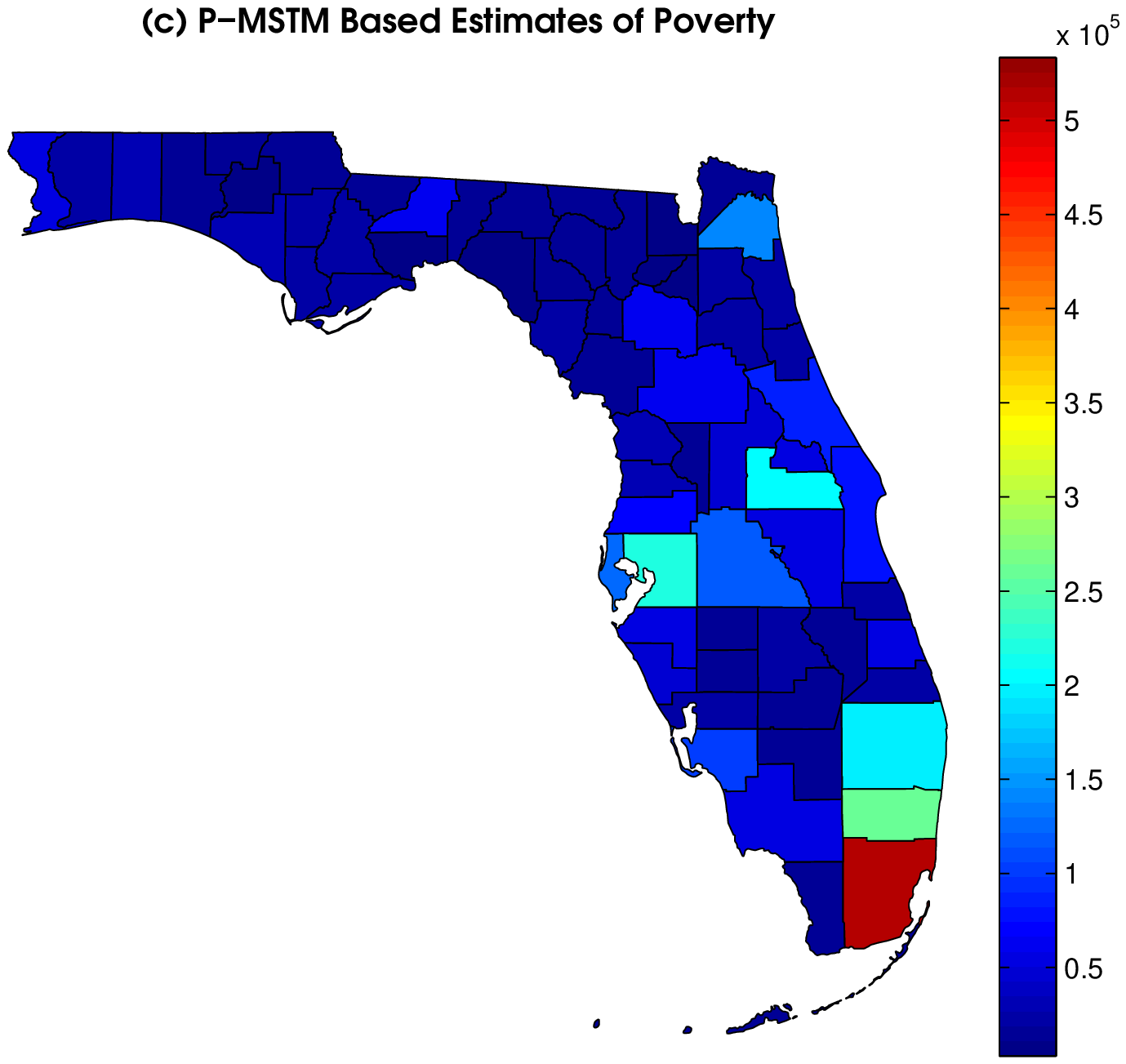}&
                   \includegraphics[width=8.5cm,height=9cm]{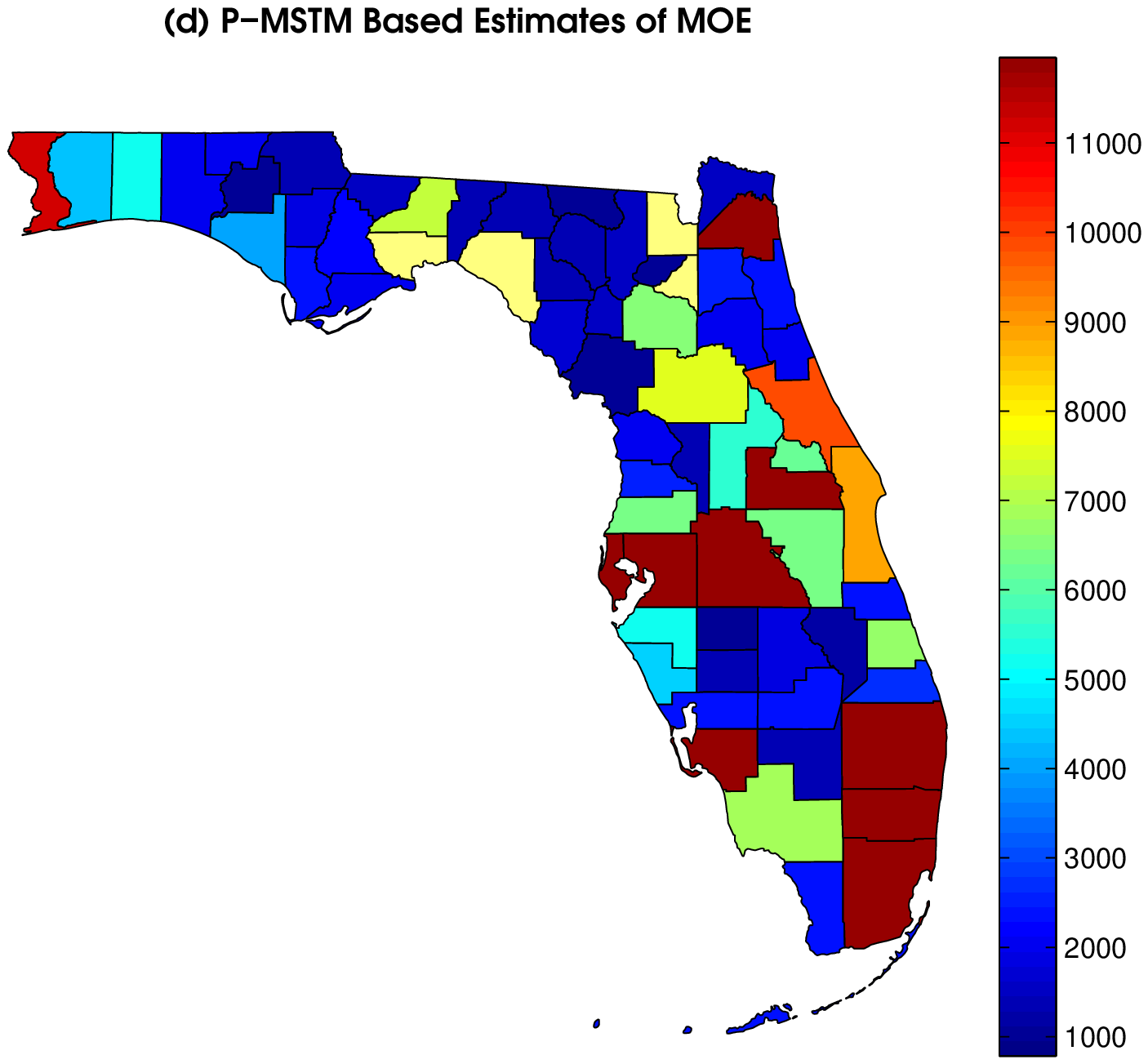}
                \end{tabular}
                \caption{\baselineskip=10pt{(a), Map of the 2013 ACS period estimates of the number of US citizens below the poverty threshold. In (b), we plot the associated ACS margin of error (MOE). In (c), we plot the P-MSTM based predictions of the mean number of US citizens that fall below the poverty threshold. In (d), we give the width of the 95$\%$ credible interval. That is, the model based MOE in panel (d) is defined to be the 97.5$\%$ quantile of the posterior distribution of $Y_{t}^{(\ell)}(\cdot)$ minus the 2.5$\%$ quantile of the posterior distribution of $Y_{t}^{(\ell)}(\cdot)$.}}
                \end{center}
                \end{figure}
\section*{Appendix B: A Spatial-Only Example: The American Community Survey} 
\renewcommand{\theequation}{B.\arabic{equation}}
\setcounter{equation}{0}
The American Community Survey (ACS) produces 1-, 3-, and 5-year period estimates of important US demographics. ACS 1-year period estimates
are provided over areal units associated with populations of 65,000 or larger, 3-year period estimates are provided over areal units associated with populations of 20,000 or larger, and 5-year period estimates are provided over all population sizes. This is done in an effort to produce estimates that strike a balance between precision (i.e., small margin of errors are associated with larger periods) and spatial coverage. 

It has recently been announced that 3-year period estimates will be discontinued starting in the 2016 fiscal year \citep{3yearperiod,bradleySTCOS}. Thus, the introduction and subsequent removal of 3-year period estimates provides a need for precise small area estimates for ACS period estimates, since the option of the more precise (compared to 1-year period estimates) 3-year period estimates will no longer be available. Furthermore, small area estimation for ACS has been a growing topic of interest \citep[e.g., see][among others]{aaronp,bradleyPCOS,bradleySTCOS}. In this section, we use this important problem to demonstrate the use of the P-MSTM in the spatial-only setting (i.e., $T = L = 1$). 

In Figures 8(a) and 8(b), we plot ACS period estimates of the number of US citizens in poverty within the last 12 months of 2013 and their associated margins of error (MOE) \citep[see][for ACS MOEs]{moeACS}. The estimates are defined over counties in Florida, and white areas represent counties that do not have ACS poverty estimates associated with it. We fit this dataset with the spatial-only version of the P-MSTM, given by,
\begin{align}\label{specialcase2}
\nonumber
&\mathrm{Data\hspace{5pt}Model:}\hspace{5pt} Z_{1}^{(1)}(A)\vert \bfbeta,\bm{\eta}_{1}, \xi_{1}^{(1)}(A) \ind \mathrm{Pois}\{\mathrm{exp}(\textbf{x}_{1}^{(1)}(A)^{\prime}\beta + \bm{\psi}_{1}^{(1)}(A)^{\prime}\bm{\eta}_{1} + \xi_{1}^{(1)}(A))\};\\
\nonumber
\nonumber
&\mathrm{Process\hspace{5pt}Model\hspace{5pt}1:}\hspace{5pt} \bm{\eta}_{1}\vert \sigma_{\mathrm{K}} \sim \mathrm{MLG}\left(\bm{0}, \textbf{W}_{1,k}^{1/2},{\alpha}_{k}\bm{1}_{r},{\kappa}_{k}\bm{1}_{r}\right);\\
\nonumber
&\mathrm{Process\hspace{5pt}Model\hspace{5pt}2:}\hspace{5pt} \bm{\xi}_{1}\vert \sigma_{\xi} \sim \mathrm{MLG}\left(\bm{0}, \alpha_{\mathrm{n}}^{k/2}\sigma_{\xi}\textbf{I}_{n_{1}},{\alpha}_{k}\bm{1}_{n_{1}},{\kappa}_{k}\bm{1}_{n_{1}}\right);\\
\nonumber
&\mathrm{Parameter\hspace{5pt}Model\hspace{5pt}1:}\hspace{5pt} \bm{\beta}\sim \mathrm{MLG}\left(\bm{0}_{p,1}, \alpha_{\mathrm{n}}^{k/2}\sigma_{\beta}\textbf{I}_{p},\alpha_{k}\bm{1}_{p},\kappa_{k}\bm{1}_{p}\right);\\
\nonumber
&\mathrm{Parameter\hspace{5pt}Model\hspace{5pt}2:}\hspace{5pt} f(\sigma_{\mathrm{K}}) = \frac{1}{U_{\mathrm{K}}};\hspace{10pt} \sigma_{\mathrm{K}}=a_{1}^{(\mathrm{K})},\ldots,a_{U_{\mathrm{K}}}^{(\mathrm{K})}\\
&\mathrm{Parameter\hspace{5pt}Model\hspace{5pt}3:}\hspace{5pt} f(\sigma_{\xi})  = \frac{1}{U_{\mathrm{\xi}}};\hspace{10pt}\sigma_{\xi}=a_{1}^{(\xi)},\ldots,a_{U_{\mathrm{\xi}}}^{(\xi)}, \hspace{10pt}A \in D_{\mathrm{O},1}^{(1)},
\end{align}
\noindent
where we let $a_{1}^{(\mathrm{K})} = a_{1}^{(\mathrm{\xi})} = 0.01, a_{2}^{(\mathrm{K})} = a_{2}^{(\mathrm{\xi})} = 0.02,\ldots,a_{U_{\mathrm{K}}}^{(\mathrm{K})} = a_{U_{\mathrm{\xi}}}^{(\xi)} = 10$. Let $\bm{\psi}_{1}^{(1)}(\cdot)$  be the MI basis function, where the weight matrix is set equal to the immediate adjacency matrix (see Appendix B). Notice that $\bm{\psi}_{t}^{(\ell)}$ can be any $r$-dimensional real-valued vector (not necessarily the MI basis functions). For illustration, let $\textbf{x}_{1}^{(1)}(A)\equiv 1$ and let $r = 7$ (roughly 10$\%$ of the available basis functions). From Proposition 3, we have the following full conditional distributions associated with the model in (\ref{specialcase2}),
  \begin{align}
  \nonumber
   & f(\bm{\beta}\vert \cdot)= \mathrm{mMLG}(\textbf{C}_{\beta,k},(\bz^{\prime},\alpha_{k}\bm{1}_{p}^{\prime})^{\prime},\textbf{c}_{\beta,k})\\
  \nonumber
  & f(\bm{\eta}_{1}\vert \cdot)= \mathrm{mMLG}(\textbf{C}_{\eta,1,k},(\bz^{\prime},\alpha_{k}\bm{1}_{r}^{\prime})^{\prime},\textbf{c}_{\eta,1,k})\\
  \nonumber
    & f(\bm{\xi}_{1}\vert \cdot)= \mathrm{mMLG}(\textbf{C}_{\xi,1,k},(\bz_{1}^{\prime},\alpha_{k}\bm{1}_{n_{1}}^{\prime})^{\prime},\textbf{c}_{\xi,1,k}),\\
    \nonumber
& f(\sigma_{\mathrm{K}}\vert \cdot) = p_{\mathrm{K}}(\sigma_{\mathrm{K}});\hspace{10pt} \sigma_{\mathrm{K}}=a_{1},\ldots,a_{U_{\mathrm{K}}},\\
  \nonumber
& f(\sigma_{\xi}\vert \cdot)= p_{\xi}(\sigma_{\xi});\hspace{10pt} \sigma_{\xi}=a_{1},\ldots,a_{U_{\mathrm{\xi}}},
  \end{align}
  \noindent
which can be used within Algorithm 1.  Since both the simulation study in Section~4.1 and the real data analysis in Section~4.2 suggested that sMLG resulted in improved prediction performance, we use the sMLG specification. The, we ran Algorithm 1 for 10,000 iterations and visually inspected trace plots to assess convergence. In 8(c) and (d), we display the posterior mean and the width of the 95$\%$ credible interval computed using the Gibbs sampler in Algorithm 1. In general, the predictions reflect the general pattern of the data. Furthermore, the estimates have complete spatial coverage and upon comparison of Figure 8(b) to 8(d) we see that we have produced MOEs on a similar order of magnitude as the ACS 3-year period estimates.

%
%

\baselineskip=14pt \vskip 4mm\noindent

\bibliographystyle{jasa}  
\bibliography{myref}

\end{document}

%% file: notations4.tex
\def\bz{\textbf{Z}}

\def\by{\textbf{Y}}

\def\bfbeta{\bm{\beta}}

%% file: PMSTM_arXiv.bbl
\begin{thebibliography}{63}
\newcommand{\enquote}[1]{``#1''}
\expandafter\ifx\csname natexlab\endcsname\relax\def\natexlab#1{#1}\fi

\bibitem[\protect\citename{Abowd et~al., }2013]{abowdlmm}
Abowd, J., Schneider, M., and Vilhuber, L. (2013).
\newblock \enquote{Differential privacy applications to Bayesian and linear
  mixed model estimation.}
\newblock {\em Journal of Privacy and Confidentiality\/}, 5, 73--105.

\bibitem[\protect\citename{Abowd et~al., }2009]{abowd}
Abowd, J., Stephens, B., Vilhuber, L., Andersson, F., McKinney, K., Roemer, M.,
  and Woodcock, S. (2009).
\newblock \enquote{The LEHD infrastructure files and the creation of the
  Quarterly Workforce Indicators.}
\newblock In {\em Producer Dynamics: New Evidence from Micro Data\/}, eds.
  T.~Dunne, J.~Jensen, and M.~Roberts,  149--230. Chicago: University of
  Chicago Press for the National Bureau of Economic Research.

\bibitem[\protect\citename{Allegretto et~al., }2013]{irle}
Allegretto, S., Dube, A., Reich, M., and Zipperer, B. (2013).
\newblock \enquote{Credible research designs for minimum wage studies.}
\newblock In {\em Working Paper Series\/},  1--63. Institute for Research on
  Labor and Employment.

\bibitem[\protect\citename{Anderson, }1958]{anderson}
Anderson, T. (1958).
\newblock {\em Introduction to Multivariate Statistical Analysis\/}.
\newblock Canada: Wiley and Sons.

\bibitem[\protect\citename{Banerjee et~al., }2015]{banerjee-etal-2004}
Banerjee, S., Carlin, B.~P., and Gelfand, A.~E. (2015).
\newblock {\em Hierarchical Modeling and Analysis for Spatial Data\/}.
\newblock London, UK: Chapman and Hall.

\bibitem[\protect\citename{Banerjee et~al., }2008]{banerjee}
Banerjee, S., Gelfand, A.~E., Finley, A.~O., and Sang, H. (2008).
\newblock \enquote{Gaussian predictive process models for large spatial data
  sets.}
\newblock {\em Journal of the Royal Statistical Society, Series B\/}, 70,
  825--848.

\bibitem[\protect\citename{Bradley et~al., }2015{\natexlab{a}}]{bradleyMSTM}
Bradley, J., Holan, S., and Wikle, C. (2015{\natexlab{a}}).
\newblock \enquote{Multivariate spatio-temporal models for high-dimensional
  areal data with application to Longitudinal Employer-Household Dynamics.}
\newblock {\em The Annals of Applied Statistics\/},  To Appear.

\bibitem[\protect\citename{Bradley et~al., }2015{\natexlab{b}}]{bradleyMSTM2}
--- (2015{\natexlab{b}}).
\newblock \enquote{Multivariate spatio-temporal models for high-dimensional
  areal data with application to Longitudinal Employer-Household Dynamics.}
\newblock {\em The Annals of Applied Statistics\/},  To Appear, arXiv preprint:
  1503.00982.

\bibitem[\protect\citename{Bradley et~al., }2015{\natexlab{c}}]{bradleyPCOS}
Bradley, J., Wikle, C., and Holan, S. (2015{\natexlab{c}}).
\newblock \enquote{Bayesian spatial change of support for count-valued survey
  data.}
\newblock {\em Journal of the American Statistical Association\/}, forthcoming.

\bibitem[\protect\citename{Bradley et~al., }2015{\natexlab{d}}]{bradleyCAGE}
--- (2015{\natexlab{d}}).
\newblock \enquote{Regionalization of multiscale spatial processes using a
  criterion for spatial aggregation error.}
\newblock {\em arXiv preprint: 1502.01974\/}.

\bibitem[\protect\citename{Bradley et~al., }2015{\natexlab{e}}]{bradleySTCOS}
--- (2015{\natexlab{e}}).
\newblock \enquote{Spatio-temporal change of support with application to
  American Community Survey multi-year period estimates.}
\newblock {\em arXiv preprint: 1508.01451\/}.

\bibitem[\protect\citename{Bradley et~al., }2014]{bradley2014_comp}
Bradley, J.~R., Cressie, N., and Shi, T. (2014).
\newblock \enquote{A comparison of spatial predictors when datasets could be
  very large.}
\newblock {\em arXiv preprint: 1410.7748\/}.

\bibitem[\protect\citename{Bradley et~al., }2015{\natexlab{f}}]{bradleyTEST}
--- (2015{\natexlab{f}}).
\newblock \enquote{Comparing and selecting spatial predictors using local
  criteria.}
\newblock {\em TEST\/}, 24, 1--28.

\bibitem[\protect\citename{Casella and Berger, }2002]{casellaBerger}
Casella, G. and Berger, R. (2002).
\newblock {\em Statistical Inference\/}.
\newblock Pacific Grove, CA: Duxbury.

\bibitem[\protect\citename{Cressie and Johannesson, }2008]{johan}
Cressie, N. and Johannesson, G. (2008).
\newblock \enquote{Fixed rank kriging for very large spatial data sets.}
\newblock {\em Journal of the Royal Statistical Society, Series B\/}, 70,
  209--226.

\bibitem[\protect\citename{Cressie and Wikle, }2011]{cressie-wikle-book}
Cressie, N. and Wikle, C.~K. (2011).
\newblock {\em Statistics for Spatio-Temporal Data\/}.
\newblock Hoboken, NJ: Wiley.

\bibitem[\protect\citename{Crooks, }2015]{Crooks}
Crooks, G. (2015).
\newblock \enquote{The Amoroso distribution.}
\newblock {\em arXiv preprint: 1005.3274\/}.

\bibitem[\protect\citename{Datta et~al., }2015]{banerjeeNN}
Datta, A., Banerjee, S., Finley, A., and Gelfand, A. (2015).
\newblock \enquote{Hierarchical nearest-neighbor Gaussian process models for
  large geostatistical datasets.}
\newblock {\em arXiv preprint: 1406.7343\/}.

\bibitem[\protect\citename{Davis et~al., }2006]{supermarket}
Davis, E., Freedman, M., Lane, J., McCall, B., Nestoriak, N., and Park, T.
  (2006).
\newblock \enquote{Supermarket human resource practices and competition from
  mass merchandisers.}
\newblock {\em American Journal of Agricultural Economics\/}, 88, 1289--1295.

\bibitem[\protect\citename{De\hspace{5pt}Oliveira, }2013]{Oliveira}
De\hspace{5pt}Oliveira, V. (2013).
\newblock \enquote{Hierarchical Poisson models for spatial count data.}
\newblock {\em Journal of Multivariate Analysis\/}, 122, 393--408.

\bibitem[\protect\citename{Demirhan and Hamurkaroglu, }2011]{loggama}
Demirhan, H. and Hamurkaroglu, C. (2011).
\newblock \enquote{On a multivariate log-gamma distribution and the use of the
  distribution in the Bayesian analysis.}
\newblock {\em Journal of Statistical Planning and Inference\/}, 141,
  1141--1152.

\bibitem[\protect\citename{Diggle et~al., }1998]{diggle}
Diggle, P.~J., Tawn, J.~A., and Moyeed, R.~A. (1998).
\newblock \enquote{Model-based geostatistics.}
\newblock {\em Journal of the Royal Statistical Society, Series C\/}, 47,
  299--350.

\bibitem[\protect\citename{Dube et~al., }2013]{workpaper}
Dube, A., Lester, T., and Reich, M. (2013).
\newblock \enquote{Minimum wage, labor market flows, job turnover, search
  frictions, monopsony, unemployment.}
\newblock In {\em Working Paper Series\/},  1--63. Institute for Research on
  Labor and Employment.

\bibitem[\protect\citename{Finley et~al., }2010]{finley2}
Finley, A.~O., Banerjee, S., Waldmann, P., and Ericsson, T. (2010).
\newblock \enquote{Hierarchical spatial process models for multiple traits in
  large genetic trials.}
\newblock {\em Journal of the American Statistical Association\/}, 105,
  506--521.

\bibitem[\protect\citename{Finley et~al., }2009]{finley}
Finley, A.~O., Sang, H., Banerjee, S., and Gelfand, A.~E. (2009).
\newblock \enquote{Improving the performance of predictive process modeling for
  large datasets.}
\newblock {\em Computational Statistics and Data Analysis\/}, 53, 2873--2884.

\bibitem[\protect\citename{Ford and Fricker, }2009]{ford}
Ford, K. and Fricker, J. (2009).
\newblock {\em Real-Time SocioEconomic Data for Travel Demand Modeling and
  Project Evaluation. Publication FHWA/IN/JTRP-2008/22\/}.
\newblock Indiana Department of Transportation and Purdue University, West
  Lafayette, Indiana, doi: 10.5703/1288284314314.: Joint Transportation
  Research Program.

\bibitem[\protect\citename{Garthwaite et~al., }2010]{RM}
Garthwaite, P., Fan, Y., and Sisson, S. (2010).
\newblock \enquote{Adaptive optimal scaling of Metropolis-Hastings algorithms
  using the Robbins-Monro process.}
\newblock {\em arXiv preprint: 1006.3690\/}.

\bibitem[\protect\citename{Gelman and Rubin, }1992]{gr1}
Gelman, A. and Rubin, D. (1992).
\newblock \enquote{Inference from iterative simulation using multiple
  sequences.}
\newblock {\em Statistical Science\/}, 7, 473--511.

\bibitem[\protect\citename{Glaeser, }1992]{Glaeser1}
Glaeser, E. (1992).
\newblock \enquote{Is there a new urbanism? The growth of U.S. cities in the
  1990s.}
\newblock {\em Journal of Economic Perspectives\/}, 12, 139--60.

\bibitem[\protect\citename{Glaeser and Shapiro, }2009]{Glaeser2}
Glaeser, E. and Shapiro, J. (2009).
\newblock {\em Is there a new urbanism? The growth of U.S. cities in the
  1990s\/}.
\newblock Cambridge, MA: NBER Working Paper no. 8357, National Bureau of
  Economic Research.

\bibitem[\protect\citename{Griffith, }2000]{griffith2000}
Griffith, D. (2000).
\newblock \enquote{A linear regression solution to the spatial autocorrelation
  problem.}
\newblock {\em Journal of Geographical Systems\/}, 2, 141--156.

\bibitem[\protect\citename{Griffith, }2002]{griffith2002}
--- (2002).
\newblock \enquote{A spatial filtering specification for the auto-Poisson
  model.}
\newblock {\em Statistics and Probability Letters\/}, 58, 245--251.

\bibitem[\protect\citename{Griffith, }2004]{griffith2004}
--- (2004).
\newblock \enquote{A spatial filtering specification for the auto-logistic
  model.}
\newblock {\em Environment and Planning A\/}, 36, 1791--1811.

\bibitem[\protect\citename{Griffith and Tiefelsdorf, }2007]{griffith2007}
Griffith, D. and Tiefelsdorf, M. (2007).
\newblock \enquote{Semiparametric filtering of spatial autocorrelation: The
  eigenvector approach.}
\newblock {\em Environment and Planning A\/}, 39, 1193--1221.

\bibitem[\protect\citename{Hobbs and Hooten, }2015]{Hooten}
Hobbs, N. and Hooten, M. (2015).
\newblock {\em Bayesian Models: A Statistical Primer for Ecologists\/}.
\newblock Princeton University Press.

\bibitem[\protect\citename{Holan and Wikle, }2015]{holan_glm}
Holan, S. and Wikle, C. (2015).
\newblock \enquote{Hierarchical dynamic generalized linear mixed models for
  discrete-valued spatio-temporal data.}
\newblock In {\em Handbook of Discrete--Valued Time Series\/}. To Appear.

\bibitem[\protect\citename{Hughes and Haran, }2013]{hughes}
Hughes, J. and Haran, M. (2013).
\newblock \enquote{Dimension reduction and alleviation of confounding for
  spatial generalized linear mixed model.}
\newblock {\em Journal of the Royal Statistical Society, Series B\/}, 75,
  139--159.

\bibitem[\protect\citename{Johnson and Wichern, }1999]{Johnson}
Johnson, R. and Wichern, D. (1999).
\newblock {\em Applied Multivariate Statistical Analysis, 3rd ed.\/}.
\newblock Englewood Cliffs, New Jersey: Prentice Hall, Inc.

\bibitem[\protect\citename{Jones et~al., }2006]{batch2}
Jones, G., Haran, M., Caffo, B., and Neath, R. (2006).
\newblock \enquote{Fixed-width output analysis for Markov chain Monte Carlo.}
\newblock {\em Journal of the American Statistical Association\/}, 101,
  1537--1547.

\bibitem[\protect\citename{Kotz et~al., }2000]{multgamma}
Kotz, S., Balakrishnan, N., and Johnson, N. (2000).
\newblock {\em Continuous Multivariate Distributions, Volume 1: Models and
  Applications\/}.
\newblock New York, NY: Wiley.

\bibitem[\protect\citename{Lee and Nelder, }1974]{Lee}
Lee, Y. and Nelder, J. (1974).
\newblock \enquote{Double hierarchical generalized linear models with
  discussion.}
\newblock {\em Applied Statistics\/}, 55, 129--185.

\bibitem[\protect\citename{Lehmann, }1999]{lehman}
Lehmann, E. (1999).
\newblock {\em Elements of Large-Sample Theory\/}.
\newblock New York, NY: Springer.

\bibitem[\protect\citename{Lindgren et~al., }2011]{lindgren-2011}
Lindgren, F., Rue, H., and Lindstr\"{o}m, J. (2011).
\newblock \enquote{An explicit link between Gaussian fields and Gaussian Markov
  random fields: The stochastic partial differential equation approach.}
\newblock {\em Journal of the Royal Statistical Society, Series B\/}, 73,
  423--498.

\bibitem[\protect\citename{McCullagh and Nelder, }1989]{glm-nelder}
McCullagh, P. and Nelder, J. (1989).
\newblock {\em Generalized Linear Models\/}.
\newblock London, UK: Chapman and Hall.

\bibitem[\protect\citename{Nychka et~al., }2014]{nychkaLK}
Nychka, D., Bandyopadhyay, S., Hammerling, D., Lindgren, F., and Sain, S.
  (2014).
\newblock \enquote{A multi-resolution Gaussian process model for the analysis
  of large spatial data sets.}
\newblock {\em Journal of Computational and Graphical Statistics\/},  DOI:
  10.1080/10618600.2014.914946.

\bibitem[\protect\citename{Porter et~al., }2013]{aaronp}
Porter, A., Holan, S.~H., and Wikle, C.~K. (2013).
\newblock \enquote{Small area estimation via multivariate Fay-Herriot models
  with latent spatial dependence.}
\newblock {\em Australian $\&$ New Zealand Journal of Statistics\/}, 75,
  15--29.

\bibitem[\protect\citename{Prentice, }1974]{Prentice}
Prentice, R. (1974).
\newblock \enquote{A log gamma model and its maximum likelihood estimation.}
\newblock {\em Biometrika\/}, 61, 539--544.

\bibitem[\protect\citename{Roberts, }1996]{batch1}
Roberts, G. (1996).
\newblock \enquote{Markov chain concepts related to sampling algorithms.}
\newblock In {\em Markov Chain Monte Carlo in Practice\/}, eds. W.~Gilks,
  S.~Richardson, and D.~Spiegelhalter,  45--57. Chapman and Hall, Boca Raton.

\bibitem[\protect\citename{Roberts and Tweedie, }2011]{Tweedie}
Roberts, G. and Tweedie, R. (2011).
\newblock \enquote{Exponential convergence of Langevin distributions and their
  discrete approximations.}
\newblock {\em Bernoulli\/}, 2, 341--363.

\bibitem[\protect\citename{Royle et~al., }1999]{royle1999}
Royle, J., Berliner, M., Wikle, C., and Milliff, R. (1999).
\newblock \enquote{A hierarchical spatial model for constructing wind fields
  from scatterometer data in the Labrador sea.}
\newblock In {\em Case Studies in Bayesian Statistics\/}, eds. C.~Gatsonis,
  R.~Kass, B.~Carlin, A.~Carriquiry, A.~Gelman, I.~Verdinelli, and M.~West,
  367--382. Springer New York.

\bibitem[\protect\citename{Rue et~al., }2009]{rue}
Rue, H., Martino, S., and Chopin, N. (2009).
\newblock \enquote{Approximate Bayesian inference for latent Gaussian models
  using integrated nested Laplace approximations.}
\newblock {\em Journal of the Royal Statistical Society, Series B\/}, 71,
  319--392.

\bibitem[\protect\citename{Sengupta et~al., }2012]{aritrajsm}
Sengupta, A., Cressie, N., Frey, R., and Kahn, B. (2012).
\newblock \enquote{Statistical modeling of MODIS cloud data using the Spatial
  Random Effects model.}
\newblock In {\em Proceedings of the Joint Statistical Meetings\/},
  3111--3123. Alexandria, VA: American Statistical Association.

\bibitem[\protect\citename{Shaby and Wells, }2011]{Shaby}
Shaby, B. and Wells, M. (2011).
\newblock \enquote{Exploring an adaptive Metropolis algorithm.}
\newblock In {\em Technical Report\/}. Department of Statistics: Duke
  University.

\bibitem[\protect\citename{Spiegelhalter et~al., }2002]{spiegel-2002}
Spiegelhalter, D.~J., Best, N.~G., Carlin, B.~P., and {Van Der Linde}, A.
  (2002).
\newblock \enquote{Bayesian measures of model complexity and fit.}
\newblock {\em Journal of the Royal Statistical Society, Series B\/}, 64,
  583--616.

\bibitem[\protect\citename{Stein, }2014]{steinr}
Stein, M. (2014).
\newblock \enquote{Limitations on low rank approximations for covariance
  matrices of spatial data.}
\newblock {\em Spatial Statistics\/}, 8, 1--19.

\bibitem[\protect\citename{Thompson, }2009]{thompson}
Thompson, J. (2009).
\newblock \enquote{Using local labor market data to re-examine the employment
  effects of the minimum wage.}
\newblock {\em Industrial and Labor Relations Review\/}, 63, 343--366.

\bibitem[\protect\citename{US\hspace{5pt}Census\hspace{5pt}Bureau,
  }2008]{moeACS}
US\hspace{5pt}Census\hspace{5pt}Bureau (2008).
\newblock \enquote{What General Data Users Need to Know.}
\newblock
  http://www.census.gov/content/dam/Census/library/publications/2008/acs/ACSGeneralHandbook.pdf.

\bibitem[\protect\citename{US\hspace{5pt}Census\hspace{5pt}Bureau,
  }2015]{3yearperiod}
--- (2015).
\newblock \enquote{Census Bureau statement on American Community Survey 3-Year
  statistical product.}
\newblock
  http://www.census.gov/programs-surveys/acs/news/data-releases/2014/release.html.

\bibitem[\protect\citename{Waller et~al., }1997]{stcar}
Waller, L., Carlin, B., Xia, H., and Gelfand, A. (1997).
\newblock \enquote{Hierarchical spatio-temporal mapping of disease rates.}
\newblock {\em Journal of the American Statistical Association\/}, 92,
  607--617.

\bibitem[\protect\citename{Wikle et~al., }2001]{wikle2001}
Wikle, C., Milliff, R., Nychka, D., and Berliner, L. (2001).
\newblock \enquote{Spatiotemporal hierarchical Bayesian modeling tropical ocean
  surface winds.}
\newblock {\em Journal of the American Statistical Association (Theory and
  Methods)\/}, 96, 382--397.

\bibitem[\protect\citename{Wikle and Anderson, }2003]{wikleanderson}
Wikle, C.~K. and Anderson, C.~J. (2003).
\newblock \enquote{limatological analysis of tornado report counts using a
  hierarchical Bayesian spatio-temporal model.}
\newblock {\em Journal of Geophysical Research-Atmospheres\/}, 108, 9005.

\bibitem[\protect\citename{Wolpert and Ickstadt, }1998]{wolpert}
Wolpert, R. and Ickstadt, K. (1998).
\newblock \enquote{Poisson/gamma random field models for spatial statistics.}
\newblock {\em Biometrika\/}, 85, 251--267.

\bibitem[\protect\citename{Wu et~al., }2013]{wuJABES}
Wu, G., Holan, S.~H., and Wikle, C.~K. (2013).
\newblock \enquote{Hierarchical Bayesian Spatio-Temporal Conway-Maxwell Poisson
  Models with Dynamic Dispersion.}
\newblock {\em Journal of Agricultural, Biological, and Environmental
  Statistics\/}, 18, 335--356.

\end{thebibliography}
